\documentclass[twocolumn]{aastex631}
\usepackage{amssymb}
\usepackage{amsmath}
\usepackage{xcolor}

\newcommand{\msun}{\mathrm{M_\odot}}
\newcommand{\rsun}{\mathrm{R_\odot}}

\begin{document}
\makeatletter
\let\frontmatter@title@above=\relax
\makeatother

\newcommand\lsim{\mathrel{\rlap{\lower4pt\hbox{\hskip1pt$\sim$}}
\raise1pt\hbox{$<$}}}
\newcommand\gsim{\mathrel{\rlap{\lower4pt\hbox{\hskip1pt$\sim$}}
\raise1pt\hbox{$>$}}}

\title{\Large A census of massive eclipsing binaries in the Whirlpool Galaxy}

\shorttitle{A census of massive eclipsing binaries in M51}
\shortauthors{Shariat et al.}

\author[0000-0003-1247-9349]{Cheyanne Shariat}
\affiliation{Department of Astronomy, California Institute of Technology, 1200 East California Boulevard, Pasadena, CA 91125, USA}

\author[0000-0002-6871-1752]{Kareem El-Badry}
\affiliation{Department of Astronomy, California Institute of Technology, 1200 East California Boulevard, Pasadena, CA 91125, USA}

\author[0000-0003-3747-1394]{Maude Gull}
\affiliation{Department of Astronomy, California Institute of Technology, 1200 East California Boulevard, Pasadena, CA 91125, USA}
\affiliation{The Observatories of the Carnegie Institution for Science, 813 Santa Barbara Street, Pasadena, CA 91101, USA}

\author[0000-0002-6442-6030]{Daniel~R.\ Weisz}\affil{Department of Astronomy, University of California, Berkeley, CA 94720-3411, USA}\affil{Miller Institute for Basic Research, University of California Berkeley, Berkeley, CA, 94720, USA}

\author[0000-0002-1590-8551]{Charlie Conroy}
\affiliation{Center for Astrophysics | Harvard \& Smithsonian, 60 Garden St, Cambridge, MA 02138, USA}

\author[0000-0002-1445-4877]{Alessandro Savino}
\affiliation{Department of Astronomy, University of California, Berkeley, CA 94720-3411, USA}

\correspondingauthor{Cheyanne Shariat}
\email{cshariat@caltech.edu}

\begin{abstract}
We present a census of massive eclipsing binary (EB) candidates in the Whirlpool Galaxy (M51), a face-on spiral galaxy at 7.5 Mpc with $Z\sim Z_\odot$. Using 34 epochs of red-optical Hubble Space Telescope (HST) photometry and a single epoch of broadband UV-optical photometry, we identify 173 massive EB candidates ($>5~{\rm M_\odot}$) with orbital periods $P_{\rm orb}$=1--30 days. The sample traces young stellar populations across the spiral arms of M51. Joint light-curve and spectral energy distribution fits for 136 systems yield binaries spanning $T_{\rm eff}=10{,}000$--$50{,}000$ K, $R=7$--$66~{\rm R_\odot}$, and masses of $5$--$120~{\rm M_\odot}$.
We use detailed injection-recovery simulations to quantify the survey's completeness and infer intrinsic properties of the massive binary population. At the population level, we infer a close-binary fraction of $f_{\rm close}=27.4\pm7.1\%$ for $2\leq P_{\rm orb}/{\rm d}<20$ and $0.1\leq q\leq1$. Over the same period range, the intrinsic period distribution is consistent with being log-uniform, ${\rm d}N/{\rm d}\log P_{\rm orb}\propto P_{\rm orb}^{\pi}$ with $\pi=0.29\pm0.32$. 
These demographics broadly match those of massive-star populations in the Milky Way, LMC, and SMC, with tentative evidence for a lower close-binary fraction at lower metallicity.
Across the $\sim0.2$--$1.1\,Z_\odot$ range spanned by these galaxies, the overall similarities suggest that the formation of close massive binaries depends only weakly on metallicity.
Our survey establishes EBs as a powerful tool for measuring massive-binary demographics beyond the Local Group and provides a rich panchromatic census of $\sim$100,000 luminous stars and variables in M51.
\end{abstract}

\section{Introduction}\label{sec:introduction}

Massive stars ($\gtrsim8~\msun$) live short yet consequential lives.  They dominate the UV and ionizing output of young stellar populations, enrich their surroundings through winds and supernovae, and leave behind neutron stars and black holes \citep[e.g.,][]{Langer12,Eldridge2022,Marchant2024}.  
Binary evolution is paramount to the evolution of massive stars.
Effectively all massive stars are born in binaries or higher-order multiples \citep[e.g.,][]{MoeDiStefano17,Offner2023}, and the majority have companions close enough to exchange mass during their evolution
\citep[e.g.,][]{Sana12,MoeDiStefano17,Moe25LMC,Sana25SMC}. 

Many of the transients and remnants associated with massive star evolution --- stripped-envelope supernovae, X-ray binaries, gravitational-wave mergers, and kilonovae --- require or are strongly shaped by binary interaction \citep[e.g.,][]{Fragos13,deMink2015,Tauris17,Abbott17,Metzger20}. 
What links this variety of phenomena is that they generally trace back to young massive stars in close binaries. 
Thus, to connect massive-star outcomes to their progenitors, it is essential to constrain the demographics of close binaries early in their lives (i.e., before or during the first interaction).  
Measuring the initial conditions -- close-binary fraction, orbital period distribution, component masses, and mass ratio distribution -- sets the context for interpreting the rates of Roche-lobe overflow, contact binaries, mergers, stripping, and various compact object populations formed \citep{DeMink14,Eldridge17,MoeDiStefano17,Eldridge2022,Henneco24}.

Studying massive star populations in the Milky Way is difficult. Massive stars are intrinsically rare 
because they are disfavored by the initial mass function and have relatively short evolutionary timescales. Building large Galactic samples requires extending searches to large -- and hence typically uncertain -- distances. 
Because we are observing from within the Galactic disk,
this usually requires looking through heavily extincted sightlines.
Massive stars are also born inside dusty star-forming regions and emit much of their flux in the UV, further worsening the impact of extinction and making these stars difficult to access from ground-based facilities.
Existing massive-star samples therefore combine spectroscopy, photometric variability, astrometry, and targeted follow-up over different volumes \citep[e.g.,][]{Mason2009,Sana12,Kobulnicky2014}. 
Although several surveys have carefully modeled these selection effects, samples with well-characterized selection functions remain relatively small, particularly for the most massive stars.

Studying massive stars in nearby galaxies, instead of the Milky Way, provides a natural workaround to several of these limitations. A single pointing can monitor a large stellar population that lies at an (effectively) common distance. The uniformity of such surveys makes it easier to cleanly characterize the selection function. In many cases, studies of external galaxies also do not require looking through the Milky Way's disk, thus diminishing foreground extinction. Given these benefits, the Magellanic Clouds have become local laboratories for massive-binary demographics.  
Homogeneous photometric and spectroscopic surveys in the LMC and SMC have measured close-binary fractions, period distributions, and mass ratio distributions for early-B and O-type stars \citep[e.g.,][]{MoeDiStefano13,Sana2013,Almeida2017,Shenar2024,Sana25SMC,Villasenor25BLOeM}, providing some of the cleanest current benchmarks for modeling massive star evolution. However, these galaxies have a lower mass, star formation rate, and average metallicity than the Milky Way; a comparable census in a massive, solar-metallicity spiral galaxy is still needed.

The Whirlpool Galaxy (M51) is a particularly well-suited target that shares many similarities with the Milky Way.
M51 is a face-on, grand-design spiral galaxy with stellar mass $M_\star=3\times10^{10}~\msun$, luminosity $L\approx L^\star$, recent star formation rate ${\rm SFR}=3-4~\msun~{\rm yr^{-1}}$, roughly solar metallicity across its star-forming disk, and importantly, a well-measured distance of $D=7.50\pm0.24$ Mpc \citep{Calzetti2005,Csornyei2023,Croxall2015, Wei2021}.
The face-on orientation makes internal reddening less severe, mitigating the challenge of looking through the Milky Way disk.
M51 is also large enough to contain a substantial population of young massive stars \citep[including a recent starburst $\sim400-500$~Myr ago;][]{MentuchCooper2012}, but compact enough on the sky that a large fraction of its luminous stellar population can be monitored with a single HST pointing.
Compared to the LMC, M51's current SFR is $\sim20\times$ larger, implying a better-sampled high-mass tail of the IMF and $\sim20\times$ more of the {\it most massive} stars.

\begin{figure}
    \centering
    \includegraphics[width=0.98\columnwidth]{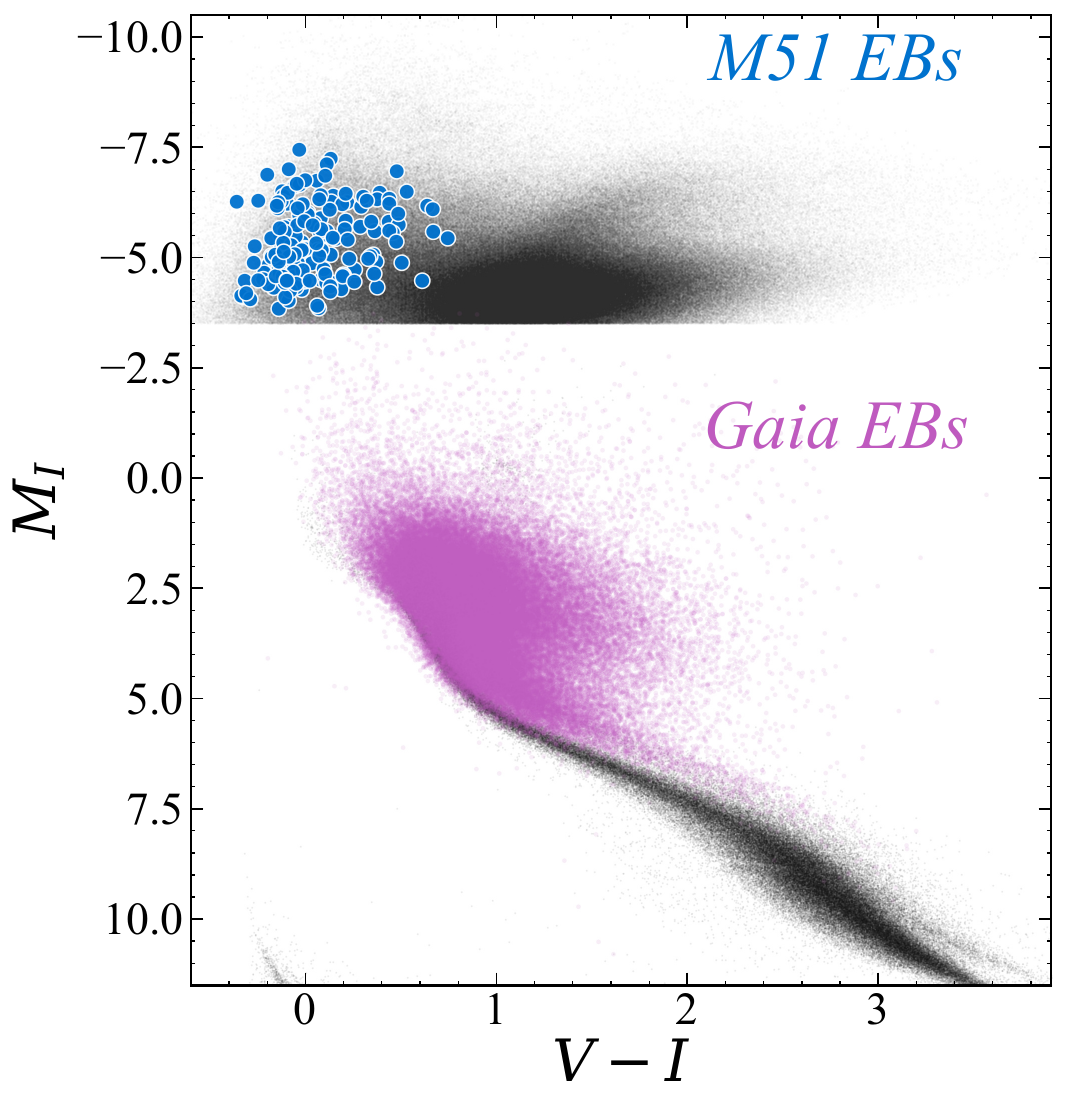}
    \caption{Optical color--absolute magnitude diagram comparing eclipsing binaries in M51 to those in the local {\it Gaia} sample.  The upper region shows luminous HST/ACS sources in M51 (black) and the EB candidates detected among them (blue, this work). The lower region shows nearby {\it Gaia} sources (black) and all eclipsing binary candidates with $\varpi/\sigma_{\varpi} \geq 10$ \citep[magenta;][]{Gaia2023,Mowlavi2023}. M51 EBs are exclusively among the most massive and luminous binaries relative to local samples.}
    \label{fig:m51_gaia_eb_cmd}
\end{figure}

\citet{Conroy18} studied the variable stars in M51 by obtaining $34$ epochs of ACS/WFC imaging in each of F606W and F814W over a $345$~d baseline, covering $\sim40\%$ of the integrated M51 F814W light (program 14704; PI Conroy). This produces light curves with up to $68$ photometric measurements per source across the galaxy.
At $7.5$ Mpc, luminous stars in the spiral arms are crowded on parsec scales, so stable, high-resolution space-based imaging is needed to measure the same sources repeatedly while mitigating blending.  Their catalog provides precise relative photometry for $73{,}000$ stars brighter than $M_{\rm I,814} = -5$ and sensitivity to variability amplitudes as low as $0.02$~mag \citep{Conroy18}.
These data revealed a large population of luminous variables across the color-magnitude diagram (CMD), including Cepheids, luminous blue variables, and eclipsing binary (EB) candidates.
More broadly, \citet{Conroy18} found that roughly half of the luminous stars in M51 show evidence for variability, with light curves spanning a wide range of amplitudes, timescales, and morphologies.

However, for the purposes of studying the massive star population, these data face a major limitation because they were obtained only in two red optical filters. For hot stars with $T_{\rm eff}\gtrsim10^4$~K, F606W and F814W sample the Rayleigh--Jeans tail of the SED, making it challenging to accurately measure temperatures, extinctions, radii, and bolometric luminosities.
We therefore supplement the ACS time series with a new single epoch of HST/WFC3 imaging (program 17200; PI El-Badry), which obtained photometry in 5 UV-optical filters in 2023: F275W, F336W, F475W, F606W, and F814W. 

In this work, we combine the optical ACS time series and the WFC3 broadband photometry to build a census of massive EB candidates in M51.
Compared to local EB samples, the EBs detected here are significantly more luminous and host exclusively massive stars (Figure~\ref{fig:m51_gaia_eb_cmd}).
Most importantly, we leverage the homogeneity of our survey to characterize the selection function of our catalog and constrain the {\it intrinsic} demographics of the massive star population.

The remainder of the paper is organized as follows. Section~\ref{sec:obs_reduction} describes the observations and photometric reduction.
Section~\ref{sec:EB_selection_modeling} outlines the EB selection, parameter inference, and selection function modeling. Section~\ref{sec:basic_properties} presents the observed properties of the EB candidates, and Section~\ref{sec:intrinsic_dem} presents the inferred intrinsic properties of close massive-star binaries. Section~\ref{sec:comparison} compares the M51 binary population to massive binaries in the Milky Way, LMC, and SMC. Section~\ref{sec:discussion} discusses the main limitations and the interpretation of the catalog. Finally, Section~\ref{sec:conclusion} summarizes the main conclusions.

\section{Observations and Data Reduction}\label{sec:obs_reduction}

\subsection{Observations}\label{sec:observations}

We use epoch imaging of the Whirlpool Galaxy (M51) from two different cameras on the Hubble Space Telescope (HST): the Advanced Camera for Surveys (ACS) and Wide Field Camera 3 (WFC3). The ACS imaging provides multi-epoch time-domain photometry used to produce light curves from which we search for periodic variables, including eclipsing binaries.  The WFC3 imaging provides a single epoch of broadband photometry from UV to red-optical wavelengths that helps constrain the spectral energy distribution (SED) of each source. 

The ACS time series data are from HST GO Cycle 24 program 14704 (PI: Conroy), presented in \citet{Conroy18}.  The program obtained $34$ visits between October 2016 and September 2017 in F606W and F814W filters (total baseline of 345 days).  Each visit used a standard 4-point dither pattern and a total exposure time of $2200$~s in each filter.  The cadence is pseudo-random, with visit separations chosen to preserve sensitivity to periods from days to months. The minimum and maximum separations between adjacent visits are $4$ and $24$~d \citep{Conroy18}.  The ACS footprint contains about $40\%$ of the integrated M51 F814W flux \citep{Conroy18}.  

The second set of observations is WFC3/UVIS imaging from HST GO program 17200 (PI: El-Badry) in Cycle 30, meant to supplement the earlier ACS time series.  These observations provide $1$ epoch in 2023 in five filters spanning the UV-to-red optical with four dithered exposures in each of F275W, F336W, F475W, F606W, and F814W, with total exposure times of 2468, 2466, 2481, 2492, and 2486~s, respectively.

\begin{figure*}
    \centering
    \includegraphics[width=0.98\textwidth]{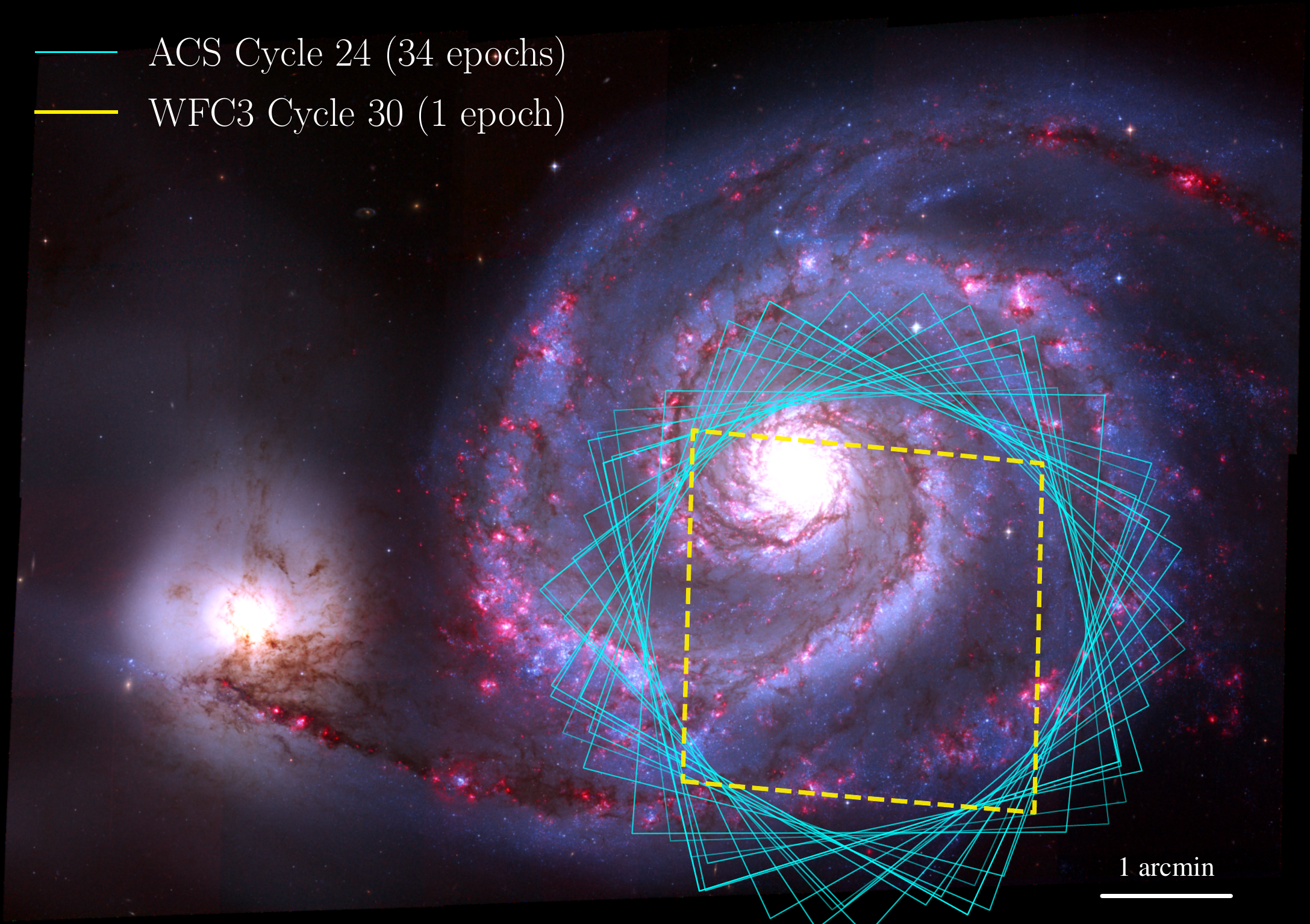}
    \caption{HST footprint of our M51 survey, shown atop an optical image of M51.  Cyan outlines show the 34 ACS/WFC visits that provide the F606W and F814W time series, and the yellow outline shows the Cycle 30 WFC3/UVIS pointing. 
    The background RGB image is reconstructed using F814W+F658N (R), F555W (G), and F435W (B) from the HST program of \citet{mutchler2005m51}.}\label{fig:m51_hst_footprint}
\end{figure*}

Figure~\ref{fig:m51_hst_footprint} shows the spatial footprint of the observations used in this paper, overlaid on an HST color image of M51. 
The background image is constructed from the Hubble Heritage ACS/WFC mosaic of M51
(program 10452, PI Beckwith; \citealt{mutchler2005m51})\footnote{\url{https://archive.stsci.edu/prepds/m51/}}, where F435W, F555W, and F814W+F658N serve as the blue, green, and red pseudocolors, respectively. 
Each image was scaled with an asinh stretch.
The cyan outlines show the 34 ACS epochs from Cycle 24 used for the time-domain search.  
The yellow outline shows the WFC3 Cycle 30 footprint. 
The new WFC3 image is roughly $85\%$ the area of each ACS image.

\subsection{Data Reduction}\label{subsec:data_reduction}

We measured the HST photometry in each camera/filter using DOLPHOT v3.1 \citep{Dolphin00,Dolphin16}, with the May 2026 ACS/WFC and WFC3/UVIS TinyTim 7.5 PSF  \citep{Krist1995} and pixel area map libraries.  The reduction follows the crowded field strategy used for previous resolved stellar population surveys \citep[e.g.,][]{Dalcanton12,Williams14,Conroy18,Weisz24}. We used CTE-corrected \texttt{flc} exposures for photometry and drizzled \texttt{drc} images only as a reference for astrometric alignment. We then ran the standard DOLPHOT pre-processing routines.
For ACS/WFC we ran \texttt{acsmask} (\texttt{wfc3mask} for WFC3/UVIS), \texttt{splitgroups}, and \texttt{calcsky} to mask bad pixels and apply the pixel area maps, split the image into its respective chips, and calculate the sky map, respectively \citep{Dolphin16}. 
In both cameras, the sky images were computed with inner and outer sky radii of 15 and 35 pixels (e.g., Table \ref{tab:dolphot_parameters}).

The DOLPHOT parameter choices are based on those adopted in the Panchromatic Hubble Andromeda Treasury (PHAT) survey \citep{Dalcanton12, Williams14}, which provides a tested standard for crowded field photometric reductions. From these defaults, we apply three main changes motivated by the new DOLPHOT 3.1 version and its specific application to the M51 time domain.  First, we use \texttt{SubResRef=1}, \texttt{SecondPass=1}, and \texttt{RCombine=1.5}. In DOLPHOT, \texttt{SubResRef} sets the subpixel grid used for reference-image source finding and source merging, \texttt{SecondPass} controls how many additional source-finding passes are made after subtracting the current model, and \texttt{RCombine} sets the radius within which candidate detections are combined.
Second, because our inputs are already CTE-corrected \texttt{flc} images, we disabled DOLPHOT's internal CTE corrections with \texttt{ACSuseCTE=0} and \texttt{WFC3useCTE=0}.  Third, we used the TinyTim-style PSF libraries, \texttt{ACSpsfType=0} and \texttt{WFC3UVISpsfType=0}, so that the production run used the latest PSFs provided in May 2026 (A. Dolphin, private communication). The full DOLPHOT parameters are listed in Table~\ref{tab:dolphot_parameters}.

\subsubsection{ACS time series photometry}\label{subsubsec:ACS_photometry}

We re-reduced the full ACS/WFC time series from HST program 14704 used by \citet{Conroy18}. The choice to re-reduce the data, rather than cross-match the new WFC3/UVIS sources with the old table, was made after cross-matches revealed ACS-WFC3 systematic magnitude offsets for the same sources in the same filter. Furthermore, since the original 2016 observations, DOLPHOT has experienced several quality upgrades\footnote{\url{http://americano.dolphinsim.com/dolphot/dolphot.pdf}}.
The data were retrieved from the MAST archive\footnote{\url{https://mast.stsci.edu/search/ui/\#/hst}} and consist of 34 visits in F606W and F814W, with four exposures per filter per visit.  After splitting the two ACS/WFC chips, this gives 272 chip-level science images in each filter.  To keep the DOLPHOT runs tractable, we reduced each filter into three batches containing 99, 99, and 74 chip-level science images.  The F606W batches used \texttt{jd8f01010\_f606w\_drc.chip1} as the reference image, and the F814W batches used \texttt{jd8f01020\_f814w\_drc.chip1}.
The F606W batches produced $661553$, $634935$, and $632566$ sources; the F814W batches produced $683901$, $658633$, and $652882$ sources.   The median alignment residuals are $0.10$ pixels for the three F606W batches and $0.08$--$0.09$ pixels for the three F814W batches, with maximum residuals of $0.10$--$0.13$ pixels. The total ACS photometry took $\sim350$ CPU hours to complete. $98\%$ of these sources overlap with the old \citet{Conroy18} catalog within $0.05\arcsec$.

\subsubsection{WFC3 photometry}\label{subsubsec:WFC3_photometry}

The supplementary WFC3/UVIS imaging is a single 2023 epoch from HST program 17200 (PI: El-Badry) in five filters spanning the UV--optical: F275W, F336W, F475W, F606W, and F814W.  Each filter has $4$ FLC exposures, or eight chip-level science images, so the five-filter DOLPHOT run contains \texttt{Nimg=40}.  The total exposure times are $2468$~s, $2466$~s, $2481$~s, $2492$~s, and $2486$~s in F275W, F336W, F475W, F606W, and F814W, respectively. We used the F606W drizzled image, \texttt{ieya01010\_f606w\_drc.chip1}, as the reference.

The WFC3 reduction simultaneously reduces all five filters.
It uses the same local sky, PSF fitting, aperture correction, and forced photometry settings as the ACS reductions.
This run chooses \texttt{PSFres=0}, which stops DOLPHOT from solving for the PSF residual image.  We adopted this setting after early runs showed that disabling residual PSF corrections preserved the alignment and source counts while improving the F814W ACS--WFC3 optical comparison relative to the \texttt{PSFres=1} run.  The adopted WFC3 run produced $423252$ sources.  The overall median alignment residual is $0.08$ pixels.  By filter, the optical residuals are $0.06$ pixels in F606W and $0.08$ pixels in F814W; the UV residuals are larger, $0.235$ pixels in F275W and $0.175$ pixels in F336W.  The total WFC3 photometry took $\sim20$ CPU hours to complete.
Table~\ref{tab:dolphot_parameters} lists the input parameters for the ACS and WFC3 production photometry.

\begin{deluxetable}{lc}
\tablecaption{DOLPHOT parameters used for the ACS/WFC and WFC3/UVIS photometric reductions.\label{tab:dolphot_parameters}}
\tablehead{
\colhead{Parameter} & \colhead{Value}
}
\startdata
\texttt{img\_RAper} & \texttt{3} \\
\texttt{img\_RChi} & \texttt{2.0} \\
\texttt{img\_RSky} & \texttt{15 35} \\
\texttt{img\_RPSF} & \texttt{10} \\
\texttt{img\_apsky} & \texttt{15 25} \\
\texttt{UseWCS} & \texttt{1} \\
\texttt{ACS/WFC3useCTE} & \texttt{0/0} \\
\texttt{PSFPhot} & \texttt{1} \\
\texttt{PSFPhotIt} & \texttt{2} \\
\texttt{FitSky} & \texttt{2} \\
\texttt{SkipSky} & \texttt{2} \\
\texttt{SkySig} & \texttt{2.25} \\
\texttt{SecondPass} & \texttt{1} \\
\texttt{SearchMode} & \texttt{1} \\
\texttt{SigFind} & \texttt{5.0/3.0} \\
\texttt{SigFindMult} & \texttt{0.85} \\
\texttt{SigFinal} & \texttt{5.0/3.5} \\
\texttt{MaxIT} & \texttt{25} \\
\texttt{NoiseMult} & \texttt{0.10} \\
\texttt{FSat} & \texttt{0.999} \\
\texttt{FlagMask} & \texttt{4} \\
\texttt{ApCor} & \texttt{1} \\
\texttt{Force1} & \texttt{1} \\
\texttt{Align} & \texttt{4} \\
\texttt{AlignTol} & \texttt{4} \\
\texttt{AlignStep} & \texttt{2} \\
\texttt{Rotate} & \texttt{1} \\
\texttt{RCentroid} & \texttt{1} \\
\texttt{PosStep} & \texttt{0.1} \\
\texttt{dPosMax} & \texttt{2.5} \\
\texttt{RCombine} & \texttt{1.5} \\
\texttt{SigPSF} & \texttt{3.0} \\
\texttt{PSFres} & \texttt{0} \\
\texttt{SubResRef} & \texttt{1} \\
\texttt{ACS/WFC3UVISpsfType} & \texttt{0/0} 
\enddata
\end{deluxetable}

\subsubsection{Catalog construction}\label{subsubsec:catalog_construction}

After running DOLPHOT, we converted the ACS source pixel positions to sky coordinates using the WCS coordinate system in the header of the reference image's {\tt .fits} file and merged the ACS batches into a common source table keyed by \texttt{m51\_source\_id} (which we adopt as the unique source identifier throughout this paper).  The WFC3 sources were then combined with the ACS catalog by performing a positional cross-match with a tolerance of $0.05\arcsec$ ($\sim1$ pixel). The median matched separations are $0.0093\arcsec$ in F606W and $0.0102\arcsec$ in F814W.

DOLPHOT reports several diagnostics for each source, including the PSF fit statistic $\chi$, shape parameters such as \texttt{sharp} and \texttt{round}, an object type, and the crowding parameter.  The crowding value is measured in magnitudes and quantifies how much brighter the source would have been if neighboring stars had not been fit simultaneously.
We apply different cuts for the ACS time series catalog and the single-epoch WFC3 catalog.  For ACS sources, we require object type 1 or 2, \texttt{SNR}$>5$, \texttt{crowd}$<0.4$, $\texttt{sharp}^2<0.1$, \texttt{mag}$<27.5$, and \texttt{flag}$\leq3$ in the relevant filter.  These cuts leave $1.9\times10^5$ F606W rows and $1.75\times10^5$ F814W rows per batch. 
For WFC3 sources, we make similar cuts to ACS, requiring object type 1 or 2, \texttt{SNR}$>4$, \texttt{crowd}$<0.6$, $\texttt{sharp}^2<0.15$, \texttt{magerr}$<0.5$ mag, \texttt{mag}$<29$, and \texttt{flag}$\leq3$ in each filter. These cuts leave $1.3\times10^5$ sources.

For the ACS light curves, we next apply an empirical correction to combat spatial and temporal magnitude offsets during the $345$~day observing baseline, following \citet{Conroy18}. These can be caused by low-level noise that varies over time and differently throughout the detector. In each ACS filter, high-quality stars (i.e., those passing the DOLPHOT quality cuts) are grouped into $200\times200$ pixel cells in the drizzled reference frame.  For each exposure chip, we compute the difference between the mean magnitude of a star and its magnitude in that exposure, take the median offset in each cell, and fit a quadratic function of time.  We require at least $1000$ total valid measurements in a cell, at least $20$ stars for an individual cell/exposure correction, and at least $8$ time points for the quadratic smoothing.  The smoothed correction is added to the DOLPHOT magnitude. To the DOLPHOT magnitude uncertainty, we also add $0.02$~mag in quadrature as a systematic floor \citep{Conroy18}.
The corrections have typical values of $0.013$--$0.024$~mag, with tails reaching roughly $-0.11$ to $+0.10$~mag in the most impacted cells.  These amplitudes are similar to the position- and time-dependent corrections found by \citet{Conroy18}. The correction is applied only to the ACS time series, since the WFC3 imaging consists of a single epoch.

\begin{figure*}
    \centering
    \includegraphics[width=0.99\textwidth]{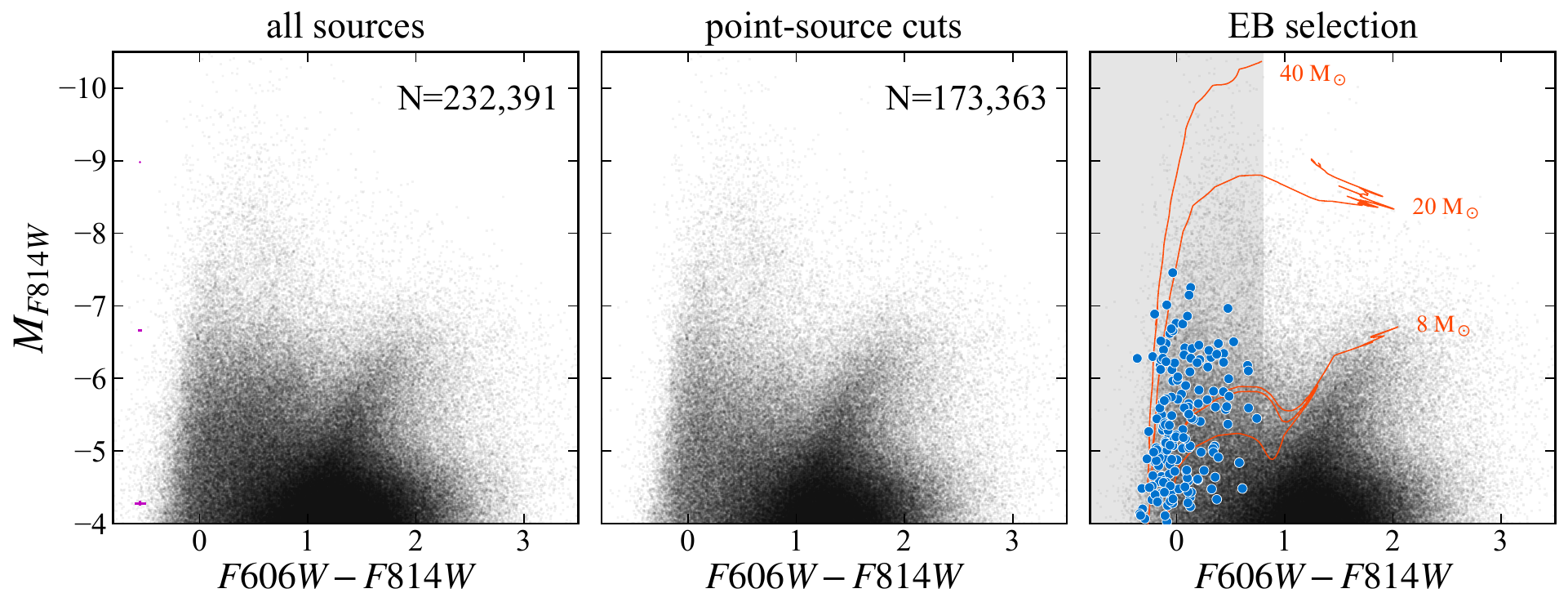}
    \caption{Selection of point sources from HST/ACS photometry. 
    {\bf Left:} all DOLPHOT sources identified in M51, with median photometric errors shown in purple crosses.
    {\bf Middle:} only sources that pass the point-source DOLPHOT quality cuts (outlined in Section \ref{subsec:data_reduction}). 
    {\bf Right:} the eclipsing binary candidates (blue) identified among these point sources. The gray shaded region shows the parent sample of blue sources from which we searched for EBs. We also overlay solar-metallicity MIST II evolutionary tracks for $8,20,40~\msun$ stars.  All magnitudes are corrected for foreground extinction.}
    \label{fig:cmd_selection_flow}
\end{figure*}

After this time series correction, we convert all ACS and WFC3 magnitudes reported by DOLPHOT into an infinite aperture system and correct for Galactic foreground extinction. The production DOLPHOT runs use \texttt{ApCor=1}, so DOLPHOT has already applied its image-level aperture corrections.  We then subtract Milky Way foreground extinction and add the scalar correction from the DOLPHOT aperture-corrected system to an infinite aperture.  The corrected magnitudes are
\begin{equation}
m_{\rm 0,inf}=m_{\rm input}+\Delta m_{\rm inf}-A_{\rm fg},
\end{equation}
where $m_{\rm input}$ is the reported apparent magnitude with only the aforementioned spatiotemporal corrections, $\Delta m_{\rm inf}$ is the infinite aperture correction, and $A_{\rm fg}$ is the foreground extinction.  
The adopted foreground extinctions are derived from the Schlafly--Finkbeiner dust map \citep{Schlafly11} using the CCM $R_V=3.1$ extinction law \citep{Cardelli1989}:
$A_{\rm F606W}=0.086$ and $A_{\rm F814W}=0.053$ mag for ACS, and
$A_{\rm F275W}=0.190$, $A_{\rm F336W}=0.152$,
$A_{\rm F475W}=0.111$, $A_{\rm F606W}=0.086$, and
$A_{\rm F814W}=0.055$ mag for WFC3.

The corresponding infinite-aperture terms are
$\Delta m_{\rm inf}=-0.096$ and $-0.098$ mag in ACS/F606W and ACS/F814W, respectively, using the ACS encircled energy corrections from \citet{Bohlin16}.  For WFC3/UVIS, we use the UVIS2 encircled energy corrections \citep{Calamida22,Medina22WFC3EE}, providing
$\Delta m_{\rm inf}=-0.136$, $-0.106$, $-0.086$, $-0.082$, and $-0.099$ mag in F275W, F336W, F475W, F606W, and F814W, respectively.
The reported photometric uncertainties are assumed to be unchanged by these scalar offsets. All final magnitudes are reported in the Vega system, which is the system adopted throughout the paper. 

\subsubsection{Summary}
Figure \ref{fig:cmd_selection_flow} summarizes the results of these DOLPHOT cuts on the parent sample in the ACS color-magnitude diagram (CMD). The first panel shows all DOLPHOT sources, the second applies the point-source DOLPHOT cuts in both F606W and F814W, and the third shows the luminous-blue parent region for our EB search and the selected EB candidates (see Section \ref{subsec:eb_selection}). We also show solar-metallicity MIST v2.5 evolutionary tracks \citep[][]{Dotter16, Choi16, Dotter2026} for $8$, $20$, and $40~\msun$ stars. To convert between apparent and absolute magnitudes, we adopt the M51 distance modulus $\mu=29.375$~mag, corresponding to $D=7.50\pm0.24$ Mpc \citep{Csornyei2023}. This differs slightly from the $\mu=29.67$~mag ($d=8.6\pm 0.1$~Mpc) adopted by \citet{Conroy18} using the TRGB distance of \citet{McQuinn17}.

The optical CMD already reveals the main stellar populations in our parent sample. 
The nearly vertical sequence near $F606W-F814W\approx0$ contains main-sequence and blue core-helium-burning stars.  The diagonal sequence at $1\lesssim F606W-F814W\lesssim2$ and $-7\lesssim M_{\rm F814W}\lesssim-4$ is dominated by red supergiants, while still redder sources are mostly AGB stars (e.g., compare to Figure \ref{fig:m51_gaia_eb_cmd}).  The catalog becomes increasingly incomplete at $M_{\rm F814W}\gtrsim-4$, especially for red sources.  
As emphasized by \citet{Conroy18}, F814W luminosity is not a direct proxy for initial mass among hot massive stars because the bolometric correction changes strongly during post-main-sequence evolution.  Massive stars can brighten by several magnitudes in F814W while evolving at nearly constant bolometric luminosity, whereas even very massive stars ($\gtrsim50~\msun)$ are comparatively faint in F814W on the main sequence since most of their flux is emitted in the UV.  Thus, sources brighter than roughly $M_{\rm F814W}\approx-6$ are generally evolved high-mass stars, which is relevant for interpreting the EB results. These degeneracies motivated us to obtain WFC3 UV photometry to study hot stars in M51.

We also note a source of contamination common to resolved stellar population studies beyond the Local Group: blending of unresolved sources. The angular resolution of the HST imaging corresponds to a physical scale of a few parsecs at the distance of M51: across the ACS and WFC3 filters used here, the PSF FWHM ranges from $\simeq0.067$--$0.09\arcsec$, or $\simeq2.4$--$3.3$ pc at M51's distance \citep[$7.5\pm0.24$~Mpc;][]{Csornyei2023}. The DOLPHOT point-source cuts also remove most sources extended on $\gtrsim1$~pc scales and obvious blends \citep{Thilker2022}. Moreover, HST studies of star clusters in M31 show that bright clusters have characteristic sizes of $1$--$10$ pc with only a few compact systems below $\sim1$ pc \citep{Johnson2012}, so compact clusters are unlikely to dominate the EB sample. 
Nonetheless, some candidates may include light from nearby stars, background galaxies, or bound tertiary companions; we discuss their implications in Section \ref{subsubsec:third_light_discussion}.

\vspace{-0.7cm}
\begin{deluxetable}{llr}
\tablecaption{Number of sources passing point-source quality cuts by camera and filter.}
\tablehead{
\colhead{Camera} & \colhead{Filter} & \colhead{$N$}
}
\startdata
ACS  & F606W & 189{,}520 \\
ACS  & F814W & 174{,}116 \\
WFC3 & F275W &  33{,}923 \\
WFC3 & F336W &  59{,}977 \\
WFC3 & F475W & 170{,}076 \\
WFC3 & F606W & 250{,}702 \\
WFC3 & F814W & 213{,}635 \\
\enddata
\end{deluxetable}

\subsection{Artificial Star Tests}\label{subsec:ASTs}

We use artificial star tests (ASTs) to estimate photometric completeness and empirical uncertainties of the final catalogs.  
ASTs are a standard tool in crowded-field photometry \citep[e.g.,][]{Stetson1987, Stetson1988}.
The ASTs inject artificial point sources with known input positions and magnitudes into the real science images using the same PSFs, then attempt to recover them with the same DOLPHOT setup used for the science photometry \citep[e.g.,][]{Dolphin00,Dolphin16,Weisz24}.
These tests are required because the formal uncertainties do not fully capture the effects of crowding, spatially varying background, and the catalog-level quality cuts.  The comparison between the input and recovered artificial stars gives an empirical measurement of photometric completeness, bias, and uncertainty.

\subsubsection{Setup}
We run ASTs for both the ACS and WFC3 images.  For ACS, we inject $10^5$ artificial stars into each F606W batch and $10^5$ artificial stars into each F814W batch, totaling $3\times10^5$ artificial stars per ACS filter.  For WFC3, we inject $10^5$ artificial stars into the five-filter production run.  The WFC3 artificial stars are assigned magnitudes in F275W, F336W, F475W, F606W, and F814W, so the same input catalog is used to characterize the completeness in all five WFC3 filters.  In total, the AST campaign contains $7\times10^5$ artificial stars injected into DOLPHOT.

\begin{figure}
    \centering
    \includegraphics[width=0.99\columnwidth]{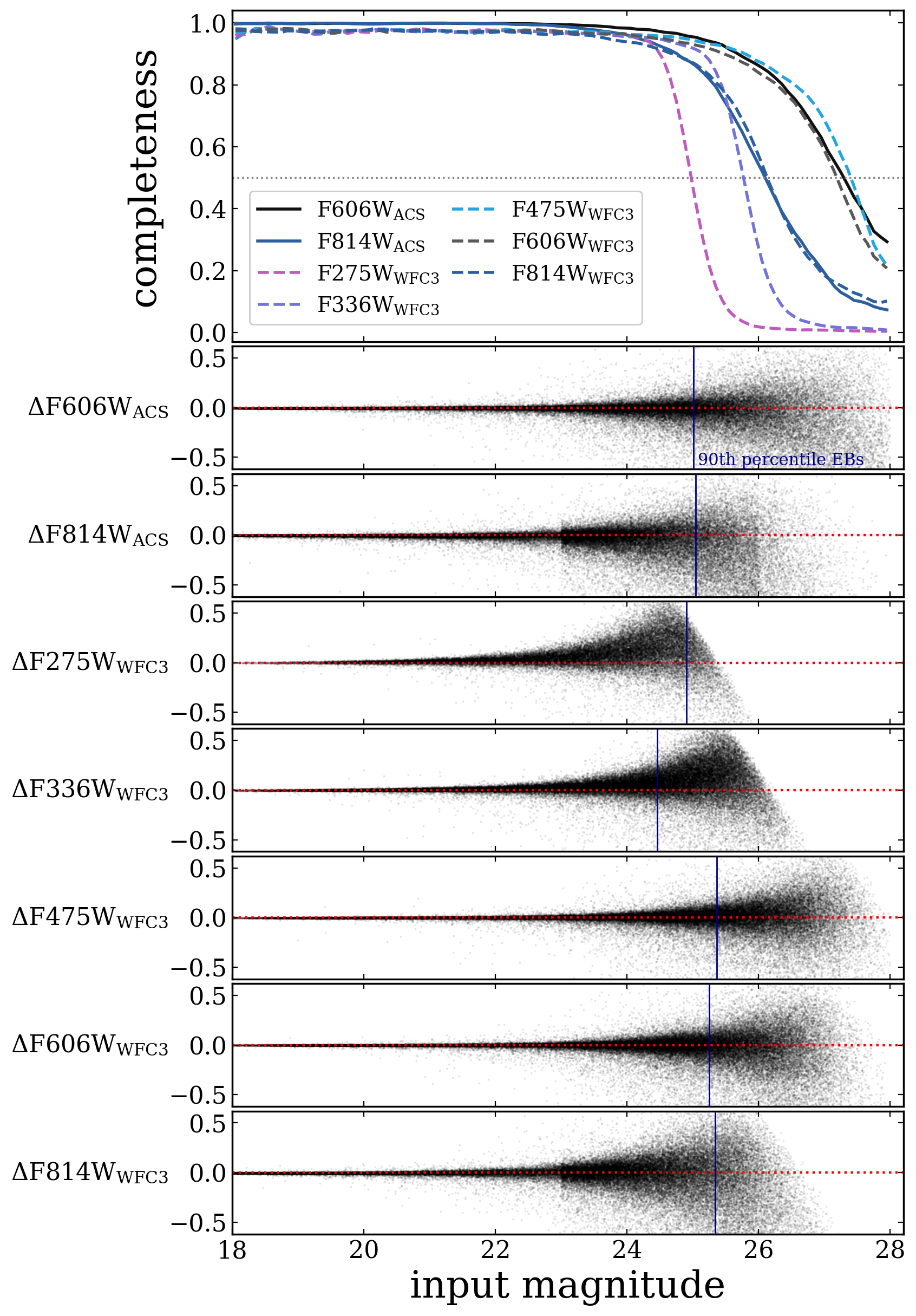}
    \caption{Completeness and photometric uncertainty for the M51 HST photometry computed from artificial star tests.  The top panel shows the completeness function for each filter.  The bottom panels show the difference between recovered and input magnitudes for ASTs that pass the same quality cuts applied to the science photometry. The 90th percentile magnitudes for EB candidates are marked in vertical lines.
    The ridge-like features are due to the input stellar catalog (see text).
    The $50\%$ completeness limits are $m_{\rm F606W}=27.10$ and $m_{\rm F814W}=25.96$ for ACS, and $m_{\rm F275W}=24.76$, $m_{\rm F336W}=25.56$, $m_{\rm F475W}=27.23$, $m_{\rm F606W}=27.05$, and $m_{\rm F814W}=25.87$ for WFC3.}
    \label{fig:AST_completeness}
\end{figure}

The input artificial star catalogs were designed to sample both the bright stars relevant for the eclipsing binary search and the fainter stars needed to define the edge of the completeness limit.  Input magnitudes span $m_{\rm input}=18$--$28$ mag, with $30\%$ of the artificial stars drawn from $18<m<23$, $50\%$ from $23<m<26$, and $20\%$ from $26<m<28$.  The input positions were drawn from a mixture of two spatial distributions: $55\%$ were placed at the location of real catalog sources, and $45\%$ were placed uniformly over the reference-frame footprint.  This sampling captures both the crowded regions where most of the massive-star population is located and the lower-density regions.  For ACS, the F606W and F814W artificial star catalogs are paired by sky position and have a defined input F606W$-$F814W color.  For WFC3, the input catalog contains five-filter SEDs drawn from the observed five-band SEDs.

\vspace{-0.8cm}
\begin{deluxetable*}{lccccccc}
\tablecaption{Summary of artificial star tests\label{tab:ast_summary}}
\tablehead{
\colhead{Camera} &
\colhead{$N_{\rm AST}$} &
\colhead{$m_{\rm input}$ range} &
\colhead{F275W} &
\colhead{F336W} &
\colhead{F475W} &
\colhead{F606W} &
\colhead{F814W} \\
\colhead{} &
\colhead{} &
\colhead{[mag]} &
\colhead{$m_{50}$} &
\colhead{$m_{50}$} &
\colhead{$m_{50}$} &
\colhead{$m_{50}$} &
\colhead{$m_{50}$}
}
\startdata
ACS/WFC   & $3\times10^5$ & 18--28 & \nodata & \nodata & \nodata & 27.10 & 25.96 \\
WFC3/UVIS & $1\times10^5$            & 18--28 & 24.76   & 25.56   & 27.23   & 27.05 & 25.87 \\
\enddata
\tablecomments{
$m_{\rm input}$ denotes the magnitude of the injected star and
$m_{50}$ indicates the magnitude where the fraction recovered (completeness) is $50\%$. 
}
\end{deluxetable*}

DOLPHOT injects one artificial star at a time to preserve the crowding, background, and noise properties of the original images.  This makes the ASTs computationally expensive: the total AST campaign required $\approx3000$ CPU hours, $\sim10\times$ longer than the photometric data reduction.
After injection, each artificial star is measured as if it were a real source.  After identifying recovered artificial star measurements, we applied the same quality cuts used to select stars in the science images (see Section \ref{subsec:data_reduction}).

For a given range of observable quantities (e.g., magnitude and position), we define the photometric completeness as
\begin{equation}
    {\rm completeness} = \frac{N_{\rm recovered}}{N_{\rm injected}}, \label{eq:completeness}
\end{equation}
where $N_{\rm injected}$ is the number of injected artificial stars in the bin and $N_{\rm recovered}$ is the number that are recovered {\it and} pass the corresponding DOLPHOT quality cuts.  The magnitude residual is then $\Delta m = m_{\rm rec}-m_{\rm input}$
for the recovered artificial stars.  The distribution of $\Delta m$ measures the empirical photometric scatter and bias, including the effects of crowding and blending.

\subsubsection{Completeness}

Figure~\ref{fig:AST_completeness} shows the AST completeness functions in each observed filter for both cameras. For ${\rm F606W}_{\rm ACS}$, ${\rm F606W}_{\rm WFC3}$, and ${\rm F475W}_{\rm WFC3}$, our sample is virtually complete down to 26th magnitude. In ${\rm F814W}_{\rm ACS}$ and ${\rm F814W}_{\rm WFC3}$, the sample is complete down to $\approx25.5$~mag. 
The $50\%$ completeness limits for optical filters are $m_{\rm F606W}=27.10$ and $m_{\rm F814W}=25.96$ for ACS, and $m_{\rm F475W}=27.23$, $m_{\rm F606W}=27.05$, and $m_{\rm F814W}=25.87$ for WFC3.  The HST UV bands are shallower, with 50\% completeness limits  $m_{\rm F275W}=24.76$ and $m_{\rm F336W}=25.56$, likely due to lower throughput\footnote{see \url{https://www.stsci.edu/hst/instrumentation/wfc3/performance/throughputs} for transmission curves}.

For unevolved single stars, the $50\%$ photometric completeness limits in the UV and blue-optical filters correspond to masses of approximately $17~\msun$, $14~\msun$, $13~\msun$, and $14~\msun$ in F275W, F336W, F475W, and F606W, respectively, assuming $E(B-V)=0.2$. Despite the UV bands having shallower imaging, they reach lower main-sequence masses because hot stars emit a larger fraction of their flux at UV wavelengths.
The WFC3/F814W limit corresponds to a higher unevolved main-sequence mass of approximately $29~\msun$, because hot main-sequence stars emit only a small fraction of their flux at red-optical wavelengths. Lower-mass stars are therefore recovered at the F814W limit primarily after evolving off the main sequence and brightening at red wavelengths.
These completeness limits are fainter than the CMD region occupied by most of the massive stars in our EB sample selected from ACS optical light curves (e.g., Figure~\ref{fig:multi_cmd_wfc3}). 
The main exception is for heavily extincted or cooler systems, where incompleteness in F275W and F336W becomes important.

The lower panels of Figure~\ref{fig:AST_completeness} show the difference between recovered and input magnitude for ASTs that pass the same quality cuts as the science catalog.  The residuals are centered near $\Delta m=0$ over the high-completeness regime, indicating little mean photometric bias for the stars that enter the catalog.  The scatter increases toward fainter magnitudes, as expected when sources become more affected by photon noise, crowding, and local background structure.  The optical filters retain small scatter over the magnitude range containing luminous massive stars, while the UV filters show a sharper loss of precision near their completeness limits.
The vertical lines mark the 90th percentile magnitude of the EB candidates (
25.01,
25.04,
24.91,
24.46,
25.36,
25.25,
25.34, from top to bottom), 
showing that they generally reside in the $\approx81$--$95\%$ recovery regions in all filters except F275W, where it falls to $\approx58\%$. Thus, the survey is relatively complete over most of the observed EB magnitude range.
A summary of our AST campaign and the photometric completeness is provided in Table~\ref{tab:ast_summary}.

\begin{figure*}
    \centering
    \includegraphics[width=0.98\textwidth]{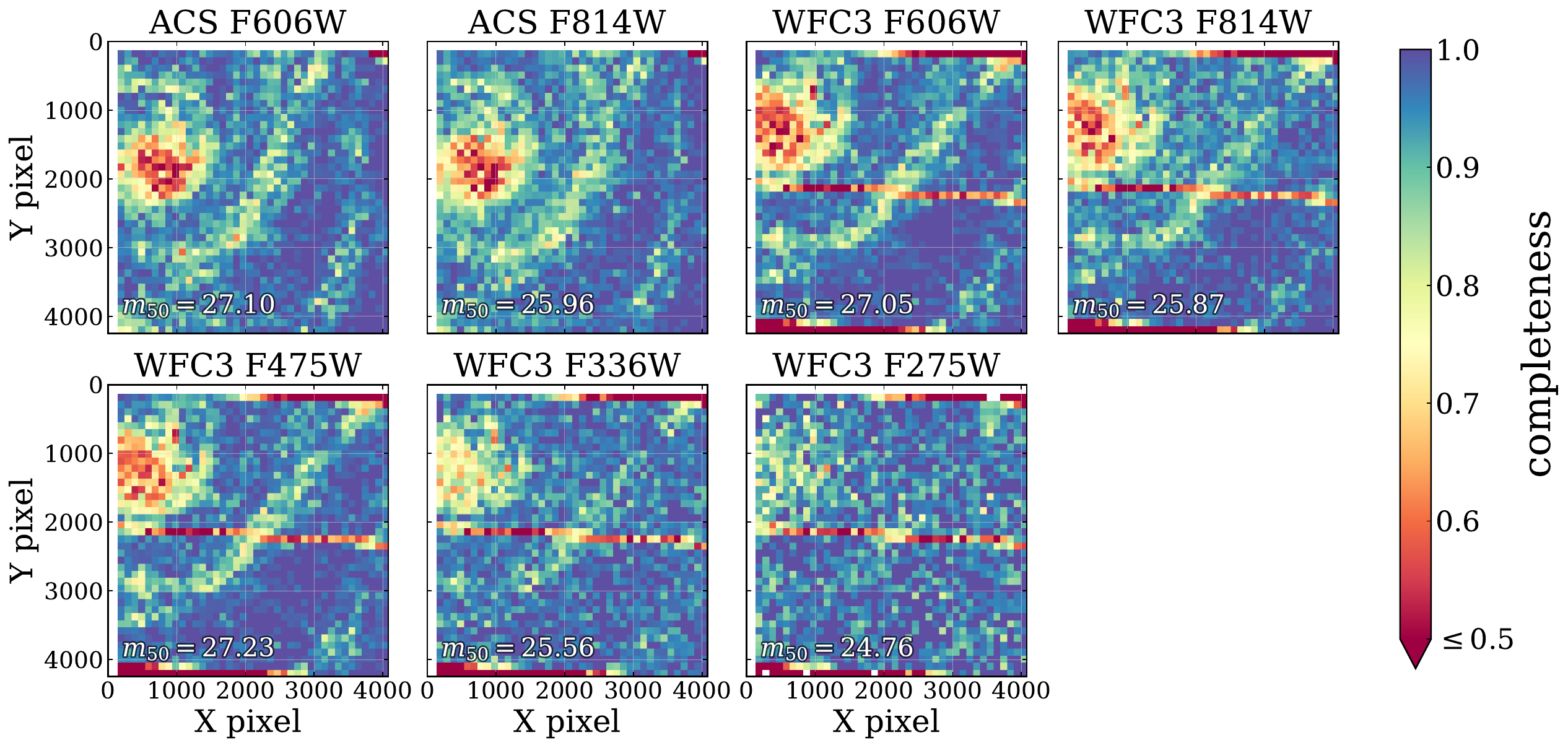}
    \caption{Spatial completeness maps derived from artificial star tests. For each filter, we show the recovered fraction (i.e., completeness) in $40\times40$ spatial bins among artificial stars brighter than that filter's $50\%$ completeness magnitude ($m_{\rm 50}$; Table \ref{tab:ast_summary}). The WFC3 panels also feature low-completeness regions associated with the chip gaps (middle line) and detector edges. 
    All panels share the same color scale, with X and Y pixels calculated from their respective reference images (Section \ref{subsec:data_reduction}).  The low-completeness regions are mainly associated with crowded regions, such as M51's spiral arms and galactic nucleus.}
    \label{fig:AST_spatial_completeness}
\end{figure*}

Figure~\ref{fig:AST_spatial_completeness} shows the photometric completeness across the M51 spatial footprint, as measured by ASTs.  For each filter, we divide the reference-frame coordinates into $40\times40$ spatial bins and compute the recovery fraction, $N_{\rm pass}/N_{\rm input}$, using only artificial stars with $m_{\rm input}\le m_{50}$ for that filter (Table \ref{tab:ast_summary}).  
The lowest-completeness regions generally coincide with the most crowded regions, where such systems fail the quality cuts used during catalog construction. Equivalently, the fraction of recovered systems decreases along the spiral arms and nucleus of M51, where crowding and background surface brightness are both higher. In the WFC3 filters, completeness also decreases due to chip gaps (middle line) and detector edges (corners).

\subsubsection{Calibrating uncertainties}

We use the AST residuals to calibrate the photometric uncertainties used for the point-source catalog. The formal DOLPHOT errors are smaller than the empirical scatter of recovered artificial stars, especially for faint and crowded sources \citep[][]{Dolphin00,Dolphin16,Williams14,Jang2023, Savino2024}.  For each science source and filter, we assign an AST-derived excess uncertainty using artificial stars matched in filter, magnitude, and DOLPHOT crowding.  We define
\begin{equation}
    \sigma_{\rm AST,extra}^2
    =
    \max\left(0, 
    \sigma_{\rm AST}^2
    -
    \widetilde{\sigma}_{\rm DOLPHOT}^2
    \right)
    ,
\end{equation}
where $\sigma_{\rm AST}=(P_{84}-P_{16})/2$ is the width of the AST magnitude residuals, and $\widetilde{\sigma}_{\rm DOLPHOT}$ is the median reported DOLPHOT uncertainty of the same matched ASTs.  The uncertainty assigned to a science measurement ($\sigma_{\rm phot}$) is then
\begin{equation}
    \sigma_{\rm phot}^2
    =
    \sigma_{\rm DOLPHOT}^2
    +
    \sigma_{\rm AST,extra}^2.
\end{equation}
This adds only the empirical scatter that is not already represented by the formal DOLPHOT errors.

We assign these calibrated uncertainties to the $1.60\times10^6$ measurements across the ACS and WFC3 catalogs. For the EB sample, the median calibrated uncertainties are $0.083$ and $0.141$ mag in ACS F606W and F814W, and $0.097$, $0.064$, $0.077$, $0.099$, and $0.175$ mag in WFC3 F275W, F336W, F475W, F606W, and F814W, respectively.  These are larger than the median formal catalog uncertainties by factors of $\sim1.5-9$ in the ACS time series filters and $\sim1.3$--$3.9$ in the WFC3 filters for the final EB sample. While the error inflation for ACS can be significant, the original formally reported errors were small (median $0.009$~mag).

In summary, our DOLPHOT ASTs quantify the completeness of the photometry and provide a method of reporting accurate photometric uncertainties. We use them to inflate the DOLPHOT-reported uncertainties and, in later sections, to quantify the selection function of our binary search and bias-correct the observed massive star binary population in M51.
The total selection function for our eclipsing binary search includes the DOLPHOT completeness (derived from ASTs) along with the biases associated with selecting eclipsing binaries from the ACS light curves (discussed in Section \ref{subsec:selection_function}).

\section{Eclipsing Binary Selection and Modeling}\label{sec:EB_selection_modeling}

\subsection{Eclipsing Binary Selection}\label{subsec:eb_selection}

We identify EB candidates from the ACS light curves in three steps.  We first search for periodic variability among sources with $\gtrsim20$ epochs of ACS photometry.  We then restrict the periodic sample to luminous blue sources in the ACS optical CMD, where massive binaries are expected to lie. Finally, we visually inspect the folded light curves to select systems with coherent, EB-like morphologies. We also visually inspect the folded light curves of periodic sources in redder regions of the CMD and find no additional convincing EB candidates.

\begin{figure*}
    \centering
    \includegraphics[width=0.95\textwidth]{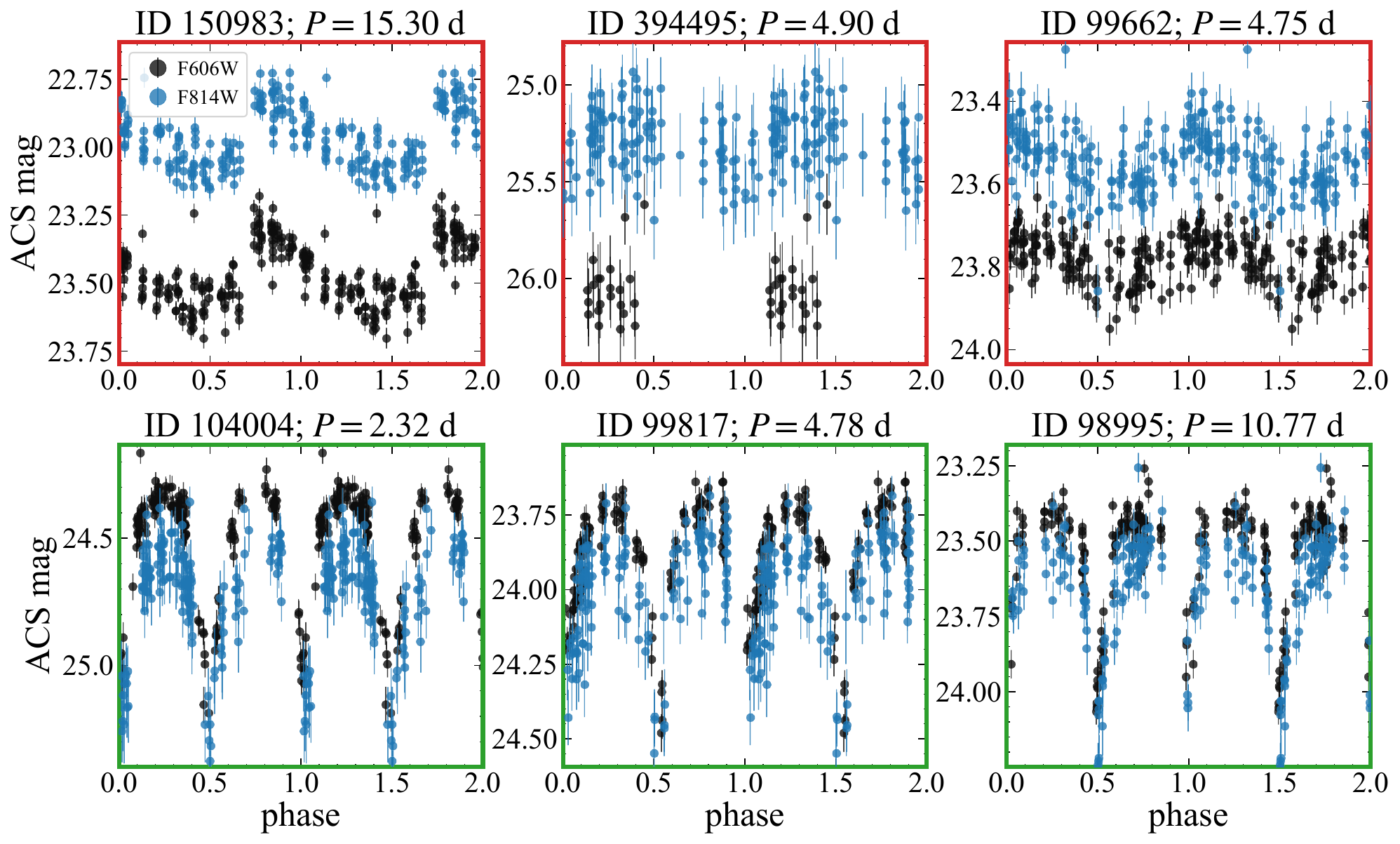}
    \caption{Examples of rejected and accepted EB candidate light curves. The top row shows HST/ACS light curves for three examples of rejected candidates. The first system shows sawtooth-like variability consistent with a pulsator, the second is likely an alias, and the third shows shallow sinusoidal variability. The bottom row shows three examples of selected EB candidate light curves at different periods. These show coherent, smooth variability with two apparent eclipses per orbital cycle.
    }
    \label{fig:visual_selection_examples}
\end{figure*}

We normalize and concatenate the F606W and F814W light curves before running the period search, using only measurements with uncertainties $<0.35$~mag. For true EBs, the periodic signal should appear coherently in both filters, so the combined light curve serves to strengthen the common periodic signal.  
We search $100{,}000$ uniformly spaced frequencies from $1/400$ to $1/1~{\rm d}^{-1}$ with a Lomb--Scargle periodogram \citep{Lomb76,Scargle82,VanderPlas18}.  We denote the resulting periodogram power by $z_{\rm LS}$ and retain sources with strong candidate periods, defined as $z_{\rm LS}>15$.

We then apply two cuts before visually inspecting each light curve. First, we only consider sources satisfying F606W$-$F814W$\leq0.8$ to exclude highly evolved stars and most of the instability strip (Figure \ref{fig:cmd_selection_flow}, right panel; \citealt{Conroy18}).  
Second, we only consider sources with best-fit periods in the range $1<P/{\rm d}<80$, above which the eclipse probability ($\propto P_{\rm orb}^{-2/3}$) becomes negligible. In practice, we find no eclipsing binaries with $P_{\rm orb}/{\rm d}>30$.
After these cuts, $2411$ sources remain.  We visually inspect each source folded on its top three periodogram peaks and common aliases.  We determine the orbital period of the binary as the one that produces two local minima per orbital phase, as expected for EBs where both stars contribute comparable flux. In practice, we find that the orbital period is usually twice the highest-power Lomb-Scargle period.

During visual inspection, we select EB candidates as sources showing smooth, eclipse-like variability with sufficient sampling throughout the orbital phase. We refer to the final set of high-confidence candidates retained after this review as the Gold EB sample. Figure~\ref{fig:visual_selection_examples} shows examples of rejected (top) and selected (bottom) candidates. Among the rejected candidates, \texttt{150983} shows sharp sawtooth-like modulation, consistent with being a Cepheid or other pulsator. \texttt{394495} does not phase coherently at the candidate period and is likely an alias, while \texttt{99662} shows shallow, approximately sinusoidal variability rather than discrete eclipses. 
We exclude such sources from the Gold sample because we cannot confidently identify eclipses. Some may still be contact or ellipsoidal binaries, but this loss is included in the visual-selection term of our selection function and therefore accounted for in the demographic analysis.

The bottom panel of Figure \ref{fig:visual_selection_examples} shows examples of selected EB candidates spanning a range of periods. \texttt{104004} is a short-period EB with large amplitude and coherent periodicity between the two filters, \texttt{99817} shows a cleaner primary/secondary eclipse pair at nearly half-phase separation, and \texttt{98995} is a longer-period object also showing clear eclipses.
We identify $N=173$ EB candidates in the Gold sample, whose light curves are shown in Appendix \ref{app:all_lightcurves}.

\subsection{Parameter Inference}\label{subsec:parameter_estimates}

For EB candidates with WFC3 photometry ($N=136$), we jointly fit the ACS light curves and WFC3 SEDs. For a given set of binary parameters, a normalized light curve is generated using \texttt{ellc} \citep{Maxted2016}.  The stellar fluxes are then derived using the sampled $T_{\rm eff}$, $\log g$, and $A_V$ with the \texttt{pystellibs}\footnote{\url{https://github.com/mfouesneau/pystellibs}} package using TLUSTY non-LTE atmosphere models \citep{Lanz2003, Lanz2007} where available and BaSeL LTE atmospheres \citep{Lejeune1997,Lejeune1998} for regions of the grid TLUSTY does not cover.
The models are reddened and integrated through the HST passbands using \texttt{pyphot}\footnote{\url{https://mfouesneau.github.io/pyphot/}} to generate mock photometry.
Both the light curves and SEDs are constructed with the same models: i.e., a sampled $T_1$, $T_2$, $R_1$, $R_2$, $A_V$, inclination, and mass ratio impact both the ACS eclipse morphology and the WFC3 UV--optical fluxes.

For each system, the orbital period is fixed, and the sampled parameters are
\begin{align}
\theta = \big(&M_1,\ q,\ R_1/a,\ R_2/R_1,\ \cos i,\ T_1,\ T_2/T_1,
\nonumber\\
&A_V,\ \phi_0,\ \Delta t_{\rm WFC3}\big).
\end{align}
Here $q=M_2/M_1$, $a$ is the semi-major axis, $i$ is the orbital inclination, $A_V$ is the extinction applied to the unresolved binary after subtracting the Milky-Way foreground contribution (Section~\ref{subsubsec:catalog_construction}), $\phi_0$ sets the phase of primary eclipse, and $\Delta t_{\rm WFC3}$ is a shared timing offset for the WFC3 epoch.  We include $\Delta t_{\rm WFC3}$ because the WFC3 photometry was obtained $\sim7$ years after the ACS time series, so small uncertainties in the adopted period can accumulate into a significant phase offset.  
This single parameter moves the WFC3 filters coherently in orbital phase relative to the ACS ephemeris, while the relative phase differences among the individual WFC3 filters are held fixed by their observation times reported in MJD. 
Note that the primary (subscript 1) and secondary (subscript 2) labels correspond to the hotter and cooler stars.

We precompute and store a grid of SEDs in $\log T_{\rm eff}$, $\log g$, and $A_V$.  For each point in this grid, we generate a spectrum at solar metallicity, redden it with a Cardelli--Clayton--Mathis (CCM) extinction law \citep{Cardelli1989}, and project it through the HST/ACS and HST/WFC3 passbands with {\tt pyphot}.  
For each filter $f$, \texttt{ellc} computes the normalized binary light curve using the sampled $R_1/a$, $R_2/a$, $i$, $q$, fixed quadratic limb darkening coefficients, and the surface brightness ratio implied by the SED model.  We fit each light curve using both {\tt roche} and {\tt sphere} shapes, then identify the best-fit model. The former allows for tidal distortion of the stars at close separations \citep{Maxted2016}.  We reject samples in which either star exceeds its Roche lobe, using the analytic Roche-lobe approximation of \citet{Eggleton1983}.

\begin{figure*}
    \centering
     \includegraphics[width=0.99\textwidth]{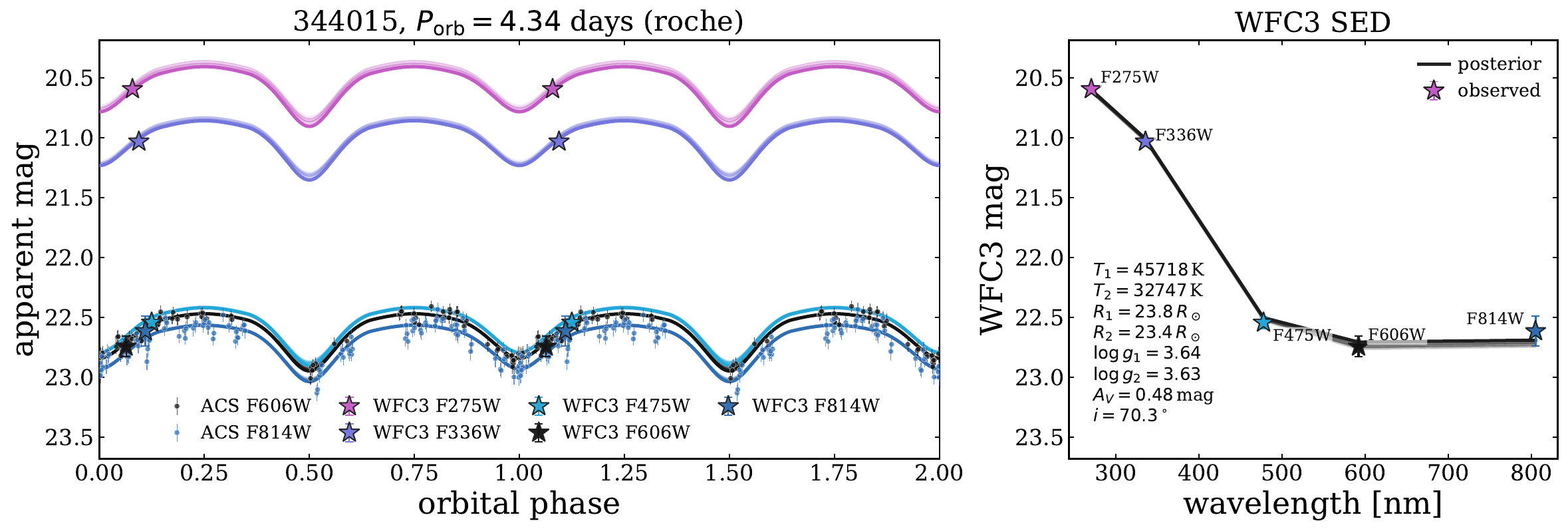}
    \caption{Example light curve and SED joint fit.  The left panel shows the ACS F606W/F814W light curves and phased WFC3 points folded on the fixed orbital period.  Thin curves show posterior draws, and thick curves show the maximum-posterior model.  The right panel shows the WFC3 SED evaluated with the same $T_1$, $T_2$, $R_1$, $R_2$, $A_V$, inclination, and WFC3 phase offset used in the light curve model. }
    \label{fig:example_fit}
\end{figure*}

The model is evaluated at the actual observation times.  For the ACS data, the phase of an observation at time $t$ is
\begin{equation}
    \phi_{\rm ACS} =
    \left[\frac{t-T_{\rm ref}}{P_{\rm orb}}+\phi_0\right]\bmod 1 .
\end{equation}
For the WFC3 data, each filter is evaluated at its own exposure time, but all five filters share one timing offset from the ACS epoch,
\begin{equation}
    \phi_{\rm WFC3} =
    \left[\frac{t-T_{\rm ref}+\Delta t_{\rm WFC3}}{P_{\rm orb}}+\phi_0\right]\bmod 1 .
\end{equation}
The likelihood is computed by comparing observed to predicted magnitudes from the model.  For each ACS or WFC3 point $j$, the residual is $m_j-m_j^{\rm mod}$ and
\begin{equation}
\begin{aligned}
    \ln \mathcal{L}
    =
    -\frac{1}{2}
    \sum_{j\in{\rm ACS, WFC3}}
    \left(\frac{m_j-m_j^{\rm mod}}{\sigma_j}\right)^2.
\end{aligned}
\end{equation}
The uncertainties $\sigma_j$ are the AST-calibrated photometric uncertainties described in Section~\ref{subsec:ASTs}.  

\vspace{-0.3cm}
\begin{deluxetable}{ll}
\tablecaption{Priors and bounds for the joint light curve+SED fits\label{tab:ellc_fit_priors}}
\tablehead{
\colhead{Parameter} &
\colhead{Prior}
}
\startdata
$M_1$ & $p(M_1)\propto M_1^{-2.35}$; $5.5<M_1/{\rm M_\odot}<120$ \\
$q$ & $\mathcal{U}(0.08,2.00)$ \\
$R_1/a$ & $\mathcal{U}(0.035,0.55)$ \\
$R_2/R_1$ & $\mathcal{U}(0.18,2.50)$ \\
$\cos i$ & $\mathcal{U}(0,\cos30^\circ)$ \\
$T_1$ & $\mathcal{U}(9000\,{\rm K},45000\,{\rm K})$ \\
$T_2/T_1$ & $\mathcal{U}(0.30,2.00)$ \\
$A_V$ & $\mathcal{N}(0.35,0.85^2)$; $0<A_V<3.5$ \\
$\phi_0$ & $\mathcal{U}(0,1)$ \\
$\Delta t_{\rm WFC3}$ & $\mathcal{N}(0,(0.12P_{\rm orb})^2)$; $|\Delta t_{\rm WFC3}|<0.25P_{\rm orb}$ \\
\enddata
\end{deluxetable}

Component masses are required to fall within $5.5$--$120~{\rm M_\odot}$ with an IMF-informed mass prior \citep[][]{Kroupa01} applied to both stars, $p(M_1)p(M_2)\propto M_1^{-2.35}M_2^{-2.35}$, with a corresponding Jacobian to match the sampled variables $(M_1,q)$.  In the prior, we also include a weak consistency term that compares each component to MIST track points in $(M,T_{\rm eff},R)$, with broad widths of $0.2$ dex in $\log M$, $0.06$ dex in $\log T_{\rm eff}$, and $0.2$ dex in $\log R$.  This term suppresses unphysical combinations of temperature--radius--extinction but does not require both stars to be coeval. All prior choices are provided in Table~\ref{tab:ellc_fit_priors}.

For each target, we initialize the MCMC walkers broadly from the prior, and the posterior is then sampled with \texttt{emcee} \citep{ForemanMackey2013}.  The production run uses $144$ walkers and $1800$ steps, discarding the first $600$ steps. 
We fit all EB candidates with WFC3 photometry, yielding $N_{\rm fit}=136$ joint light-curve+SED fits. The $111$ systems with good fits ($\chi^2_{\rm reduced}<3$) are used for the population analysis. The remaining systems do not yield good fits for the eclipse profiles and SED simultaneously, which could reflect contamination from third light or limitations of the adopted binary and stellar-atmosphere models (see Section \ref{subsubsec:third_light_discussion} and \ref{subsubsec:models_discussion}).

Figure~\ref{fig:example_fit} shows an example joint fit result. The optical light curve provides reasonably good constraints on the eclipse morphology, including the relative radii, inclination, and phase of primary eclipse, while the WFC3 photometry constrains the temperatures, radii, and extinction through the UV--optical SED.  The WFC3 points are evaluated at their observed phases after applying the fitted $\Delta t_{\rm WFC3}$ offset, so the same binary model matches both the folded light curve and the single-epoch SED.  For this system, the posterior draws cluster around the maximum-posterior model, whose parameters are: 
$T_1 = 32747~{\rm K}$,
$T_2 = 45718~{\rm K}$,
$R_1 = 23.4~\rsun$,
$R_2 = 23.8~\rsun$,
$\log g_1 = 3.63$,
$\log g_2 = 3.64$,
$A_V = 0.48~{\rm mag}$, and
$i = 70.3^\circ$.

\subsection{Selection Function}\label{subsec:selection_function}

To be selected, the EB candidates in our catalog must have satisfied (a) the point-source quality cuts in DOLPHOT photometry and (b) visual appearance as an EB in the ACS light curves. The selection function can be written as the product of these two terms,
\begin{equation}\label{eq:full_selection}
    \mathcal{P}_{\rm EB} = \mathcal{P}_{\rm phot}~\mathcal{P}_{\rm LC},
\end{equation}
where $\mathcal{P}_{\rm phot}$ is the probability that the unresolved source is detected by DOLPHOT and passes the catalog quality cuts, and $\mathcal{P}_{\rm LC}$ is the probability that its light curve is selected as an EB. 
The first term depends on position in M51, crowding, background, intrinsic luminosity, color, and the DOLPHOT cuts described in Section~\ref{subsec:data_reduction}.  $\mathcal{P}_{\rm phot}$ is already partially quantified by the artificial star tests in Section~\ref{subsec:ASTs}.  
The second term, $\mathcal{P}_{\rm LC}$, depends on the intrinsic binary properties (e.g., orbital period, inclination, stellar properties) and properties of the HST light curves (e.g., phase coverage, photometric noise, period aliases).
Here, we characterize $\mathcal{P}_{\rm LC}$ to model the full selection function ($\mathcal{P}_{\rm EB}$) and correct the observed EB sample for observational biases to infer properties of the intrinsic binary population.

To quantify $\mathcal{P}_{\rm EB}$, we perform injection--recovery tests using a synthetic binary population. The construction of this binary population is outlined in Appendix \ref{app:binary_pop}.
Drawing from this population ensures that the injected eclipse morphologies, radii, temperatures, and extinction values are tied to a physical binary population.  
The selection function is quantified by 
(a) injecting the synthetic binaries throughout M51, 
(b) simulating their HST observations, and
(c) determining those that pass the photometric selection ($\mathcal{P}_{\rm phot}$) and are recovered as EB candidates based on their mock HST/ACS light curves ($\mathcal{P}_{\rm LC}$).

\subsubsection{Photometric selection function }\label{subsubsec:selection_function_sphot}

For the simulated binary population, we calculate $\mathcal{P}_{\rm phot}$ by placing each system throughout M51 and identifying which are recovered.
We draw a position and internal extinction from the model described in Appendix~\ref{app:binary_pop}. Briefly, the spatial distribution assumes an exponential disk with a spiral-arm component. This places most simulated massive binaries along the spiral arms, where crowding and background levels are highest, without assigning them positions from already-detected EBs. The dust model accounts for the radial and vertical distributions of stars and dust in M51.

We then compute the unresolved HST photometry at $7.50$~Mpc and use the local artificial star tests to determine whether each source would be recovered and pass the same DOLPHOT quality cuts as the observed catalog.

A mock EB is considered photometrically recovered if it satisfies the same requirements used for the observed catalog: object type 1 or 2, ${\rm S/N}>5$, crowding $<0.4$ mag, ${\rm sharp}^{2}<0.1$, magnitude uncertainty $<0.5$ mag, recovered magnitude $<27.5$, and photometric flag $\leq3$. These cuts reject extended, blended, poorly measured, and strongly contaminated
detections and follow the DOLPHOT-based catalog construction described in Section~\ref{subsubsec:catalog_construction} \citep{Dolphin00,Dolphin16}. We require the same injected star to pass in both F606W and F814W. This spatial injection recovery quantifies $\mathcal{P}_{\rm phot}$.
From an intrinsic $42{,}556$ massive binaries ($m_{\rm F814W}<26$ and
$m_{\rm F606W}-m_{\rm F814W}<0.8$) with $P_{\rm orb}=1$--$30$~d, $10{,}579$ are recovered and enter the ACS point source catalog. The photometric recovery fraction across the full intrinsic population is therefore $24.9\%$.

\subsubsection{Light curve selection function}\label{subsubsec:selection_function_slc}
For each simulated binary that is recovered photometrically (e.g., passes $\mathcal{P}_{\rm phot}$), we 
now quantify $\mathcal{P}_{\rm LC}$ by determining which ones are identified as EBs based on their light curve.
We generate synthetic light curves for each synthetic binary with \texttt{ellc} and inject only those that are geometrically eclipsing:
\begin{equation}\label{eq:eclipse_criteria}
    \cos i < \frac{R_1+R_2}{a}.
\end{equation}
The fraction and properties of the non-eclipsing binaries in our mock sample are still considered in the selection function to study all close binaries in the period range. 

We inject the synthetic EB signals into the ACS light curves of real stars in the catalog, which we refer to as ``carrier'' stars.
We select carriers from ACS sources that pass the F606W and F814W photometric quality cuts (Section \ref{sec:obs_reduction}) and exclude known EBs.  To avoid injecting into strongly variable stars, we also require photometric amplitudes $A_{\rm F606W}<0.5$ mag and $A_{\rm F814W}<0.5$ mag.  
For each injection, we choose up to $10$ carrier stars that are close to the simulated source position and have similar photometry, requiring $\Delta{\rm F814W}<0.25$ mag, $\Delta({\rm F606W-F814W})<0.2$ mag, and a separation within $20\arcsec$ or, if too few carriers are available, within $80\arcsec$.
Choosing carriers from the same parent catalog allows us to reproduce the observing conditions of the EB search, such as the actual cadence, photometric uncertainties, crowding, and background.

For each synthetic EB--carrier pair, we add the phase-dependent EB variability to the carrier light curve. In each ACS filter, we evaluate the \texttt{ellc} model at the carrier observation times, subtract its median flux, and add the remaining flux variation to the observed carrier fluxes. Each injection is assigned a random orbital phase. We then pass the injected light curves through the same joint F606W+F814W Lomb--Scargle search used for the real sample, using $100{,}000$ trial frequencies between $1/1$ and $1/400~{\rm d}^{-1}$ (Section~\ref{subsec:eb_selection}). We count an injection as recovered if the strongest periodogram peak lies within $5\%$ of $P_{\rm orb}$, $P_{\rm orb}/2$, or $2P_{\rm orb}$, matching the real EB search where systems can be selected at either the orbital period or at the (more common) half-period alias.

Finally, we arrive at the most laborious part of characterizing the selection function: visually inspecting each injected source whose period is recovered.
As for the real EB search, a recovered period is not sufficient by itself; the folded light curve must also show an EB-like morphology with enough phase coverage to be classified confidently by the inspector (C.S.). In total, $29{,}000$ geometrically eclipsing systems were injected from the synthetic binary population (Appendix \ref{app:binary_pop}), of which $8689$ had periods recovered by the LSP search.  We divided the recovered injections into $1$~d period bins and inspected systems until each bin contained at least 20 accepted EBs.  
This required $1{,}769$ manual visual classifications. Of these, $533$ ($30.1\%$) were kept as EB-like and $1{,}236$ ($69.9\%$) were rejected.
The rejected injections were mainly low-inclination systems with shallow or ambiguous eclipse morphologies, or systems whose eclipses were poorly sampled by the ACS cadence. They also included binaries with periods near aliases, where the aliases were actually recovered.

For an intrinsic close-binary population that has already entered the photometric catalog, the selection function can be described by
\begin{equation}\label{eq:LC_selection}
    \mathcal{P}_{\rm LC}=p_{\rm geom}\,p_{\rm LSP}\,p_{\rm visual}.
\end{equation}
Here $p_{\rm geom}$ is the probability that a randomly oriented binary eclipses, $p_{\rm LSP}$ is the probability that a geometrically eclipsing system is recovered by the Lomb-Scargle period search, and $p_{\rm visual}$ is the probability that a recovered folded light curve is visually classified as an EB.  

\subsubsection{Summary}\label{subsubsec:selection_function_summary}

\vspace{-0.5cm}
\begin{deluxetable}{lrr}
\tablecaption{Selection cascade for the synthetic close-binary population with
$2\leq P_{\rm orb}/{\rm d}<20$.\label{tab:selection_cascade}}
\tablehead{
\colhead{Selection step} &
\colhead{$N_{\rm exp}$} &
\colhead{Retained}
}
\startdata
intrinsic binary population             & $95{,}635$ & \nodata \\
close binaries             & $31{,}828$ & $33.2\%$ \\
recovered in the photometric catalog & $8{,}608$  & $27.0\%$ \\
luminous-blue CMD region       & $5{,}434$  & $63.1\%$ \\
geometrically eclipsing              & $2{,}362$  & $43.5\%$ \\
period recovered                     & $848$    & $35.9\%$ \\
visually-selected EB       & $289$    & $34.0\%$
\enddata
\tablecomments{retained fractions are relative to the preceding row.}
\end{deluxetable}

Table~\ref{tab:selection_cascade} provides the expected number of binaries remaining at each selection step. 
From the initial 
$95{,}635$ massive binaries in our synthetic population, 
$31{,}828$ are in close orbits ($P_{\rm orb}=2$--$20$~d), 
$8{,}608$ are recovered as HST sources passing the DOLPHOT cuts,
$5{,}434$ are in the luminous-blue CMD region, 
$848$ have their periods recovered in the LSP search, 
and $289$ have their light curves visually selected as EBs.
Thus, we arrive at an estimate for the {\it full} selection fraction of our survey: only $0.9\%$ ($289/31{,}828$; Table \ref{tab:selection_cascade}) of all close binaries in M51 are identified as EBs in our HST sample.

We point out that $289$ is a factor of $1.7\times$ larger than the observed population of $\sim173$ EBs in our sample, which can be attributed to the simple assumptions made for the initial massive binary population, such as their total size, spatial distribution, dust extinction in M51, and binary physics. 
In Section \ref{sec:intrinsic_dem} we compare the properties of these simulated binaries after they are passed through the selection function to our observed sample.

\begin{figure}
    \centering
    \includegraphics[width=0.99\columnwidth]{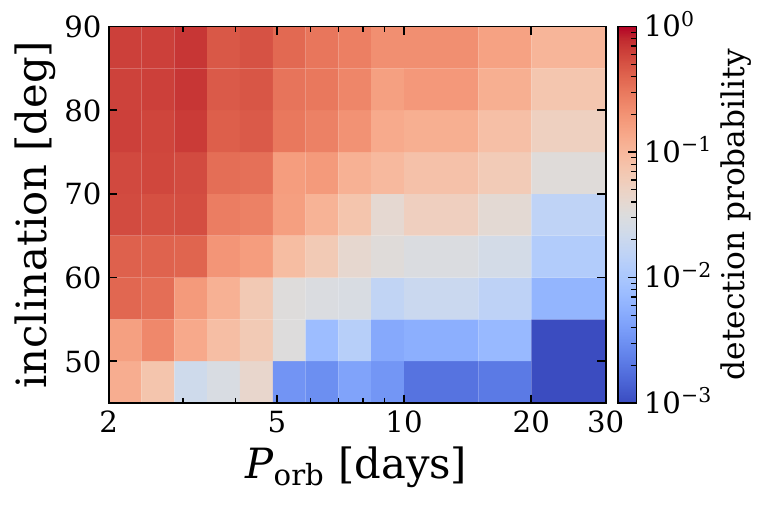}
    \caption{Sensitivity of our binary search, as determined by injection--recovery tests.  The color shows the probability that a close binary with a given orbital period, inclination, and properties similar to the observed EBs is recovered in our HST/ACS light curves ($\mathcal{P}_{\rm LC}$; Equation \eqref{eq:LC_selection}). Low-inclination systems have small detection probabilities because they rarely eclipse, and long-period systems are harder to recover because the eclipse duty cycle and phase coverage decrease.}
    \label{fig:lc_selection_period_inclination}
\end{figure}

Figure~\ref{fig:lc_selection_period_inclination} shows the sensitivity of our HST light curves to close binaries as a function of their periods and inclinations. Each grid cell is colored by the fraction of synthetic binaries in the photometric catalog that are recovered ($\mathcal{P}_{\rm LC}$), where the injected systems have parameters reflecting the real EB sample.
The sample is nearly complete at short periods and nearly edge-on orientations.  At $P_{\rm orb}\approx2$--$3$ d, systems above $i\approx75^\circ$ are recovered $30$--$100\%$ of the time, while the same inclinations fall to $\sim3$--$10\%$ recovery by $P_{\rm orb}\approx20$--$30$ d. At fixed stellar radii, the eclipse probability declines as $P^{-2/3}$ (e.g., Equation \ref{eq:eclipse_criteria}), explaining this dropoff. After marginalizing over inclination and stellar parameters, $\mathcal{P}_{\rm LC}$ can be closely approximated by a power law with slope $-1.1$ over the $2$--$30$~d period range, which combines the effects of geometric eclipse probability and period recovery in ACS light curves.

We use these detection efficiencies in Section \ref{sec:intrinsic_dem} to constrain the intrinsic close binary population. The observed periods constrain the intrinsic period distribution, while the observed EB count constrains the close-binary fraction. In summary, the total selection function of our EB catalog is that (1) the source is detected in HST and passes the survey quality cuts and (2) is identified as an EB from the ACS light curve.

\section{Basic Properties}\label{sec:basic_properties}

\subsection{Spatial distribution}\label{subsubsec:basic_properties_spatial}

\begin{figure}
    \centering
    \includegraphics[width=0.99\columnwidth]{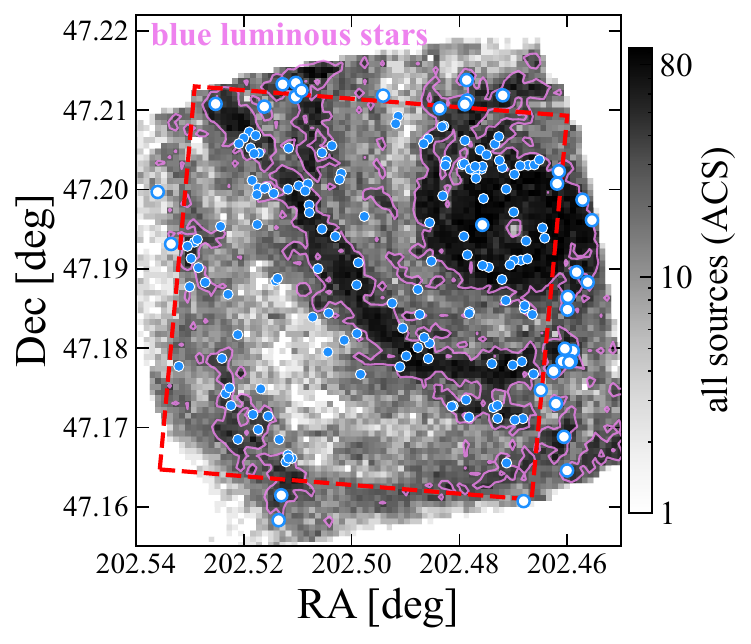}
    \caption{Spatial distribution of the massive eclipsing binary candidates.  The grayscale image shows the density of all luminous sources $m_{\rm F814W,0}<26$, while the pink contours show the density of blue luminous sources $m_{\rm F606W,0}-m_{\rm F814W,0}<0.8$, from which the EB candidates were selected.  Blue points mark EB candidates; the larger white-filled symbols are EBs with only ACS photometry, and the filled symbols have WFC3 photometry as well.  The dashed red outline shows the WFC3 footprint. EB candidates trace the star-forming spiral arms of M51, which coincide with the blue massive star population.}
    \label{fig:spatial_distribution}
\end{figure}

Figure~\ref{fig:spatial_distribution} shows the spatial distribution of EB candidates in M51. We compare the EBs to two different reference populations. First, we compare all luminous ACS point sources with $m_{\rm F814W,0}<26$ (grayscale background),  including blue massive stars, red supergiants, AGB stars, and other evolved luminous sources \citep[e.g.,][]{Conroy18}.  This full luminous catalog traces the overall resolved stellar population in the ACS footprint.  
The pink contours show the subset of blue luminous sources, $m_{\rm F606W,0}-m_{\rm F814W,0}<0.8$, which is the parent population from which the EB candidates were selected.  This blue population is concentrated along the spiral arms and inner star-forming regions, consistent with maps of M51 showing that FUV, H$\alpha$, and mid-infrared star formation tracers are enhanced in the arms and nucleus \citep{Calzetti2005}.  
These same regions contain structured dust and molecular gas, so they are both where young massive binaries are expected to form and where crowding and extinction can affect detectability \citep{Messa2018,FaustinoVieira2023}.  M51 has also experienced enhanced star and cluster formation over the past $\sim50$--$500$ Myr, probably associated with recent passages of its companion NGC~5195 \citep{Gieles2005,Lee2005,Dobbs2009}.

We quantify this comparison by dividing the ACS footprint into $85\times85$ spatial cells. We define the fraction of all luminous sources that are luminous blue stars in a given cell:
\begin{equation}
    f_{\rm blue,local}=\frac{N_{\rm blue}}{N_{\rm all}}.
\end{equation}
The EB candidates preferentially occupy cells with large $f_{\rm blue,local}$.  The median value at EB positions is $0.61$, compared to $0.50$ for the full luminous parent sample, and $84/173\approx50\%$ of the EBs lie in the top quartile of the $f_{\rm blue,local}$ distribution.  Expressed as an occurrence rate, the number of detected EBs per $10^4$ luminous sources rises from $4.3\pm1.4$ in the lowest $f_{\rm blue,local}$ quartile to $21.0\pm2.2$ in the highest quartile. This factor of $\approx5$ increase shows that the EB candidates are preferentially associated with regions where the luminous population is dominated by young blue stars.

HST+JWST observations reveal a substantial population of very young clusters in M51 that remain embedded in their natal gas and are absent from optical cluster catalogs \citep{Pedrini2026}. These emerging clusters and optically visible clusters younger than $10$ Myr retain similarly hierarchical spatial distributions \citep{Scoville2001,Lee2005,Lapeer2026}. Because our EB search requires detection in ACS F606W and F814W, it samples only massive binaries that have emerged sufficiently to become optically visible and misses the youngest, most heavily obscured systems.

\subsection{Color-magnitude diagram}\label{subsubsec:basic_properties_CMD}

\begin{figure*}
    \centering
    \includegraphics[width=0.99\textwidth]{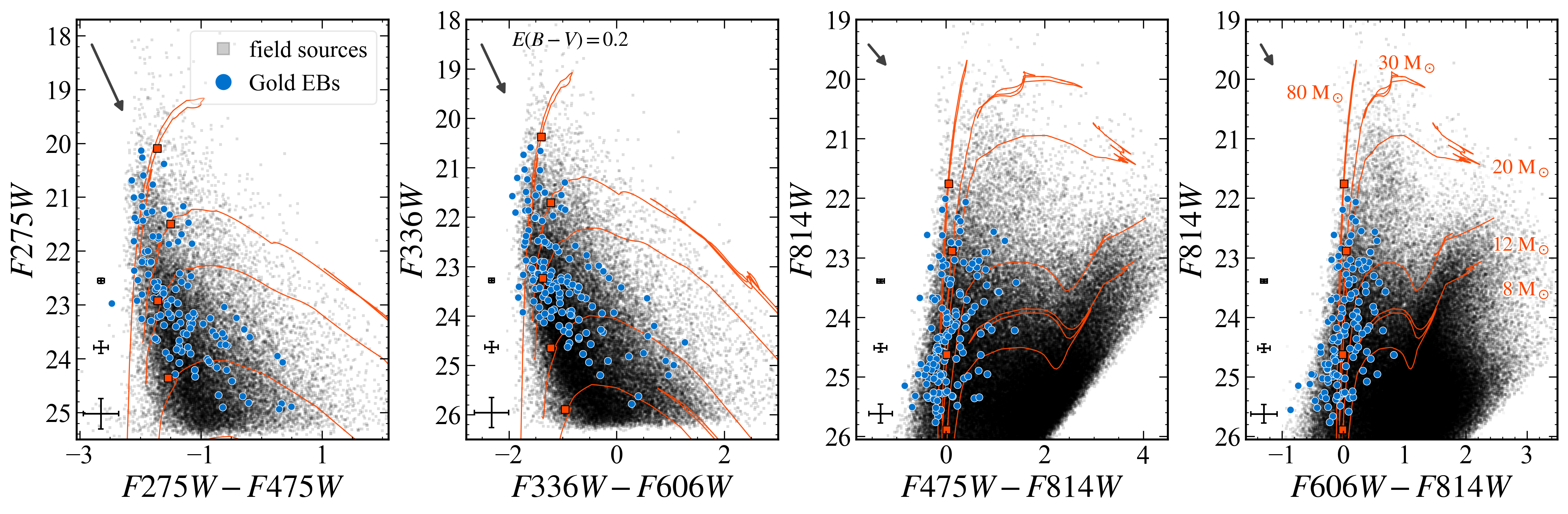}
    \caption{Panchromatic color-magnitude diagrams of the M51 sample from HST/WFC3. Gray points show field sources that pass WFC3 star cuts, and blue points show the Gold EB candidates.
    From left to right, the panels show NUV--optical (left), blue optical--red optical (left middle), optical only (right middle), and red optical only (right).
    All observed magnitudes have been corrected for Milky Way foreground extinction and placed on the infinite aperture system.
    Orange curves show example MIST evolutionary tracks at the distance of M51 for solar-metallicity stars of mass $8, 12,20, 30, 80~\msun$ and $v/v_{\rm crit} = 0.4$ with $E(B-V)=0.2$, a typical value for these sources. The squares on each track denote the terminal-age main-sequence (TAMS).
    The arrows also show the direction of increasing extinction by $E(B-V)=0.2$ ($A_V=0.62$ mag).
    The EB candidates lie preferentially along the blue luminous sequence, as expected for massive short-period binaries, with a subset extending redward where local extinction in M51 and evolved companions become more important.}
    \label{fig:multi_cmd_wfc3}
\end{figure*}

The final Gold EB sample contains $173$ candidates selected from the ACS light curves.  By construction, these systems lie in the luminous blue region of the ACS CMD (Figure~\ref{fig:cmd_selection_flow}), which is defined using only the red-optical F606W and F814W filters.  For hot massive stars ($T_{\rm eff}\gtrsim10{,}000$~K), these filters sample the Rayleigh--Jeans tail of the SED, where only a small fraction of the bolometric flux is emitted, and optical color is only weakly sensitive to temperature. The bluer WFC3 filters provide a more informative view of the hot-star population.

In Figure~\ref{fig:multi_cmd_wfc3}, we show the WFC3 CMDs of our parent sample and EB candidates, which include blue-optical and UV HST filters.
The EB candidates occupy the blue luminous sequence traced by young massive stars, extending over the region spanned by the $\sim8$--$80~\msun$ MIST tracks shown in the figure. These CMDs have been corrected for Milky Way foreground extinction using local dust maps (Section \ref{subsubsec:catalog_construction}), but not for internal extinction within M51. We also show reddening vectors corresponding to $E(B-V) = 0.2$ ($A_V=0.62$~mag), typical for M51 \citep[e.g.,][]{Wei2021}. Stars behind the face-on disk or still inside their natal gas clouds can experience substantially larger extinction.
Internal extinction shifts hot stars to redder colors and fainter magnitudes, with the largest effect in the UV--optical CMDs. Spatial variations in dust attenuation can therefore account for much of the width of the observed massive-star sequence and for EB candidates that lie redward of the unreddened stellar tracks \citep[e.g.,][]{Conroy18,Dalcanton2015,FaustinoVieira2023}.

Our WFC3 catalog is mainly sensitive to sources with $m_{\rm F275W}\lesssim25$ (Figure~\ref{fig:AST_completeness}), corresponding to main-sequence stars with $M\gtrsim8~\msun$ in the absence of internal extinction. Dust makes stars fainter and redder, raising this mass limit and leaving some sources undetected in the UV (Figure~\ref{fig:multi_cmd_wfc3}, left panel). The WFC3-detected sample therefore favors massive or evolved luminous stars with sufficiently low extinction to remain UV bright.

A red or UV-faint CMD position does not necessarily indicate a cool or low-mass star. We use the WFC3 CMDs to establish that the sample broadly traces hot, luminous stars, and the joint light curve and SED fits to estimate extinction, temperatures, radii, and binary parameters (Section~\ref{subsec:parameter_estimates}). Stellar evolution and unresolved blending can also affect the CMD positions, as discussed in Section~\ref{subsubsec:catalog_construction}.

\subsection{Period distribution}\label{subsubsec:basic_properties_periods}

\begin{figure}
    \centering
    \includegraphics[width=0.99\columnwidth]{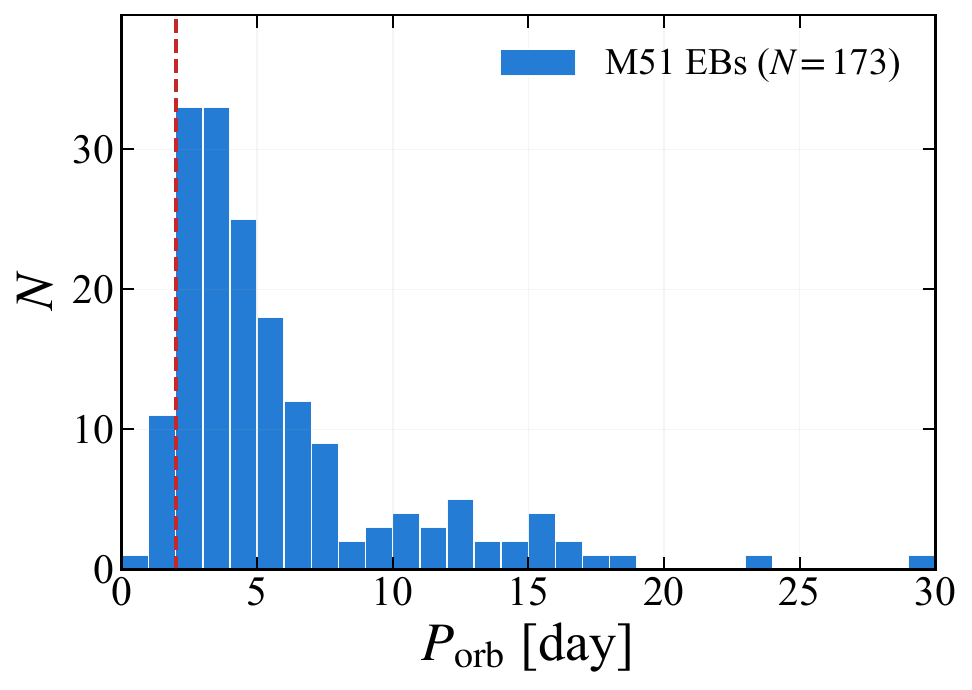}
    \caption{Observed orbital period distribution for the gold eclipsing binary sample. The blue histogram shows the EB candidates at their adopted orbital periods.  The dashed vertical line marks $P_{\rm orb}=2$~d.  The observed sample is strongly weighted toward short periods, consistent with geometric selection biases favoring close binaries.}
    \label{fig:goldv3_periods}
\end{figure}

The observed orbital period distribution is shown in Figure~\ref{fig:goldv3_periods}.  The $N=173$ Gold sample has orbital periods $P_{\rm orb}=1$--$30$~d, with a median of $4$~d.  The distribution is strongly concentrated at short periods: $83\%$ of systems have $P_{\rm orb}<10$~d and $68\%$ have $P_{\rm orb}<5$~d.  We also identify a separate set of $10$ short-period EB candidates with $P_{\rm orb}=1.3$--$2$~d (red dashed line), which are not used for the close binary demographics in the following sections due to high alias contamination in this region from the sparse ACS light curves.

The observed period distribution is affected by selection biases.
At fixed stellar radii, the geometric eclipse probability decreases with increasing orbital separation.  Longer-period binaries also spend a smaller fraction of each orbit in eclipse, making them less likely to be sampled by the ACS cadence.
Fitting the shape of the observed distributions
$dN/d\log P_{\rm orb}\propto P_{\rm orb}^{\pi}$ over
$2<P_{\rm orb}/{\rm d}<20$ gives
$\pi_{\rm obs}=-1.04\pm0.16$.
In Section \ref{subsubsec:intrinsic_period_dist}, we apply our selection function to infer the corrected (intrinsic) period distribution of binaries over the period range.

\subsection{Inferred stellar parameters}\label{subsubsec:fit_stellar_parameters}

\begin{figure*}
    \centering
    \includegraphics[width=0.8\textwidth]{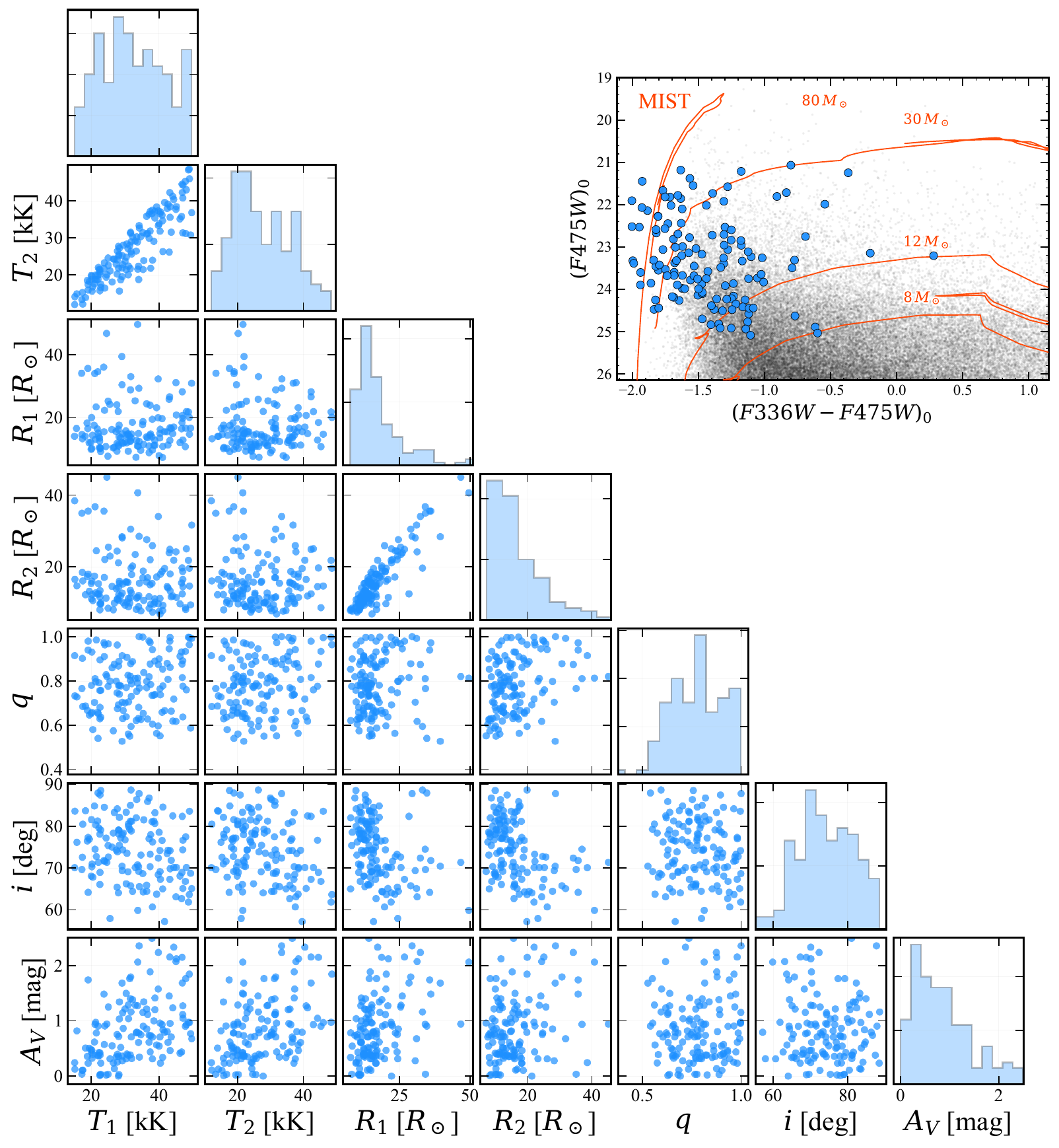}
    \caption{Stellar parameters for the EB candidates from the joint light curve+SED fits.
    Diagonal panels show the 1D distributions and off-diagonal panels show 2D parameter spaces. Each point represents the best-fit parameters of one eclipsing binary.  
    We include a dust-corrected WFC3 blue-optical CMD in the top right, which corrects each EB by its fitted local $A_V$.  Gray points show the observed WFC3 field population, and orange curves show solar-metallicity MIST tracks for $8, 12, 30, 80~\msun$ stars at the distance of M51 without internal extinction.  The fitted EB population hosts hot and luminous stars with typically similar flux and mass ratios.}
    \label{fig:fit_parameter_corner_cmd}
\end{figure*}

Figure \ref{fig:fit_parameter_corner_cmd} shows the resulting best-fit stellar parameters from our joint light curve+SED fits for the $136$ EB candidates with good WFC3 photometry (Section \ref{subsec:parameter_estimates}).
We show temperatures, radii, mass ratios, inclinations, and local extinction for each binary (off-diagonal) and the population-wide distributions (diagonal).
The fitted components are generally hot and luminous, as expected for the massive star sample.  The median temperatures are $T=32.6$ kK, with 16th--84th percentile ranges of $22$--$43$ kK.  The corresponding median radii are $R=17.3~\rsun$ with 16th--84th percentile ranges of $12$--$29~\rsun$. The components typically have similar fluxes, partly because the selection favors two detectable eclipses.
The fitted mass ratios are also concentrated near unity, with a median $q=0.90$ and 16th--84th percentile range $0.68$--$1.13$. While the masses derived from our photometric fits are uncertain \citep[e.g.,][]{Serenelli2021} and biased by EB selection, the preference for similar mass ratios is consistent with previous observations of close massive star binaries \citep[e.g.,][]{Pinsonneault2006,Sana12,MoeDiStefano17}.

The fitted systems are viewed preferentially near edge-on, with a median inclination of $i=73^\circ$ and a 16th--84th percentile range of $65^\circ$--$81^\circ$.  This is expected for an EB sample, since high-inclination systems are more likely to eclipse and generally produce deeper, more easily recognized variability in ACS light curves.  The posterior concentration toward high inclinations, despite an uninformative prior over the allowed range (Table~\ref{tab:ellc_fit_priors}), also provides a consistency check on the fits.  The fitted extinction distribution is broad, with a median of $A_V=0.75$ mag and a 16th--84th percentile range of $0.30$--$1.4$ mag, consistent with the WFC3 CMDs in Figure~\ref{fig:multi_cmd_wfc3}, where several candidates lie redward of the blue massive-star sequence before shifting back toward it after correction for their fitted internal extinction.

\subsection{Evolutionary state and coevality}\label{subsubsec:fit_ages_coevality}

\begin{figure*}
    \centering
    \includegraphics[width=0.99\textwidth]{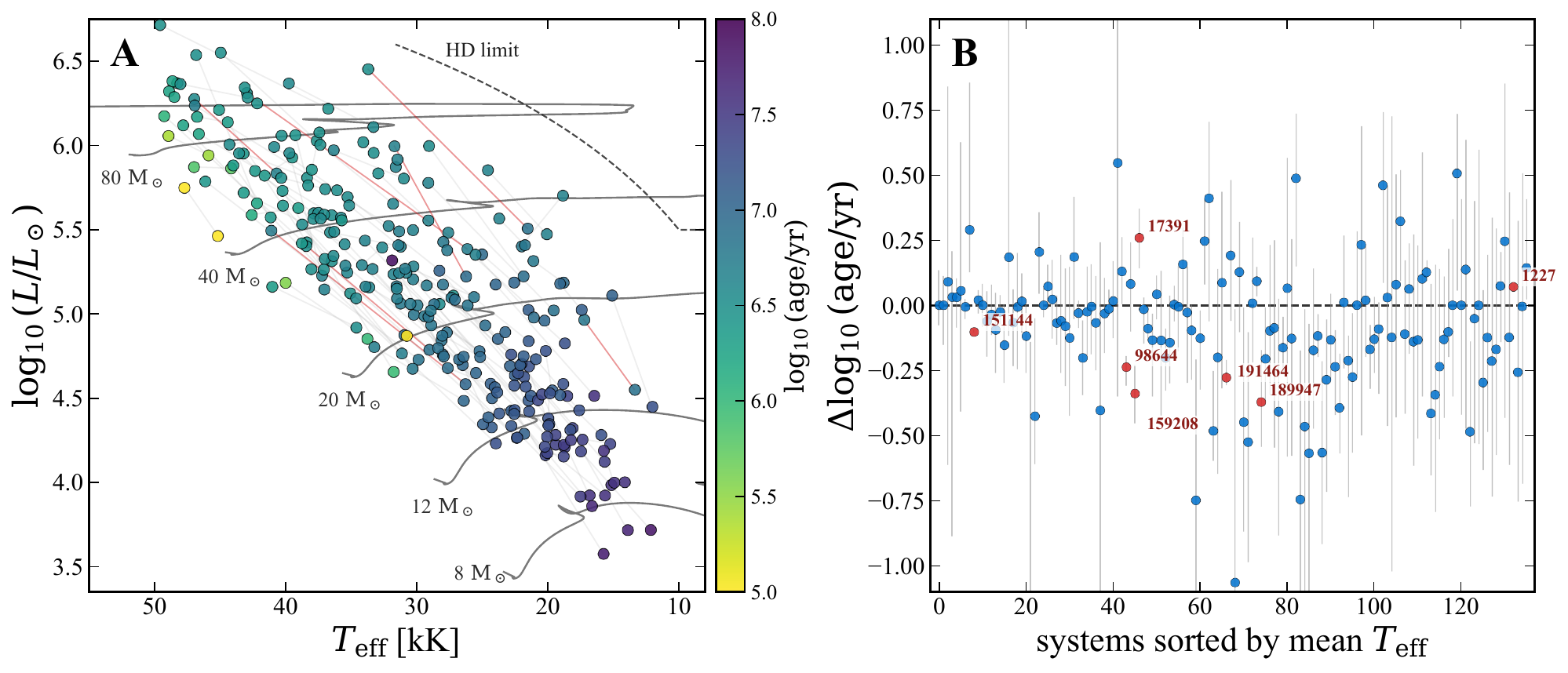}
    \caption{HR diagram and apparent ages of massive binaries in M51. {\bf Panel A} shows both components in the HR diagram, colored by the inferred MIST age, with solar-metallicity MIST tracks overplotted. The dashed black curve shows the classical Humphreys--Davidson (HD) luminosity limit \citep{Humphreys1979,Sabhahit2026}.
    {\bf Panel B} shows the difference between the two component ages for each binary, sorted by the mean $T_{\rm eff}$. Each point represents one binary with the posterior medians (blue) and $1\sigma$ range shown.  Most systems are consistent with coeval components at the current level of precision, except for $7$ that are discrepant at $95\%$ confidence, possibly due to a history of binary interaction.}
    \label{fig:HR_age}
\end{figure*}

Figure \ref{fig:HR_age} shows the 
binary candidates on an HR diagram ($T_{\rm eff}-\log L$) using the best-fit stellar parameters. The bolometric luminosity $L$ is calculated using the $T_{\rm eff}$ and $R$ for each component.
During the fitting process, we did not impose that the binary components necessarily be the same age, but coevality is generally expected for stars in the same system. We estimate the age of each component by matching its fitted temperature and surface gravity to solar-metallicity MIST II stellar tracks \citep{Dotter2026} and color the systems in the figure by their inferred age.

The left panel of Figure~\ref{fig:HR_age} shows that the stars lie in the expected part of the HR diagram for young and moderately evolved massive stars, by construction of the catalog (Section \ref{subsubsec:catalog_construction}).  
We also show the classical Humphreys--Davidson (HD) limit \citep{Humphreys1979}, an empirical upper luminosity limit of stars, as first inferred from supergiants (but see also \citealt{Davies2018,Sabhahit2021,Sabhahit2026}).  All systems lie safely below this empirical boundary.
The median inferred age is $\log_{10}({\rm age/yr})=6.68$ for both primary and secondary stars, corresponding to about $5$ Myr.  Many systems are also somewhat evolved, with inflated luminosities relative to the zero-age main sequence.  To quantify whether the two components in a given binary have mutually consistent ages, we define
$\Delta \log ({\rm age}) = \log({\rm age}_1/{\rm age}_2)$\footnote{$\log \equiv\log_{10}$ throughout the paper},
where ${\rm age}_1$ and ${\rm age}_2$ are the inferred ages of the two fitted components.
The distribution is centered near zero, with median $\Delta\log(age)=5\times10^{-4}$ dex, a 16th--84th percentile range of $-0.15$ to $0.20$ dex, and a median absolute difference of $0.10$ dex.  For reference, $0.10$ dex corresponds to an age ratio of $1.26$, while $0.30$ dex corresponds to a factor of $2$.  Thus, most systems are consistent with coevality at the precision of our data.  
Only $7/136$ fitted systems are discrepant at the formal $95\%$ level: IDs $1227$, $17391$, $98644$, $151144$, $159208$, $189947$, and $191464$.

Whether these systems are truly age-discrepant remains unclear, given that the inferred ages are highly model-dependent. Independent spectroscopy and dynamical masses would be required to calculate their $\log g$ more precisely and determine whether they are statistically discrepant. Apparent age discrepancies in close binaries can be interpreted as the result of mass transfer, which can reverse the apparent evolutionary order of the two stars, as in the classical case of Algol \citep[e.g.,][]{Paczynski1971}. Tidal distortion, rotational mixing, contact evolution, or mergers can also cause deviations from single-star tracks \citep[e.g.,][]{Sana12, deMink2013, Schneider2015, Menon2021, Tkachenko2020}.

\section{Intrinsic Demographics}\label{sec:intrinsic_dem}

\subsection{Period Distribution}\label{subsubsec:intrinsic_period_dist}

We infer the {\it intrinsic} orbital period distribution by correcting the {\it observed} EB period distribution by the selection function (Section~\ref{subsec:selection_function}), as developed using the synthetic binary population (Appendix \ref{app:binary_pop}). 
Our goal is to infer this distribution for the main-sequence star population, so we select only sources with $F336W<24$ blueward of the TAMS in the $F336W$--$F606W$ vs $F336W$ CMD. The TAMS is defined by MIST II \citep{Dotter2026} models with $Z=Z_\odot$, $v/v_{\rm crit}=0.4$, and an internal $A_V=0.62$. These filters are chosen as they provide the deepest WFC3 coverage in the UV and red-optical (Figure \ref{fig:AST_completeness}). Repeating the inference with other CMDs gives consistent results.  This leaves $3{,}616$ unresolved parent systems and $50$ Gold EBs in the adopted period interval.  

The intrinsic distribution is modeled as a power-law over $2\leq P_{\rm orb}/{\rm d}<20$:
\begin{equation}
\frac{\mathrm{d}N}{\mathrm{d}\log_{10}P_{\rm orb}}\propto P_{\rm orb}^{\pi}.
\label{eq:period_law}
\end{equation}
For each trial $\pi$, a {\tt COSMIC} system with period $P_j$ is assigned the weight
\begin{equation}
w_j(\pi)=w_{j,0}\frac{P_j^\pi}{g_0(\log_{10}P_j)},
\end{equation}
where $w_{j,0}$ is its original population weight and $g_0$ is the period distribution of the original {\tt COSMIC} population. This removes the baseline COSMIC period distribution and replaces it with the trial power law, without changing the other properties of each binary. 

We multiply each system's weight by its probability of passing the complete survey selection, including photometric recovery, CMD selection, eclipse geometry, period recovery, and visual selection. The resulting weighted periods define $p_{\rm sel}(P_{\rm orb}\mid\pi)$, the predicted period distribution of detected EBs. This also determines $S(\pi)$, the probability that a close binary in the CMD is identified as an EB (i.e., selected).
As a check, we find that the selected \texttt{COSMIC} population spans the period, mass, temperature, radius, and extinction distributions inferred for the $50$ Gold EBs used in the primary demographic analysis, although the mock systems have somewhat smaller hot-component radii and a narrower $A_V$ distribution (Appendix~\ref{app:cosmic_parameter_check}; Figure~\ref{fig:cosmic_parameter_check}).

\begin{figure}
    \centering
    \includegraphics[width=0.99\columnwidth]{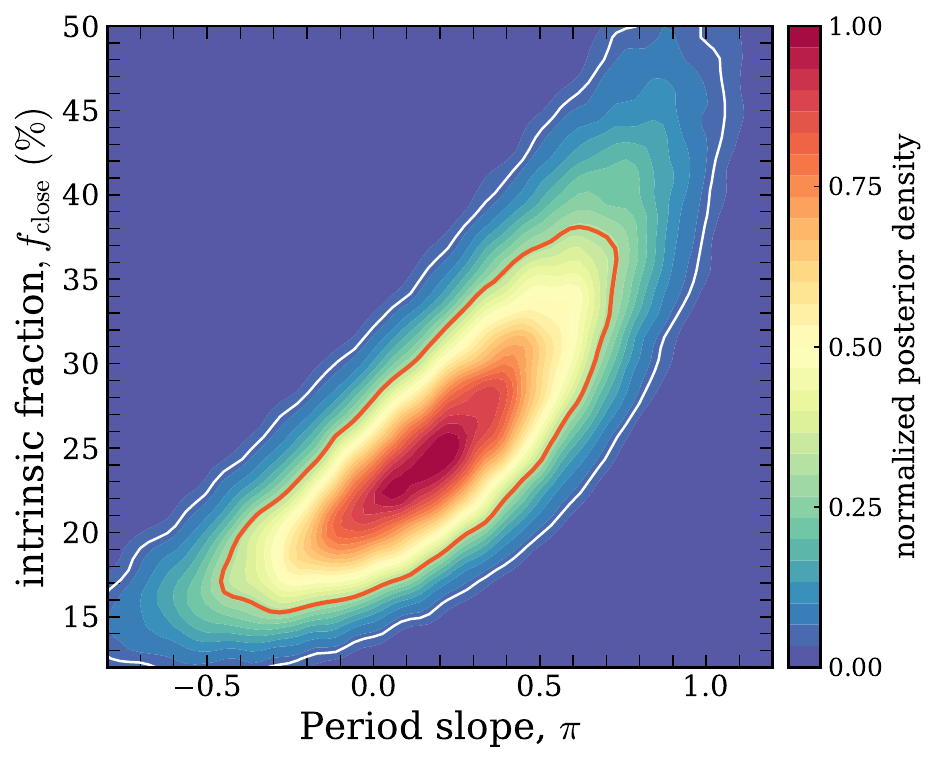}
    \caption{2D posterior for the intrinsic period slope ($\pi$) and close-binary fraction ($f_{\rm close}$).  Filled contours enclose the $68\%$ and $95\%$ density regions.  The covariance arises because longer-period-weighted populations have lower selection probabilities and therefore require larger $f_{\rm close}$.}
    \label{fig:joint_pi_fclose}
\end{figure}

We jointly infer $\pi$ and $f_{\rm close}$ using a likelihood with two terms:
\begin{equation}
\mathcal{L}(\pi,f_{\rm close})
=
\mathcal{L}_{P}(\pi)\,
\mathcal{L}_{N}(\pi,f_{\rm close}).
\label{eq:joint_demographic_likelihood}
\end{equation}
The first term compares the individual observed periods to the predicted period distribution:
\begin{equation}
\mathcal{L}_{P}(\pi)
=
\prod_{i=1}^{N_{\rm EB}}
p_{\rm sel}(P_i\mid\pi).
\label{eq:period_likelihood}
\end{equation}
The second term gives the probability of detecting $N_{\rm EB}$ EBs among the $N_{\rm parent}$ systems in the adopted CMD region:
\begin{equation}
\begin{split}
\mathcal{L}_{N}(\pi,f_{\rm close})
&=
\binom{N_{\rm parent}}{N_{\rm EB}}
\left[f_{\rm close}S(\pi)\right]^{N_{\rm EB}}\\
&\quad\times
\left[1-f_{\rm close}S(\pi)\right]^{
N_{\rm parent}-N_{\rm EB}} .
\end{split}
\label{eq:count_likelihood}
\end{equation}
This likelihood takes the form of a binomial distribution where $f_{\rm close}S(\pi)$ is the probability that a system in the parent sample contains a close binary that is detected as an EB. The observed periods primarily constrain $\pi$, while the count of $50$ EBs among $3{,}616$ total sources in the CMD region constrains $f_{\rm close}S(\pi)$. Uniform priors are adopted. The parameters are positively correlated: a larger $\pi$ places more binaries at long periods, where the recovery probability is lower, and therefore requires a larger $f_{\rm close}$ to reproduce the observed count (Figure~\ref{fig:joint_pi_fclose}).  

\begin{figure}
    \centering
    \includegraphics[width=0.99\columnwidth]{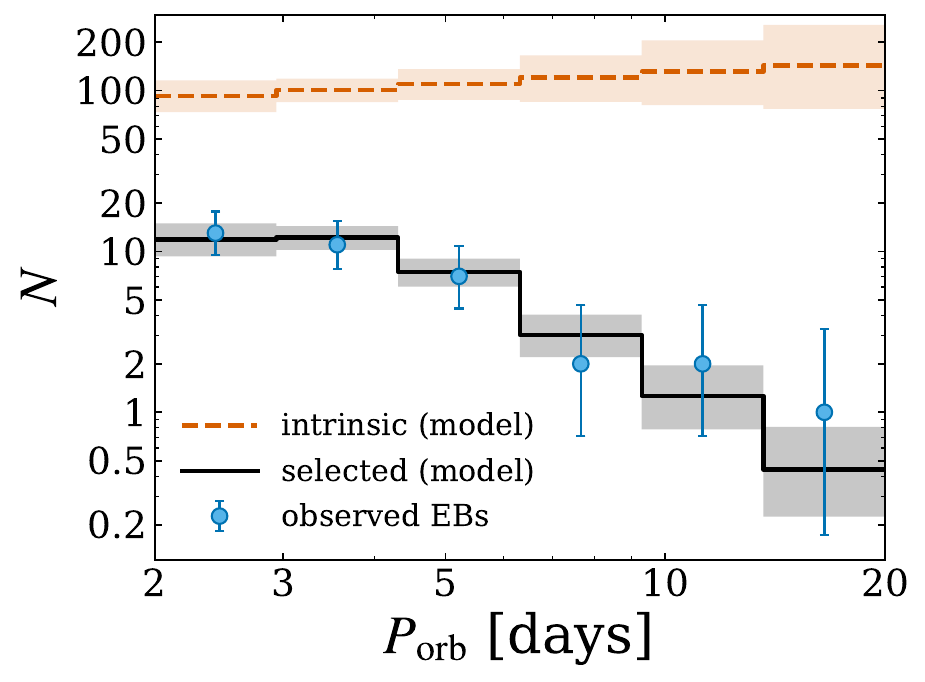}
    \caption{Observed and forward-modeled period distributions over $2\leq P_{\rm orb}/{\rm d}<20$.  The dashed orange curve shows the intrinsic {\tt COSMIC} binary population, while the solid black curve shows the same population after applying the survey selection function. Blue points show the observed $50$ Gold EBs in the primary main-sequence sample.
    The observed period distribution of massive binaries in M51 is consistent with a roughly log-flat ($\pi=0.29 \pm 0.32$) intrinsic period distribution. }
    \label{fig:intrinsic_period_dist}
\end{figure}

Marginalizing the joint posterior over $f_{\rm close}$ gives a best-fit power-law index
\begin{equation}
\pi=0.29 \pm 0.32,
\label{eq:pi_primary}
\end{equation}
where the uncertainties give the 16th--84th percentiles (Figure~\ref{fig:joint_pi_fclose}).  Figure~\ref{fig:intrinsic_period_dist} shows the inferred intrinsic distribution and compares the selected subset to the observed Gold EBs. The selection function favors short-period systems, and the selected model reproduces the observed period distribution. The inferred intrinsic distribution is consistent with being log-uniform ($\pi=0$).

\subsection{Binary Fraction}\label{subsubsec:intrinsic_binary_frac}

We detect $50$ EBs among $3{,}616$ unresolved sources in the main-sequence region of the $F336W-F606W$ vs $F336W$ CMD, corresponding to a raw EB fraction of $1.38\%$.  The likelihood in Equation \eqref{eq:joint_demographic_likelihood} constrains the fraction of recovered binaries $S(\pi)$, allowing us to infer the intrinsic binary fraction among main-sequence stars in the sample ($f_{\rm close}$).

Marginalizing the joint posterior over $\pi$ gives
\begin{equation}
f_{\rm close}=27.4 \pm 7.1\%,
\label{eq:fclose_primary}
\end{equation}
where the quoted error is a $1\sigma$ uncertainty that includes uncertainties in the selection probability.  Here, $f_{\rm close}$ is the fraction of unresolved systems in the luminous-blue CMD parent that contain a binary with $2\leq P_{\rm orb}/{\rm d}<20$ and $0.1\leq q\leq1$.  The posterior median corresponds to roughly $990$ intrinsic close-binary systems, of which $\sim5\%$ ($50$) are detected as EBs.

\section{Comparison to nearby galaxies}\label{sec:comparison}

\subsection{Period distribution}\label{subsec:comparing_periods}

\begin{figure*}
    \centering
    \includegraphics[width=\linewidth]{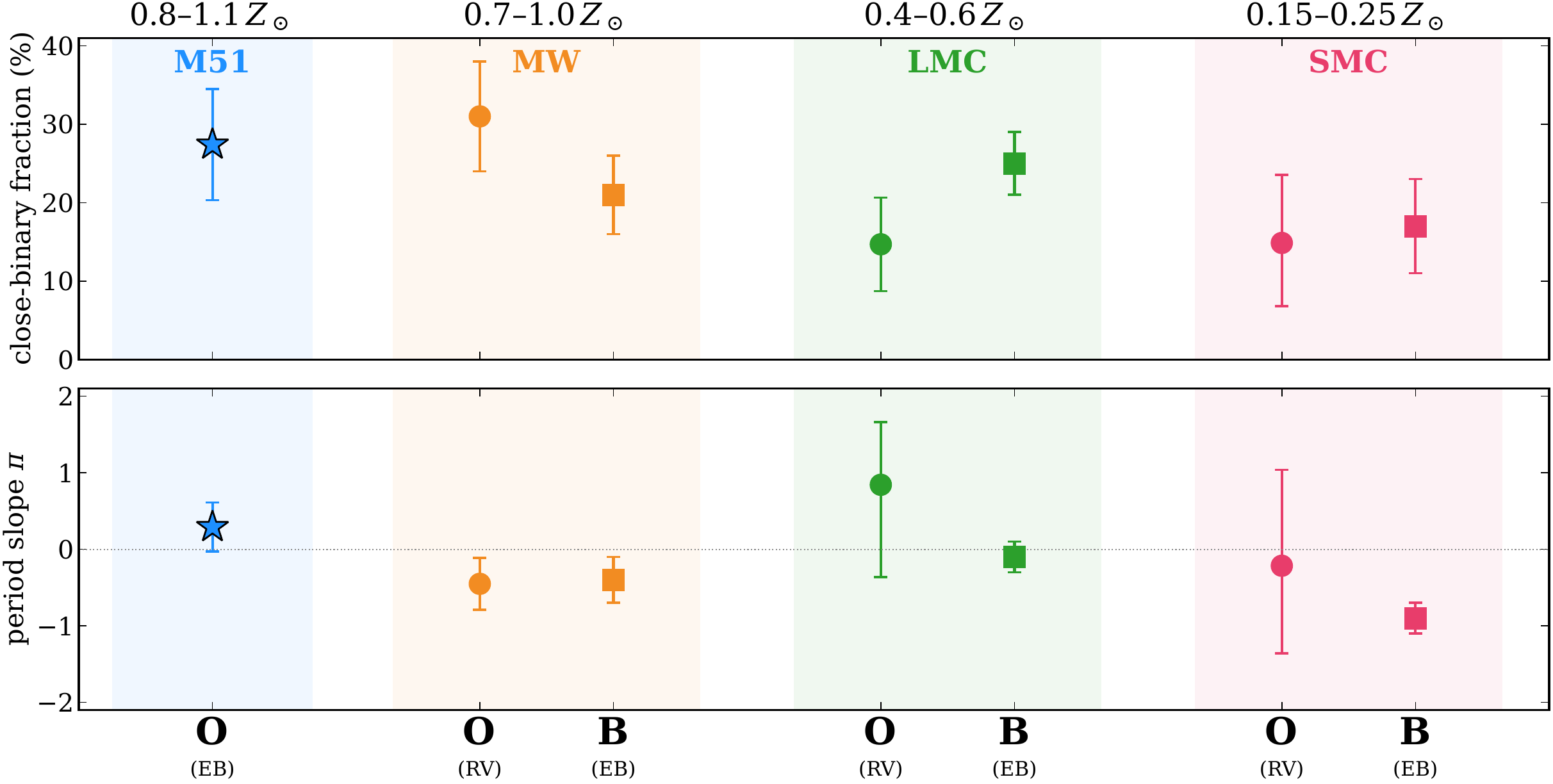}
    \caption{Close-binary fractions (top) and period slopes (bottom) for massive stars in M51 and nearby galaxies, over $P_{\rm orb}=2$--$20$~d and $q\geq0.1$. The period slope ($\pi$) is defined by $dN/d\log P_{\rm orb}\propto P_{\rm orb}^{\pi}$, with the dotted line marking a log-flat distribution ($\pi=0$). 
    `EB' and `RV' denote results from eclipsing-binary and radial-velocity surveys, whereas `O' and `B' refer to the approximate spectral types of each sample. Approximate metallicities are shown above each host galaxy \citep{Croxall2015,Russell1992,Hunter2007}. The early-B EB results and Galactic O-star close-binary fraction are from \citet{MoeDiStefano13}; the remaining RV estimates are derived in Appendix~\ref{app:rv_comparison}. Table~\ref{tab:fclose_pi} lists all values and uncertainties. M51 has a high close-binary fraction and a nearly log-flat period distribution, similar to the massive binaries in nearby galaxies.}
    \label{fig:demographic_comparison}
\end{figure*}

Figure~\ref{fig:demographic_comparison} (bottom panel) compares the period distribution in M51 to those in the Milky Way (MW), LMC, and SMC over $2$--$20$~d and $q\geq0.1$. The period slope $\pi$ is defined by Equation~\eqref{eq:period_law}, so that $\pi=0$ is uniform in log period. Our M51 sample consists of very luminous massive stars, although their individual O- or B-type classifications are uncertain because they are based on photometry. The $F336W<24$ selection (Section~\ref{subsubsec:intrinsic_period_dist}) corresponds to a single star with an initial mass of $\sim15~\msun$ at the TAMS, so we assume the majority are O stars \citep{Dotter2026}. 
We compare this sample to early-B and O-star populations in the other galaxies. The early-B comparisons use results from eclipsing-binary (EB) surveys; the O-star comparisons outside M51 use radial-velocity (RV) surveys.

For the early-B EB samples, we adopt the completeness-corrected period slopes from \citet{MoeDiStefano13}, using Hipparcos observations in the MW, OGLE-III in the LMC, and OGLE-II in the SMC. These estimates were derived over the same $2$--$20$~d interval and use the same period-law definition as our M51 analysis.
We refit the O-star RV period distributions over this interval. For Galactic O stars, we use the $21$ binaries with published orbital periods between $2$ and $20$~d from \citet{Sana12}, following the analysis of \citet{MoeDiStefano13}. The LMC O-star sample contains $360$ targets from the VLT-FLAMES Tarantula Survey \citep{Sana2013,HenaultBrunet2012}, while the SMC sample contains $139$ O stars from BLOeM \citep{Shenar2024,Sana25SMC}. 
Appendix~\ref{app:rv_comparison} describes these fits in detail, and the results are summarized in Table~\ref{tab:fclose_pi}.

The M51 slope, $\pi=0.29\pm0.32$, is consistent with a log-flat distribution. The Galactic O-star and MW/LMC early-B EB samples favor somewhat shorter periods, with slopes between approximately $-0.5$ and $-0.1$. The SMC early-B EB sample has a stronger preference for short periods, $\pi=-0.9\pm0.2$ \citep{MoeDiStefano13}. 

The O-star RV estimates in the Magellanic Clouds are less precise and allow for both log-increasing and log-decreasing period distributions.
Unlike the M51 and MW results, which use measured orbital periods, the Magellanic Cloud fits infer the period distribution from sparse RV variations \citep{Sana2013,HenaultBrunet2012,Sana25SMC}.
This motivates the use of EB surveys in other galaxies, which can monitor many more stars simultaneously and measure periods directly from eclipses. For example, despite the greater distance of M51, its $50$ EBs in this period range provide a sharp constraint on the intrinsic period distribution after correcting for the survey selection function.
The most significant difference in the period slopes is between M51 and the SMC early-B EB sample, although their different stellar masses and evolutionary states prevent attributing that difference to metallicity alone.
Altogether, the intrinsic period distribution of massive-star binaries in M51 is broadly similar to those measured in the MW, LMC, and SMC.

\subsubsection{Environmental dependence}

The samples in Figure~\ref{fig:demographic_comparison} differ in environment as well as metallicity. They include Galactic clusters and associations, the dense 30~Doradus star-forming region in the LMC, lower-mass Magellanic Cloud clusters, and the spiral-arm population of M51. Simulations suggest that multiplicity may decrease with increasing stellar density, although they primarily resolve binaries wider than the $2$--$20$~d systems studied here \citep{Guszejnov2023}. Observationally, Galactic O stars in clusters and associations have higher multiplicity fractions than field O stars \citep[e.g.,][]{Gies1986,Gies1987,Mason2009,Sana12,DorigoJones2020}. The broad-period binary fraction is likewise $\approx0.5$--$0.6$ in 30~Doradus, compared with $\approx0.7$ for lower-mass clusters in the SMC O-star sample, although the uncertainties are typically $5$--$10$ percentage points \citep{Sana2013,Almeida2017,Moe25LMC,Sana25SMC}. 
While our M51 sample accesses a range of star-forming environments, the point-source and photometric quality cuts preferentially exclude stars in the most crowded cluster cores. The sample is therefore likely weighted toward OB stars in sparse associations and the field, making field OB star measurements the most relevant comparison.

In the LMC and Milky Way, \citet{Moe25LMC} find that young dense clusters, young average clusters, sparse associations, and low-mass clusters all have consistent O-star EB fractions near $9$--$11\%$. This uniformity, they suggest, provides evidence that the formation of close massive binaries depends on small-scale gas physics -- fragmentation and migration in massive protostellar disks \citep{Kratter2006,Tokovinin2020,Offner2023,Moe25LMC}.  
By contrast, old clusters have a lower EB fraction, $5.5\pm0.9\%$. The difference is attributed to binary evolution, where close binaries have had time to interact via mass transfer, mergers, and ejections \citep{Moe25LMC}. Consistently, ejected O-stars in the field exhibit a lower close binary fraction, although most of these were likely ejected via dynamical N-body interactions and not supernova explosions \citep{Moe25LMC}.
Similar variations in massive binary properties across environments have been noted in earlier Galactic O-star samples \citep{Gies1986,Gies1987,Mason2009,DorigoJones2020}.  
Thus, interpreting any differences between massive close binaries in the MW, LMC, SMC, and M51 requires considering a combination of metallicity, age, environment, and dynamical processing. Nonetheless, the ubiquity of massive stars in close binaries across galaxies and environments emphasizes their significance in massive star evolution.

\vspace{-0.8cm}
\begin{deluxetable*}{lllccc}
\tablewidth{\textwidth}
\tablecaption{Close-binary fractions and period-distribution slopes for nearby galaxies and M51.}
\tablehead{
\colhead{Galaxy} &
\colhead{Spectral type} &
\colhead{Method} &
\colhead{$f_{\rm close}$ (\%)} &
\colhead{$\pi$} &
\colhead{Refs.}
}
\startdata
M51 & O & EB & $27.4\pm7.1$ & $+0.29\pm0.32$ & 1 \\
MW  & O        & RV & $31\pm7$ & $-0.45\pm0.34$ & 1,2,3 \\
    & early B  & EB & $21\pm5$ & $-0.4\pm0.3$ & 2 \\
LMC & O        & RV & $14.7 \pm 6$ & $+0.84 \pm 1.0$ & 1,4,5 \\
    & early B  & EB & $25\pm4$ & $-0.1\pm0.2$ & 2 \\
SMC & O        & RV & $14.9 \pm 8.4$ & $-0.22 \pm 1.19$ & 1,6 \\
    & early B  & EB & $17\pm6$ & $-0.9\pm0.2$ & 2 \\
\enddata
\tablecomments{
Values plotted in Figure~\ref{fig:demographic_comparison}, for $2$--$20$~d and $q\geq0.1$. The LMC EBs use results from OGLE-III; the SMC EBs use OGLE-II. Magellanic Cloud RV values are posterior medians with 16th--84th percentile intervals from the population fits in Appendix~\ref{app:rv_comparison}. The Galactic O-star slope uncertainty is a delete-one jackknife estimate \citep{MoeDiStefano13}. References identify data sources where values were refitted: (1) this work; (2) \citet{MoeDiStefano13}; (3) \citet{Sana12}; (4) \citet{Sana2013}; (5) \citet{HenaultBrunet2012}; (6) \citet{Sana25SMC}.
}
\label{tab:fclose_pi}
\end{deluxetable*}

\subsection{Close Binary Fraction}\label{subsec:comparing_binary_fractions}

Figure~\ref{fig:demographic_comparison} (top panel) compares the close-binary fractions across galaxies. 
The close binary fraction of O stars in M51 is consistent with the massive star populations in the Milky Way.
The median fraction is slightly higher than the O-star estimates in the Magellanic Clouds, although the $1\sigma$ error bars overlap. Interestingly, the binary fractions of O and B stars in this period range seem to be similar across the MW, LMC, and SMC.

The M51 binaries likely span a small range of near-solar metallicities ($\pm0.1$~dex) because of the galaxy's shallow radial metallicity gradient \citep{Croxall2015}. 
Together, these host galaxies span a broad metallicity range \citep[$0.2Z_\odot$--$1.1Z_\odot$;][]{Croxall2015,Russell1992,Hunter2007}. 
The O-star estimates hint at a lower close-binary fraction at lower metallicity, but the uncertainties and differences between samples prevent a firm conclusion. Overall, we find no evidence for a strong metallicity dependence, consistent with recent observational studies \citep[e.g.,][]{Sana25SMC,Villasenor25BLOeM} and radiation-hydrodynamic simulations \citep[e.g.,][]{Guszejnov2023,Chon2024}.

The high close-binary fraction and log-flat period distribution in M51 support the notion that binary interaction plays a dominant role in the evolution of massive stars. Surveys covering a broader period range suggest that $\sim60-70\%$ of massive stars interact with a binary companion during their evolution \citep[e.g.,][]{Sana12}. Given the similarities between the massive-binary properties in M51 and those in the nearest galaxies (e.g., Figure \ref{fig:demographic_comparison}), it stands to reason that M51 binaries have a similarly high interacting fraction.
Most massive stars in M51 will thus not evolve as single stars but commonly undergo early mass transfer, providing direct implications for the production of stripped stars, mergers, X-ray binaries, interacting supernovae, compact-object progenitors, ionization of their surrounding environments, and the integrated spectra of such galaxies \citep[e.g.,][]{DeMink14,Sana12,Langer12,Stanway2016, Gotberg2019,Gotberg2020, Eldridge2022,Marchant2024}.

Our forward model predicts that only $\sim5\%$ of the massive close binaries in our M51 parent sample are detected as EBs (Section~\ref{subsubsec:intrinsic_binary_frac}). Thus, the $50$ detected EBs represent approximately $990$ close binaries among the $3{,}616$ parent O stars, implying a selection correction of $\approx20$. For comparison, the Milky Way contains an estimated $1.4\times10^4$--$5\times10^4$ O-star systems \citep{MaizApellaniz2013}. Combined with observed O-star EB fractions of $\sim5\%$--$10\%$ \citep{Moe25LMC}, this implies a Galactic population of roughly $10^3$--$5\times10^3$ O-star EBs.

Evolutionary state can also affect the measured binary fraction. In the SMC BLOeM survey, \citet{Britavskiy2025} infer an intrinsic spectroscopic binary fraction of $40\pm4\%$ for $262$ B0--B3 class I--II supergiants over $1<P/{\rm d}<10^{3.5}$ and $0.1<q<1$, compared with $80\pm8\%$ for $309$ B0--B2.5 class III--V stars over the same parameter range \citep{Villasenor25BLOeM}. While our M51 EB sample does not contain red supergiants, it does include partly evolved OB stars, which could modestly decrease the inferred close-binary fraction relative to dwarfs.

A natural next step is to push resolved multiplicity surveys below the SMC metallicity. Individual massive contact binary candidates have been identified in very metal-poor Local Group galaxies such as WLM \citep[$Z\simeq0.14~Z_\odot$;][]{Gull25}. Extending a homogeneous EB census to such environments would quantify the abundance and properties of close massive binaries at the metallicities most relevant for high-redshift star formation.

\subsection{Stellar parameters}\label{subsec:comparing_stellar_parameters}

\begin{figure*}
    \centering
    \includegraphics[width=0.96\textwidth]{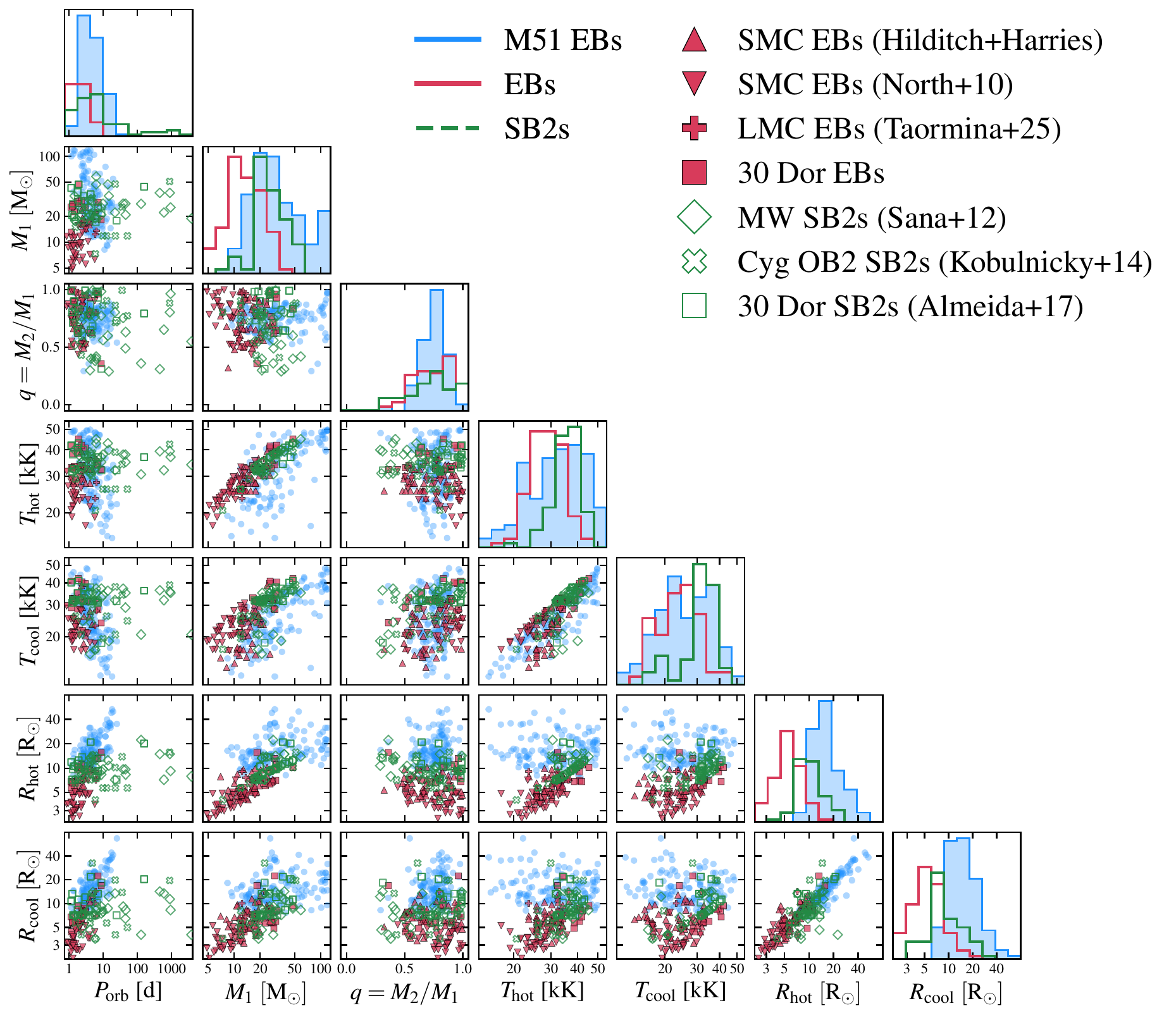}
    \caption{Comparing M51 EB parameters to massive-binary samples in the MW, LMC, and SMC.  Diagonal panels show 1D distributions; off-diagonal panels show 2D scatter plots.  
    Blue points and histograms show the M51 parameters. 
    Red markers show EB results from 30 Dor (Tarantula Nebula), LMC, and SMC samples \citep{Mahy2020,Harries2003,Hilditch2005,North2010,Taormina2024}.  Green open markers show SB2 results from the MW, Cyg OB2, and 30 Dor \citep{Sana12,Kobulnicky2014,Almeida2017}.  For all samples, $M_1$ is the more massive component and $q=M_2/M_1\leq1$.  The EB comparisons use derived component masses, radii, and temperatures.  The SB2 samples use published periods and mass ratios, with $T_{\rm eff}$, $R$, and $M$ derived from the spectral type and luminosity class of each star \citep[e.g.,][]{Martins2005,Pecaut2013}.  The M51 systems overlap the comparison samples in temperature and mass ratio, but extend to larger radii, consistent with the HST sample being weighted toward luminous, partially evolved systems.}
    \label{fig:stellar_parameter_comparison}
\end{figure*}

Figure~\ref{fig:stellar_parameter_comparison} compares the M51 EB parameters to measurements of massive binaries in the Local Group.  The M51 sample uses the posterior medians from the joint light curve+SED fits. For the EB comparison sample, we use $99$ EBs with published physical solutions: $13$ systems in the LMC's Tarantula Nebula (30 Dor) \citep{Mahy2020}, $50$ SMC EBs from \citet{Harries2003,Hilditch2005}, $33$ SMC EBs from \citet{North2010}, and $3$ LMC systems from \citet{Taormina2024}.  These works report component masses, radii, temperatures, and periods from light curve and spectroscopic modeling. 
We also compare to $74$ SB2 binaries: Galactic O-star SB2s from \citet{Sana12}, $15$ SB2s in the Milky Way's Cyg OB2 cluster from \citet{Kobulnicky2014}, and SB2s in 30 Dor from \citet{Almeida2017}.
For the SB2s, we use the published orbital periods and mass ratios, and derive $T_{\rm eff}$, $R$, and $M$ from the reported spectral types and luminosity classes using standard hot-star calibrations \citep[e.g.,][]{Martins2005,Pecaut2013}.

The samples have different selection biases. The EB comparison samples have well-measured component radii and temperatures, but are restricted to short periods ($P_{\rm orb}\lesssim30$~d) where the eclipse probability is larger.  The SB2 samples extend to $P_{\rm orb}\approx2000$~d, since these wide systems can still produce measurable radial velocity variations.  With these differences in mind, Figure~\ref{fig:stellar_parameter_comparison} shows two main features.  First, the M51 systems occupy the same broad temperature and mass ratio range as the Local Group samples.  Their median component temperatures are $T_{\rm hot}=33.5$ kK and $T_{\rm cool}=28.5$ kK, compared to $T_{\rm hot}=27.9$ kK for the published EB solutions and $T_{\rm hot}=34.9$ kK for the SB2 comparison sample.  The mass ratio distributions are also similar: the M51 median is $q=0.80$, while the EB and SB2 comparison samples both have median $q=0.77$.

The second major difference is that the M51 sample is shifted toward larger inferred radii, consistent with a population that is more evolved on average.  The M51 systems have median fitted radii $R_{\rm hot}=17.3~{\rm R_\odot}$ and $R_{\rm cool}=16.5~{\rm R_\odot}$.  The comparison EB samples are smaller in size, with median $R_{\rm hot}=5.2~{\rm R_\odot}$ and $R_{\rm cool}=5.5~{\rm R_\odot}$, and the SB2 sample has median $R_{\rm hot}=9.6~{\rm R_\odot}$ and $R_{\rm cool}=8.3~{\rm R_\odot}$.  Thus, the median hot-component radius in M51 is about $3.3\times$ larger than in the comparison EB sample and about $1.8\times$ larger than in the SB2 sample.  The high-radius tail is also much more prominent: $33\%$ of M51 systems have $R_{\rm hot}>20~{\rm R_\odot}$, compared to $0/99$ literature EBs and $4\%$ of SB2 systems.

The larger inferred radii are consistent with the selection and modeling of the M51 sample. The sample is drawn from F606W and F814W time series photometry, and massive-star luminosities in these red-optical bands depend strongly on evolutionary state (Figure~\ref{fig:cmd_selection_flow}, right panel). 
The $50\%$ completeness limits correspond to unevolved main-sequence masses of approximately $14~\msun$ in F606W and $29~\msun$ in F814W (Section~\ref{subsec:ASTs}). Requiring usable light curves in both filters therefore favors systems that are luminous in the red optical, such as evolved stars and the most luminous (massive) main-sequence stars. For example, stars with radii typical of the comparison EB samples ($\sim5~\rsun$) are difficult to detect at M51's distance.
Our catalog also likely includes contact and near-contact binaries, where isolated star models are likely inaccurate \citep{Menon2021,Henneco24}.
In addition, the M51 radii and temperatures are inferred from photometry alone, while many comparison EBs have spectroscopic and radial-velocity constraints. We therefore interpret the radius offset primarily as a selection and modeling effect.

The apparent high-mass tail ($\gtrsim50~\msun$) should be interpreted with similar caution, although some systems are likely legitimate. Among the $111$ fitted EBs in the modeled period range, $34$ ($30.6\%$) have $M_1\geq50~\msun$. Of the $50$ EBs that also lie in the blue CMD region used for the demographic inference (Section~\ref{sec:intrinsic_dem}), $13$ ($26\%$) exceed this mass. After applying the complete survey selection, the synthetic {\tt COSMIC} binary population predicts corresponding fractions of $13\%$ in total and $15\%$ within the blue CMD region. The observed population therefore contains more very massive stars than predicted, so this difference may partly reflect measurement scatter and systematic uncertainties in the fitted masses.
Nevertheless, {\tt COSMIC} still predicts approximately $14$ such systems in the sample, so some of the high-mass candidates are expected to be genuine.

This is also consistent with the high star formation rate of M51 compared to nearby galaxies.
M51 is forming stars at a $\sim10\times$ higher rate than the Magellanic Clouds, with recent estimates of $\sim2.4$--$3.4~\msun~{\rm yr^{-1}}$ over the last $10$--$100$ Myr \citep{Calzetti2005,Eufrasio2017,Messa2018}, compared to a long-term average of $\sim0.2~\msun~{\rm yr^{-1}}$ in the LMC and a lower rate in the SMC \citep[e.g.,][]{Harris2009,Bolatto2011}.  The HST footprint also contains $40\%$ of the stellar light in M51, amounting to a total stellar mass larger than the LMC. This makes the M51 survey sensitive to a large number of luminous and more massive candidates, but the individual masses are much less precise than those of spectroscopic surveys. Spectroscopy will be required to measure precise component masses, radii, and evolutionary states.

\section{Limitations}\label{sec:discussion}

\subsection{Third light}\label{subsubsec:third_light_discussion}

The light curve models used to infer binary parameters do not explicitly include third light: flux from any source other than the two eclipsing stars, which could include a bound tertiary or unbound chance alignment. Massive stars have high multiplicities, and close massive binaries are often members of higher-order systems \citep{Sana12,MoeDiStefano17,Offner2023}.  
Recently, \citet{Bordier2026} analyze $26$ Galactic hierarchical triples with O-type primaries and find that $10$ ($38\pm9\%$) host massive tertiaries with $q_{\rm out}=M_3/(M_1+M_2)>0.5$ within $\sim200$ au. 

We estimate the third-light contribution in our sample using the fitted masses of the $136$ M51 EBs. In each of $10{,}000$ Monte Carlo realizations, we draw the inner-binary masses from their fitted posteriors, assign a tertiary with probability $f_{\rm triple}=0.70$ \citep{MoeDiStefano17}, and draw
\begin{equation}
p(q_{\rm out})\propto q_{\rm out}^{-1.4},
\qquad
q_{\rm out}=\frac{M_3}{M_1+M_2},
\end{equation}
following \citet{Shariat2025} and consistent with \citet{Bordier2026}. We require $M_3\leq100~\msun$ and model the tertiary as a coextincted main-sequence star using the same atmosphere models as the binary fits. Defining the tertiary flux contribution $\ell_3=F_3/(F_1+F_2+F_3)$, we predict that $13 \pm 3\%$ of the fitted systems have a tertiary contribution of more than $20\%$ (i.e., $\ell_3 > 0.2$) of the total flux in F606W and F814W. Only $2 \pm 1.5\%$ contribute more than half of the total light.

Third light reduces an eclipse depth by a factor $1-\ell_3$. If the additional flux is attributed entirely to the two eclipsing stars at fixed temperatures, their inferred radii will be biased high by $(1-\ell_3)^{-1/2}$, corresponding to approximately $12\%$ for $\ell_3=0.20$ and $41\%$ for $\ell_3=0.50$. In the full light-curve--SED fit, this bias can also be distributed to produce inaccurate inclination, temperatures, and extinction \citep[e.g.,][for a discussion]{MoeDiStefano13}.

The properties of our survey help partially mitigate third light effects. 
HST point-source cuts remove obvious blends on $\sim1$~pc scales, and the WFC3 UV photometry limits contamination from cool companions. Hot tertiaries and unresolved young stars remain difficult to exclude. However, significant third light also dilutes the eclipses and makes strongly contaminated systems less likely to enter the EB sample. Together with our simulation, which predicts $\ell_3>0.20$ for only $13\pm3\%$ of the fitted systems, this suggests that third light may bias individual stellar parameters but is unlikely to dominate the population-level results.

\subsection{Massive star models}\label{subsubsec:models_discussion}

We use TLUSTY O/B non-LTE models to fit the atmospheric parameters of hot stars in our sample (e.g., Section \ref{subsec:parameter_estimates}). 
This grid does not include Wolf--Rayet (WR) or stripped star models, nor does it consider stellar winds. An order-of-magnitude estimate based on the Galactic WR population and its observed eclipse incidence \citep[$\sim6\%$;][]{Gamen2009} suggests that WR components should exist in only $\sim3$--$7\%$ of O-star EBs \citep{Rosslowe2015,Kroupa01,Sana12,Moe25LMC}. The fraction may be smaller in our sample because a compact WR star orbiting a luminous OB companion generally produces a shallow eclipse. 
Classical WR components should therefore affect only a small minority of the individual fits and are unlikely to alter our population-level results. 

We also compare representative $40$--$50$~kK hydrogen-free WNe and hydrogen-bearing WNL PoWR atmospheres \citep{Hamann2004,Todt2015,Hainich2019} with the TLUSTY models used in our fits. The differences in the relevant HST colors are $<0.06$~mag, smaller than the typical $0.1$--$0.15$~mag UV uncertainties. However, UV--optical colors change slowly with temperature above $\sim40$--$45$~kK \citep{Gull2026}. The hottest fitted systems are therefore likely genuinely hot, although their precise temperatures remain uncertain.

\section{Summary and Conclusions}\label{sec:conclusion}
Using 34 epochs of HST/ACS red-optical photometry and a single epoch of HST/WFC3 UV-to-optical photometry, we construct a panchromatic catalog of luminous stars in the Whirlpool Galaxy (M51): a nearly face-on, Milky Way-like spiral galaxy at $7.5$ Mpc. From this catalog, we construct a uniform census of massive eclipsing-binary candidates in M51, providing the first such catalog in a galaxy beyond the Local Group. We then characterize the selection function of our catalog to correct for incompleteness and constrain the demographics of the underlying massive binary population in M51.
Our main results are summarized as follows:

\begin{enumerate}
    \item {\it Catalog:} By combining $34$ epochs of HST/ACS light curves \citep{Conroy18} with a single epoch of broadband (UV-optical) HST/WFC3 imaging (Figure \ref{fig:m51_hst_footprint}), we construct a panchromatic census containing $\sim100{,}000$ of the most luminous stars in M51 (Figure \ref{fig:m51_gaia_eb_cmd} and \ref{fig:cmd_selection_flow}). This dataset provides a rich atlas for studying stellar variability.

    \item {\it Completeness:} We characterize the completeness of our HST survey and find $50\%$ completeness limits of $m_{\rm F606W}=27.10$ and $m_{\rm F814W}=25.96$ for ACS, and $m_{\rm F275W}=24.76$, $m_{\rm F336W}=25.56$, $m_{\rm F475W}=27.23$, $m_{\rm F606W}=27.05$, and $m_{\rm F814W}=25.87$ for WFC3 (Figure~\ref{fig:AST_completeness}). These limits comfortably include most main-sequence O stars and evolved B-type stars in M51.  Recovery is lower in the crowded spiral arms and galactic nucleus (Figure~\ref{fig:AST_spatial_completeness}).

    \item {\it EB sample:} After performing a periodicity search on all joint ACS F606W+F814W light curves and visually inspecting each candidate in the blue luminous part of the ACS CMD, we identify $173$ EB candidates with $P_{\rm orb}=1$--$30$~d (Figures~\ref{fig:visual_selection_examples} and \ref{fig:goldv3_periods}).  
    The observed period distribution is strongly weighted toward short periods due to geometric selection biases: $85\%$ of systems have $P_{\rm orb}<10$~d, and $70\%$ have $P_{\rm orb}<5$~d. The EB candidates also trace young stellar populations: their spatial distribution closely tracks luminous blue stars across M51 (Figure~\ref{fig:spatial_distribution}).

    \item {\it Stellar and binary parameters:} For the $136$ EB candidates with high-quality ACS and WFC3 photometry, we jointly fit the light curves and SEDs to constrain atmospheric and binary parameters (Figure~\ref{fig:example_fit}). 
    The fitted components span $T_{\rm eff}=12{,}000$--$50{,}000$~K (median $29{,}100$~K) and $R=7$--$66~\rsun$ (median $15~\rsun$; Figure~\ref{fig:fit_parameter_corner_cmd}). 
    Their locations in the HR diagram are consistent with luminous, often evolved hot stars, and their component ages are generally consistent (Figure~\ref{fig:HR_age}). The masses span $6$--$117~\msun$ (median $24.4~\msun$), although these model-dependent photometric masses are uncertain. The fitted mass ratios have a median of $q=0.8$ and span $q=0.3$--$1$.

    \item {\it EB selection function:} We quantify the selection function of our EB search using a forward model and extensive injection--recovery tests. Only $\sim5\%$ of massive binaries with $2\leq P_{\rm orb}/{\rm d}<20$ in the main-sequence CMD region are successfully recovered in our sample.
    The detection probability decreases steeply with period and lower inclination (Figure~\ref{fig:lc_selection_period_inclination}).

    \item {\it Intrinsic period distribution:} After applying the survey selection function, we find that the M51 close-binary period distribution is consistent with being roughly log-uniform ${\rm d}N/{\rm d}\log P_{\rm orb} \propto P_{\rm orb}^{0.29\pm0.32}$ over $P_{\rm orb}=2$--$20$~d (Figure~\ref{fig:intrinsic_period_dist}).    

    \item {\it Intrinsic binary fraction:} 
    We infer that
    $f_{\rm close}=27.4 \pm 7.1\%$
    of massive OB stars in M51 reside in binaries with $P_{\rm orb}=2$--$20$~d.

    \item {\it Comparison to nearby galaxies:} 
    We compare the close-binary fraction and period distribution in M51 to those in the Milky Way, LMC, and SMC (Figure~\ref{fig:demographic_comparison}). The close-binary fraction in M51 is consistent with that of O stars in the Milky Way and somewhat higher than estimates for O stars in the Magellanic Clouds. This may indicate a lower close-binary fraction at lower metallicity, although the current uncertainties prevent a firm conclusion.
    
    Across all four galaxies, the period distributions are consistent with being roughly uniform, or mildly declining, with $\log P_{\rm orb}$. B stars in the SMC show the strongest preference for short periods. 
    Together, these galaxies span metallicities from $\sim 0.2Z_\odot$ to $1.1Z_\odot$, suggesting that the formation of close massive binaries depends only weakly on metallicity over this range.  
    These results emphasize that binary interaction -- including mass transfer and mergers -- are a ubiquitous aspect of massive-star evolution across many environments.
    
\end{enumerate}

\section*{Acknowledgments} \label{acknowledgments}
We thank Andrew Dolphin for providing the updated DOLPHOT PSF libraries and for advice on their use.
C.S. acknowledges support from the Department of Energy Computational Science Graduate Fellowship.
This material is based upon work supported by the U.S. Department of Energy, Office of Science, Office of Advanced Scientific Computing Research, under Award Number DE-SC0026073. This research was supported by NSF grant AST-2540180.
C.C. acknowledges support through grant HST-GO-17200.003. K.E. acknowledges support through grant HST-GO-17200.004. 
Maude Gull is a Carnegie-Caltech Brinson Fellow supported by The Brinson Foundation.

Support for this work was provided by NASA through HST program GO-17200, administered by the Space Telescope Science Institute (STScI). STScI is operated by the Association of Universities for Research in Astronomy, Inc., under NASA contract NAS5-26555.
The computations presented here were conducted in the Resnick High Performance Computing Center, a facility supported by Resnick Sustainability Institute at the California Institute of Technology.

\section*{Data Availability}

The HST observations used in this work are publicly available from the
Mikulski Archive for Space Telescopes under programs 14704 and 17200.
The source and eclipsing-binary catalogs and calibrated ACS F606W and F814W light curves are available on Zenodo \citep{M51data}\footnote{\url{https://zenodo.org/records/22727123}} and described in Appendix~\ref{app:data_release}.

\software{
This work made use of \texttt{OverCite} \citep{Shariat2026}, an in-editor citation tool for \LaTeX.
This work made use of the following software packages: \texttt{Jupyter} \citep{2007CSE.....9c..21P,kluyver2016jupyter}, \texttt{matplotlib} \citep{Hunter:2007}, \texttt{numpy} \citep{numpy}, \texttt{pandas} \citep{mckinney-proc-scipy-2010,pandas_17806077}, \texttt{python} \citep{python}, \texttt{scipy} \citep{2020SciPy-NMeth,scipy_17467817}, \texttt{Cython} \citep{cython:2011}, \texttt{emcee} \citep{ForemanMackey2013,emcee_10996751}, \texttt{h5py} \citep{collette_python_hdf5_2014,h5py_7560547}, \texttt{schwimmbad} \citep{schwimmbad}, \texttt{seaborn} \citep{Waskom2021},  and \texttt{tqdm} \citep{tqdm_14231923}.
Software citation information aggregated using \texttt{\href{https://www.tomwagg.com/software-citation-station/}{The Software Citation Station}} \citep{software-citation-station-paper,software-citation-station-zenodo}.
}

\appendix
\twocolumngrid

\twocolumngrid

\section{Simulated Binary Population}\label{app:binary_pop}
We construct a synthetic massive binary population in M51 and forward model it through the survey's selection function, then compare it to the observed EB catalog. This connects the observed EB sample to an underlying population with physical masses, radii, luminosities, periods, and evolutionary states.

\subsection{Binary properties}
We generate the initial population with \texttt{COSMIC} \citep{Breivik2020}, a rapid binary population synthesis code based on the rapid binary stellar evolution prescriptions of \citet{Hurley2000,Hurley2002}.  Our fiducial model assumes a constant star formation rate of $3~\msun~{\rm yr^{-1}}$ \citep{Calzetti2005} over the past $100$ Myr\footnote{Different star formation tracers provide different SFR estimates, which are also spatially variable across M51 \citep{ThainaBatista2026}. The range of SFR over the past $100$~Myr in M51 is $2.5$--$3.5~\msun~{\rm yr^{-1}}$ \citep{Calzetti2005, Schuster2007, Eufrasio2017, Messa2018}.}, chosen to cover the lifetime of the lowest-mass stars in our luminous EB sample in M51.
We then consider that the ACS footprint covers only $\sim40\%$ of the total stellar light of M51 \citep{Conroy18}, giving a formed stellar mass of
\begin{equation}
    M_{\rm form} = \left(\frac{3~\msun}{{\rm yr}}\right)\times (10^8~{\rm yr})\times 0.40
    = 1.2\times10^8~\msun .
\end{equation}
The initial binaries are drawn with the \texttt{multidim.py} sampler native to {\tt COSMIC}, which uses the multidimensional binary distributions of \citet{MoeDiStefano17}, thus ensuring consistency with observed binary properties and multiplicity statistics. 
From the total $1.2\times10^8~\msun$ of formed stellar mass, the sampler generated $1.204\times10^8$ single stars and $4.15\times10^7$ initial binaries.

We only evolve binaries with initial primary mass $M_{1,{\rm init}}>5.5~\msun$, initial secondary mass $M_{2,{\rm init}}>5.5~\msun$, and initial orbital period $1\leq P_{\rm init}<1000$ d to keep it relevant for our sample of massive stars in close binaries.  This leaves $450{,}000$ binaries for \texttt{COSMIC} binary evolution.  Lower-mass binaries, wider binaries, and single stars are still used to construct summary statistics in later sections.  We assume solar metallicity \citep[e.g.,][]{Croxall2015}, the \citet{Vink2011} massive-star wind prescription (\texttt{windflag=3}), and common-envelope efficiency $\alpha_{\rm CE}=1$.  These settings are held fixed for the forward model comparison.
We also re-evaluated the binary evolution after varying several prescriptions -- common-envelope efficiency, non-conservative mass-transfer efficiency, critical mass ratio prescription, wind prescription, and accretion limit -- finding that these choices produce no discernible impact on our final conclusions presented in Section \ref{sec:intrinsic_dem}.

We only consider the binaries that do not merge, do not contain compact objects, and are within the observed sample's period range ($1\leq P_{\rm orb}/{\rm d}<30$) at the present day.  
For each binary, synthetic HST photometry is derived by interpolating the same hybrid TLUSTY/BaSeL atmosphere grid used for the EB fitting (Section \ref{subsec:parameter_estimates}), where the unresolved system magnitude is obtained by summing the two component fluxes. 

\subsection{Spatial Distribution}

We draw positions across the observed ACS footprint with probability
\begin{equation}
p(R)\propto \exp(-R/R_\star),
\qquad R_\star=3.27~{\rm kpc}.
\end{equation}
The adopted scale length $R_\star$ is based on fits to observed $R$-, $I$-, and $K$-band images of M51 \citep[$90\arcsec$;][]{Beckman1996}. We use projected circular radii because M51 is nearly face-on.

\begin{figure}
    \centering
    \includegraphics[width=0.99\columnwidth]{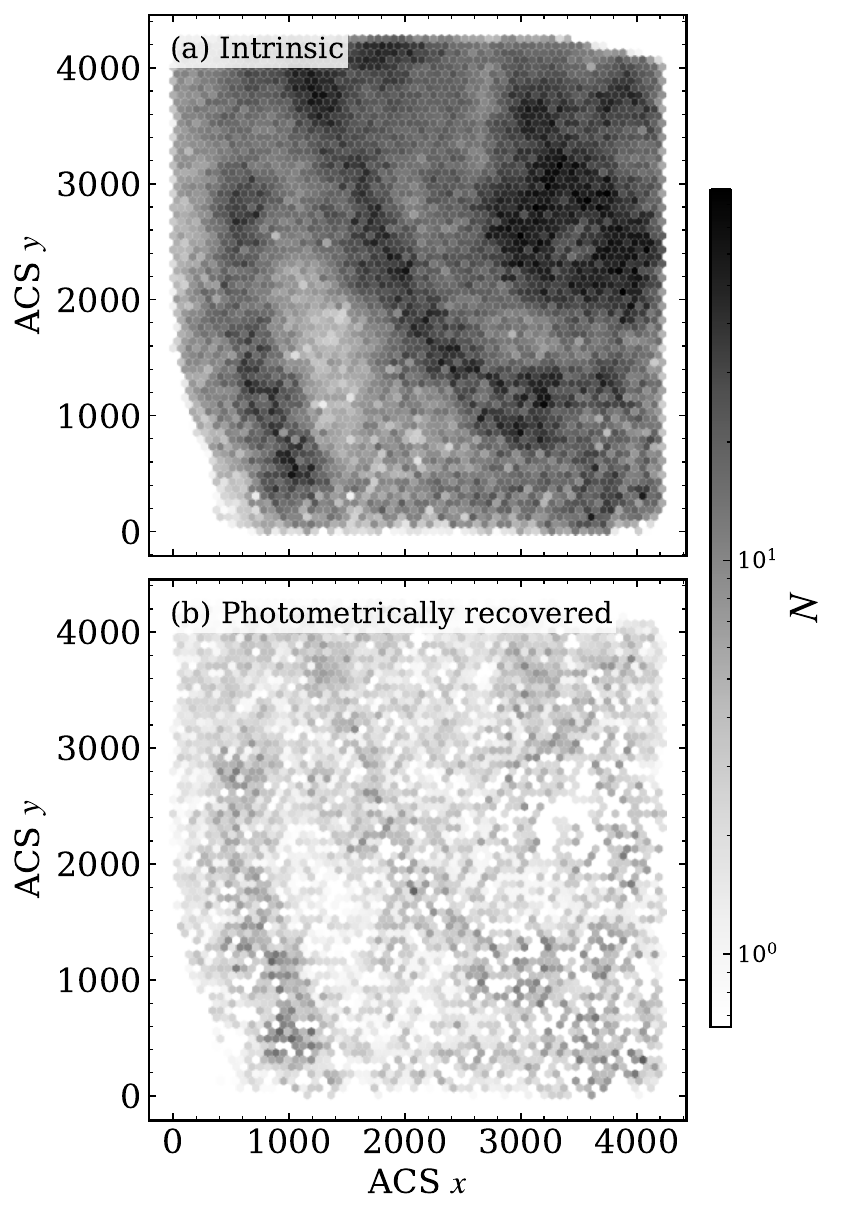}
    \caption{Spatial distribution of the simulated close binaries before (top) and after (bottom) photometric selection. The intrinsic systems follow an exponential disk + spiral-arm stellar distribution.}
    \label{fig:spatial_distribution_selection}
\end{figure}

We draw the vertical position of each massive binary from the young stellar disk,
\begin{equation}
p(z_\star)=\frac{1}{2h_{z,\rm young}}
\exp\left(-\frac{|z_\star|}{h_{z,\rm young}}\right),
\end{equation}
using $h_{z,\rm young}=100~{\rm pc}$.
These values are based on measurements of young stellar disks and 3D radiative transfer fits to edge-on galaxies
\citep{Bahcall1980,Xilouris1999,Bianchi2007,SchechtmanRook2013,DeGeyter2014, DeLooze2014}.
Lastly, we convolve this exponential profile with a Gaussian kernel density estimate (KDE) constructed using the blue population from our ACS sample and a physical length scale $200$~pc. This concentrates more sources near the spiral arms and nucleus, where most of the young stellar population resides (Figure \ref{fig:spatial_distribution_selection}, top).

For an observer on the positive-$z$ side of the disk, the fraction of the vertical dust column in front of a source is
\begin{equation}
f_d(z_\star)=
\begin{cases}
\frac{1}{2}\exp(-z_\star/h_{z,d}), & z_\star\geq0,\\
1-\frac{1}{2}\exp(z_\star/h_{z,d}), & z_\star<0.
\end{cases}
\end{equation}
We adopt a scale height $h_{z,d}=225$~pc and assign the internal extinction according to
\begin{equation}
A_{V,\rm int}
=
c\,f_d(z_\star)\,\sec i\,A_{V,10}
\exp\left[-\frac{R-10~{\rm kpc}}{R_{\rm dust}}\right],
\end{equation}
where $\ln c\sim\mathcal{N}(-\sigma_{\ln c}^{2}/2,\sigma_{\ln c}^{2})$ introduces unresolved spatial variation and $i=20^\circ$ for M51 \citep[][]{Hu2013}. We adopt $A_{V,10}=1.5$ mag, $R_{\rm dust}=5$ kpc, and
$\sigma_{\ln c}=0.5$, and add Milky Way foreground extinction of $A_{V,\rm MW}=0.086$ mag. The choice of parameterization is again motivated by 3D radiative transfer modeling and resolved extinction maps of M51 \citep{DeLooze2014,FaustinoVieira2023}. This prescription produces a median total extinction of $A_V=0.79$ mag, with a $1\sigma$ range of $0.40$--$1.52$ mag, for the intrinsic massive close-binary population.

The broadband HST photometry for each synthetic binary is derived by summing the component fluxes at the adopted distance of M51 ($7.50$~Mpc).
For each simulated source, we then identify the $K=25$ nearest ASTs in the four-dimensional space of spatial pixel ($x, y$), magnitude ($m_{\rm F814W}$), and color ($m_{\rm F606W}-m_{\rm F814W}$).
Including position allows the ASTs to capture the strongly spatially varying effects of crowding and background, while the magnitude and color describe the source-dependent detection probability. Matching the synthetic binaries to ASTs also assigns to them realistic DOLPHOT quality metrics.
Figure~\ref{fig:spatial_distribution_selection} shows the intrinsic spatial distribution and the subset recovered after photometric selection.

We then assign an inclination to each binary from an isotropic distribution, inject them into ACS light curves using the carrier stars, perform a Lomb-Scargle period search, and visually inspect each source (see Section \ref{subsec:selection_function} for details).

\begin{figure*}
    \centering
    \includegraphics[width=0.98\textwidth]{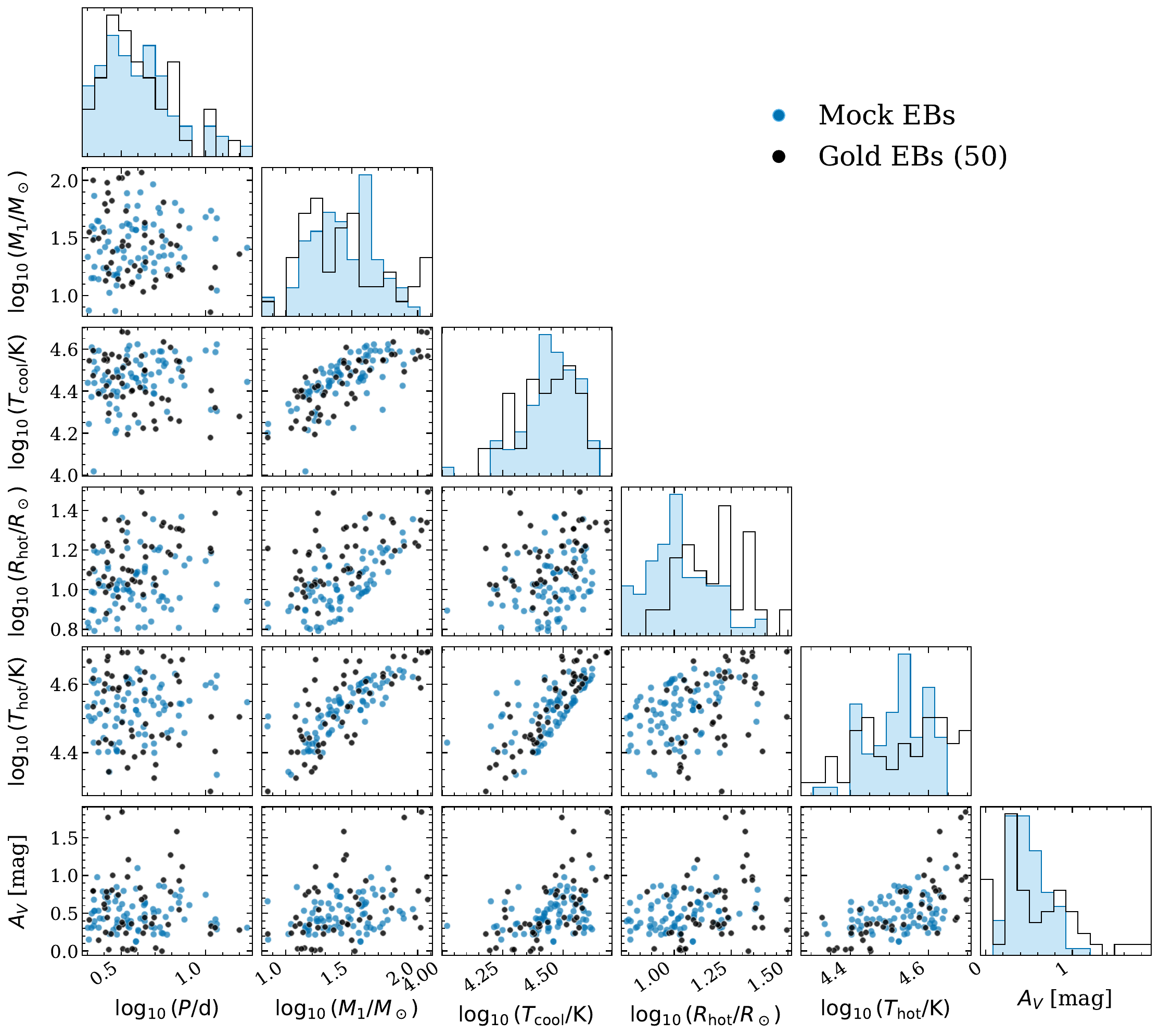}
    \caption{Comparison between the forward-modeled
    binary population (blue) and the $50$ Gold EBs used for the primary demographic inference (black). All systems satisfy
    $2\leq P_{\rm orb}/{\rm d}<20$, $F336W<24$, and lie blueward of the TAMS in the $F336W-F606W$ vs. $F336W$ CMD. Diagonal panels show normalized 1D distributions. Off-diagonal panels show 2D scatter plots. The model spans similar parameter ranges to the observed sample, although it favors somewhat smaller radii.}
    \label{fig:cosmic_parameter_check}
\end{figure*}

\subsection{Comparison with the observed sample}
\label{app:cosmic_parameter_check}

As a check on the forward model, we compare the fully selected
\texttt{COSMIC} population with the same $50$ Gold EBs used for the primary demographic inference in Figure~\ref{fig:cosmic_parameter_check}. These systems lie blueward of the TAMS on the $F336W-F606W$ vs. $F336W$ CMD.
We show binaries from our synthetic population that would be selected in our survey, meaning that they pass the photometric, CMD, eclipse geometry, period recovery, and visual selection stages. 

The selected \texttt{COSMIC} population spans the observed period, mass, temperature, radius, and extinction distributions. The mock and observed samples have median periods of $3.63$ and $3.31$~d, primary masses of $24.7$ and $27.1~\msun$, and hot-component temperatures of $33.2$ and $35.3$~kK, respectively. Their median extinctions are nearly identical
($A_V=0.424$ and $0.422$~mag), although the simulated distribution is slightly narrower. The largest difference is in the hot-component radius, with medians of $10.0$ and $13.6~\rsun$ for the simulated and observed populations. The 16th--84th percentile intervals nevertheless overlap for every parameter shown. The similarities support the use of the forward model for demographic inference.

\section{Temperature--extinction covariance}\label{app:av_teff}

\begin{figure}
    \centering
    \includegraphics[width=0.96\columnwidth]{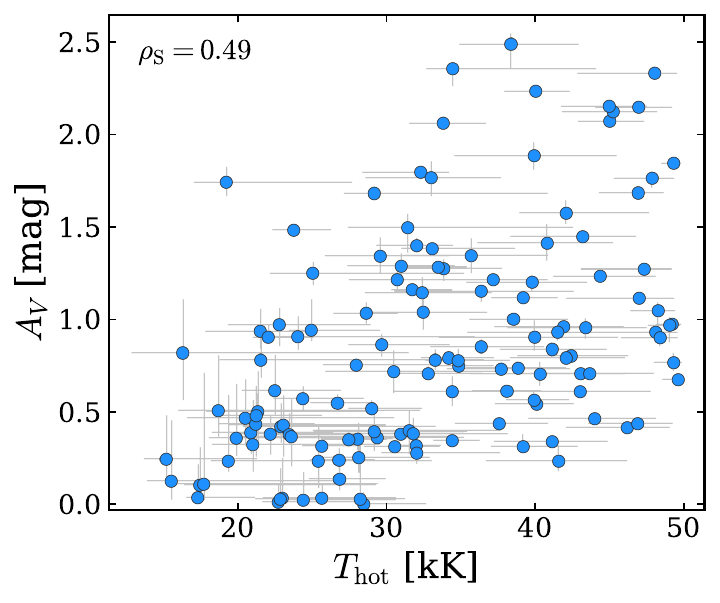}
    \caption{Extinction as a function of the temperature of the hotter component for the M51 EB candidates. Points show the posterior medians, with error bars spanning the 16th--84th percentiles. The weak correlation ($\rho_{\rm S}=0.49$) and the presence of hot systems across a broad range of $A_V$ indicate that the highest fitted temperatures are likely intrinsic, and not exclusively driven by large extinction.}
    \label{fig:av_teff_appendix}
\end{figure}

Broad-band SED fits can produce correlated temperature and extinction estimates, because a hotter, more reddened spectrum can reproduce similar observed colors to a cooler, less-attenuated one. Figure~\ref{fig:av_teff_appendix} tests whether this degeneracy is responsible for the highest temperatures inferred for the M51 EB candidates. We compare the fitted extinction with the temperature of the hotter component for all systems with joint light curve+SED fits.

The fitted values show only a moderate positive correlation, with a Spearman coefficient of $\rho_{\rm S}=0.49$.  Several systems with $T_{\rm hot}\gtrsim40$ kK have modest extinction, and systems with similar temperatures span more than $2$ mag in $A_V$. Although some covariance between $T_{\rm hot}$ and $A_V$ is expected, the broad scatter and the presence of high-temperature systems at modest extinction indicate that the hottest inferred temperatures are not driven solely by the temperature--extinction degeneracy.

\section{X-ray Cross-match}\label{sec:xray}

We cross-match the $258{,}906$ ACS sources that pass the DOLPHOT point-source cuts in F606W and F814W to the $503$ Chandra/ACIS point sources reported by \citet{Kuntz16}.  For each X-ray source, we identify the nearest ACS source, finding $22$ counterparts within $1\sigma$ of the Chandra position and $86$ within $3\sigma$, with median separations of $0.047\arcsec$ and $0.094\arcsec$, respectively.  The $3\sigma$ sources span $\log_{10}(L_X/{\rm erg~s^{-1}})=36.05$--$37.90$, with a median of $36.55$.  The nearest source is frequently not unique: $56/86$ X-ray sources have at least one additional ACS source inside the same $3\sigma$ region.

We measure the chance alignment rate by shifting all 503 Chandra positions by one random $5$--$20\arcsec$ vector and repeating the match 200 times.  Each shifted catalog preserves the relative distribution of the X-ray sources and samples the same crowded ACS field, but removes any original matches.  The shifted catalogs give $22.6\pm3.9$ matches within $1\sigma$ and $86.0\pm5.5$ within $3\sigma$, consistent with the observed counts of 22 and 86.  Restricting the optical parent to UV-selected massive star candidates gives the same conclusion: 37 sources are observed within $0.5\arcsec$, while 16 fixed offset catalogs give a mean of 30.7 and a range of 18--46.  Positional agreement alone therefore does not produce a clean optical counterpart sample in the M51 field.

This source confusion is also found in previous M51 studies.  \citet{Rice2021} found at least one HST source within $0.5\arcsec$ for $173/334$ X-ray sources; $88/173$ contained multiple optical candidates, and the nearest and brightest candidates agreed for only $113/173$.    These results show that positional cross-matches for HMXBs are often confounded by chance superposition. The underlying Chandra catalog contains LMXBs, background AGN, foreground stars, supernova remnants, and compact features in the diffuse hot gas in addition to HMXBs \citep{Kuntz16,Lehmer2017}.  \citet{Lehmer2017} reduced the soft-source contamination by selecting 86 sources detected at 2--7 keV and estimated that approximately 6--10 objects over the M51 footprint were background sources.  A nearby blue star can therefore be an unrelated neighbor of an LMXB or AGN; X-ray spectra and variability, improved relative astrometry, or optical spectroscopy are required to establish an HMXB counterpart.

No EB candidate lies inside the formal $3\sigma$ position error of its nearest Chandra source.  Sources 149974 and 107202 are separated from X280 and X367 by $0.335\arcsec$ and $0.374\arcsec$, corresponding to $4.19\sigma$ and $4.15\sigma$, and both X-ray positions have a closer ACS source.  Adding an illustrative, not measured, $0.1\arcsec$ relative astrometry uncertainty in quadrature brings both EBs inside an expanded $3\sigma$ region.  Their status therefore depends on the relative HST--Chandra frame registration.  Source 3608 is $0.951\arcsec$ from X282, or $19\sigma$, and is excluded.  We retain sources 149974 and 107202 as lower-confidence follow-up targets, not X-ray-binary identifications.

\begin{figure}
    \centering
    \includegraphics[width=0.96\columnwidth]{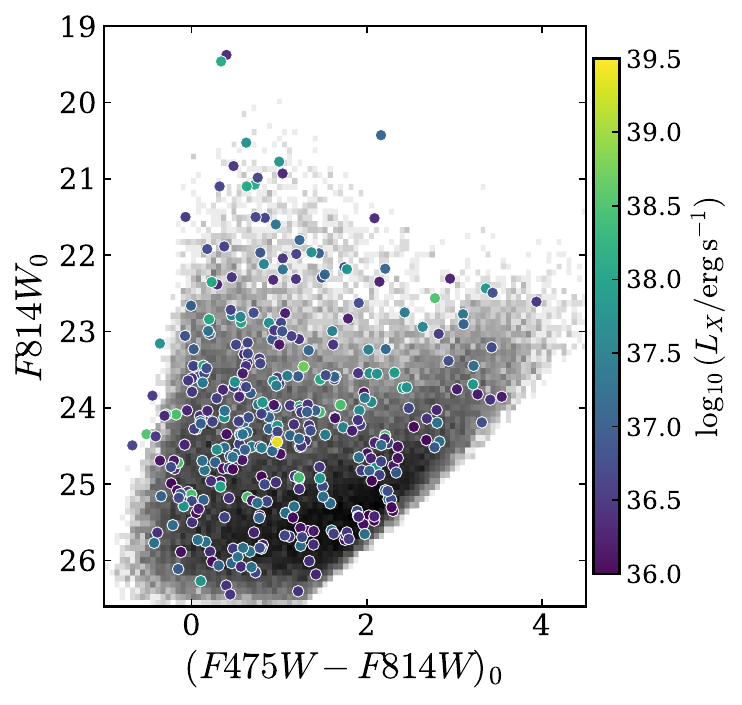}
    \caption{WFC3 CMD of point sources within $0.5\arcsec$ of a cataloged Chandra source.  The grayscale background shows all WFC3 sources that pass the DOLPHOT point-source cuts in F475W and F814W.  Colored points show the 345 ACS--Chandra matches that pass the same cuts, colored by the cataloged X-ray luminosity.  Magnitudes are corrected for Milky Way foreground extinction.  The broad CMD distribution illustrates that the X-ray matches have possible optical counterparts spanning a wide range of colors and evolutionary states.}
    \label{fig:xray_cmd}
\end{figure}

\section{Constraining close-binary demographics in nearby galaxies from RV surveys}\label{app:rv_comparison}

RV surveys generally report massive-star binary demographics over a broad period range, roughly $1$--$3000$~d. Here we use published O-star RV data to fit the period distributions in the MW, LMC, and SMC over the $2$--$20$~d interval probed by our M51 analysis. We also infer the close-binary fractions in the LMC and SMC.

\subsection{Milky Way}

For Galactic O stars, we fit the $21$ orbital periods between $2$ and $20$~d in the \citet{Sana12} sample, following the methods of \citet{MoeDiStefano13}. We assume the period distribution takes the form of Equation \eqref{eq:period_law}. A delete-one jackknife estimate gives $\pi=-0.45\pm0.34$, consistent with \citet{MoeDiStefano13}. We adopt $f_{\rm close}=31\pm7\%$ from \citet{MoeDiStefano13}.

\subsection{Magellanic Clouds}

For O stars, we use the published epoch RVs and uncertainties for $139$ SMC targets \citep{Sana25SMC} and $360$ LMC targets. The LMC sample contains $332$ stars observed with FLAMES/Medusa \citep{Sana2013} and $28$ observed with ARGUS \citep{HenaultBrunet2012}.

We apply the same RV-variability criterion to both samples. A target is counted as variable when the same pair of epochs differs by both $>20~{\rm km\,s^{-1}}$ and more than $4$ times the combined uncertainty. Each target is then assigned either to a non-detection category or to a bin in peak-to-peak RV amplitude and the shortest time separation between qualifying epochs. The amplitude-bin edges are
$0,40,80,160,\infty~{\rm km\,s^{-1}}$, and the time-separation edges are
$0,1,3,10,30,100,300,\infty$~d. This gives $28$ detection categories plus one non-detection category.

\subsection{Population likelihood and uncertainties}

We simulate RV observations to account for the observing cadence and measurement uncertainties, following the approach of \citet{Sana2013,Dunstall2015}. For each target, we calculate primary-star RVs at its observing times and add Gaussian measurement errors. We then apply the same criteria used to classify the observed RV variations, including non-detections.
These simulations give the probability $R_{ib}(c_i)$ of obtaining target $i$'s observed category $c_i$ if it is a binary in period interval $b$. We also simulate single stars with measurement noise to obtain the corresponding probability $R_{i0}(c_i)$. We infer the binary fractions by combining these possibilities for every target:
\begin{equation}
\mathcal{L}=\prod_i\left[\left(1-\sum_b f_b\right)R_{i0}(c_i)
             +\sum_b f_b R_{ib}(c_i)\right],
\label{eq:rv_comparison_likelihood}
\end{equation}
where $f_b$ is the fraction of surveyed systems that are binaries in period interval $b$. The first term accounts for single stars, and the second sums over the binary period intervals. Each set of response probabilities sums to unity over all categories, including non-detections. We fit the period intervals jointly, without assigning an orbital period to each RV-variable star.

We report the binary fraction and period slope over $2$--$20$~d, adopting $dN/d\log P_{\rm orb}\propto P_{\rm orb}^{\pi}$ with a uniform prior on $\pi\in[-2,2]$. We also model binaries at shorter and longer periods, since they can contribute to the observed RV variations. Figure~\ref{fig:demographic_comparison} shows the $2$--$20$~d results after marginalizing over these other components.
We assume isotropic orbital inclinations and draw orbital phases and arguments of periastron uniformly. The mass-ratio distribution is $p(q)\propto q^{\kappa}$. We adopt a uniform prior on $\kappa\in[-2.5,2.5]$.
We sample the posterior with \texttt{emcee} \citep{ForemanMackey2013}, marginalizing over the other period components and the mass-ratio distribution.

\section{Data release}\label{app:data_release}

The Zenodo release contains three machine-readable tables \citep{M51data}. The source catalog contains positions, photometry, uncertainties, and quality flags for 650,191 sources, including seven manually retained EB candidates outside the original quality-selected catalog. The light-curve table contains ACS F606W and F814W measurements in 34-element arrays per source and filter. The EB catalog contains adopted periods for 173 candidates and posterior parameter summaries for 136 fitted systems.

All three tables share the same source identifier. In the EB catalog, component 1 is the hotter star by median fitted temperature; component masses and radii follow these labels. All magnitudes use the infinite-aperture VEGAMAG system and are corrected for Milky Way foreground extinction. The accompanying {\tt README} describes the columns, units, and how to read the files.

\section{All Phase-folded Light Curves}\label{app:all_lightcurves}

Figures~\ref{fig:appendix_lcs_1}--\ref{fig:appendix_lcs_4} show the ACS phase-folded light curves for the 173 Gold EB candidates, sorted by orbital period.  Each panel uses the same convention, where black and blue points show empirically corrected F606W and F814W photometry, respectively, folded on the adopted orbital period, shifted so the deepest F814W eclipse is near phase 0.5, and repeated over two orbital cycles. 

\begin{figure*}
    \centering
    \includegraphics[width=0.99\textwidth]{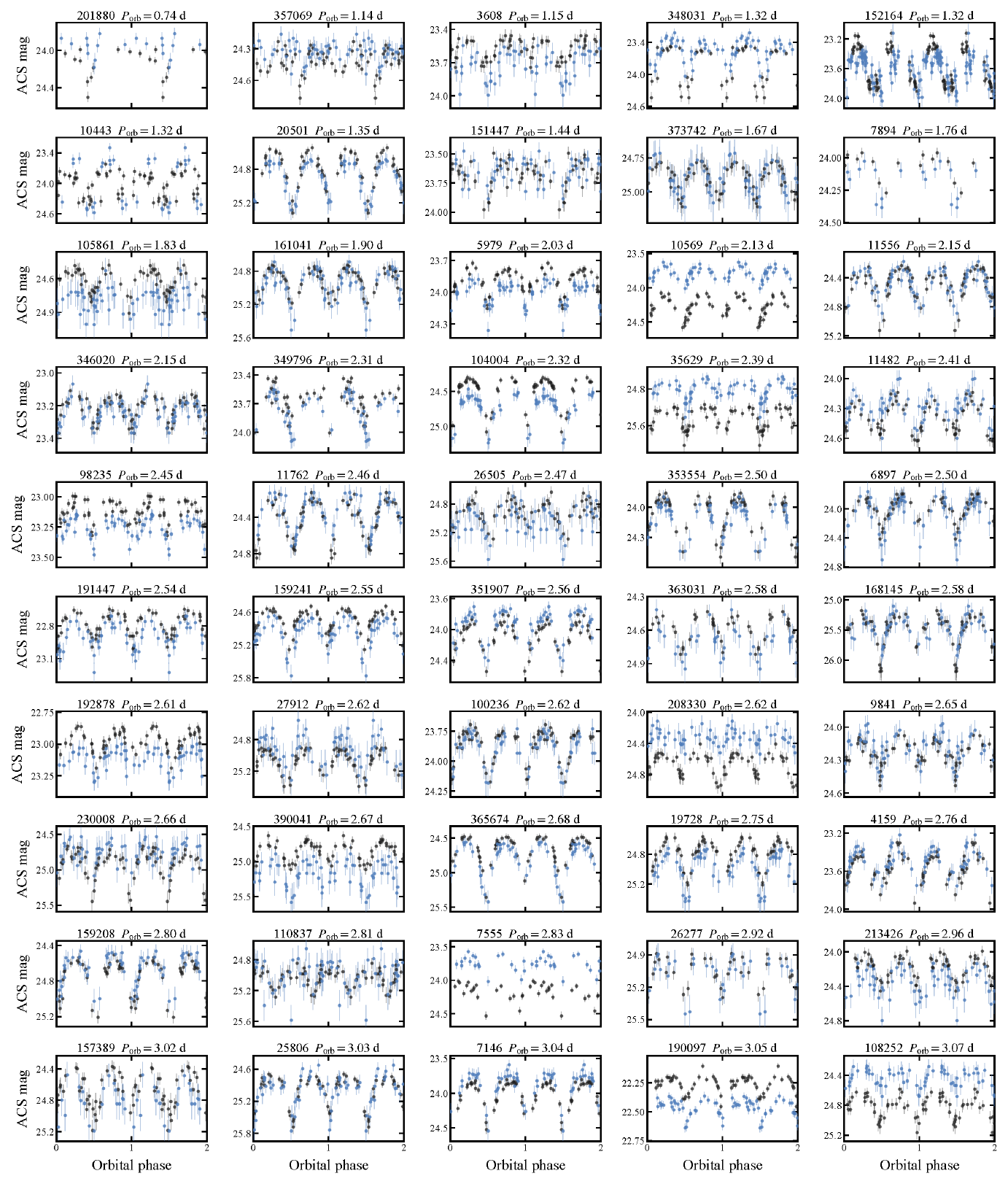}
    \caption{ACS phase-folded light curves for Gold EB candidates, sorted by increasing orbital period.  Black points show F606W and blue points show F814W, folded on the adopted orbital period and repeated over two cycles.}
    \label{fig:appendix_lcs_1}
\end{figure*}

\begin{figure*}
    \centering
    \includegraphics[width=0.99\textwidth]{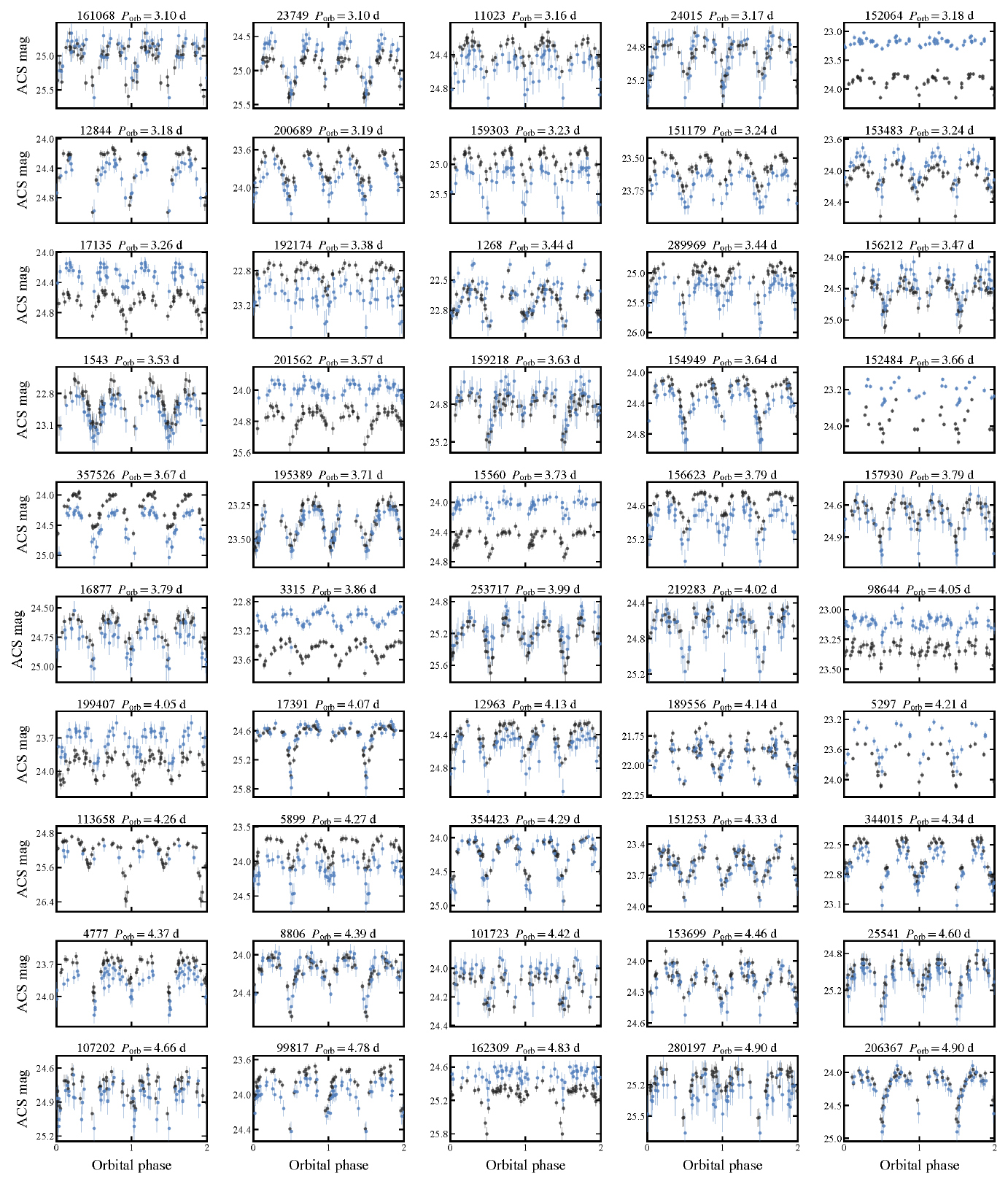}
    \caption{continued}
    \label{fig:appendix_lcs_2}
\end{figure*}

\begin{figure*}
    \centering
    \includegraphics[width=0.99\textwidth]{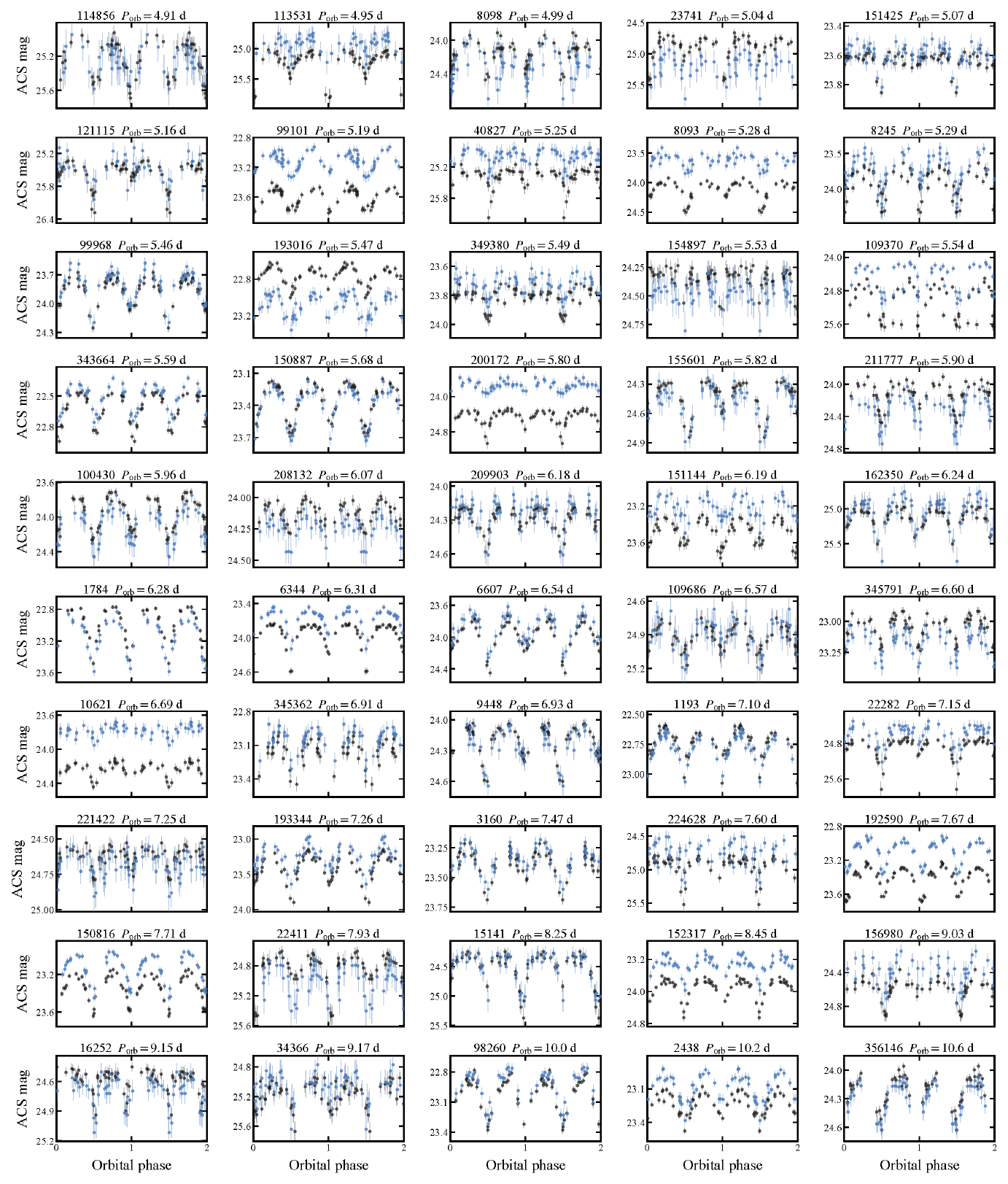}
    \caption{continued}
    \label{fig:appendix_lcs_3}
\end{figure*}

\begin{figure*}
    \centering
    \includegraphics[width=0.99\textwidth]{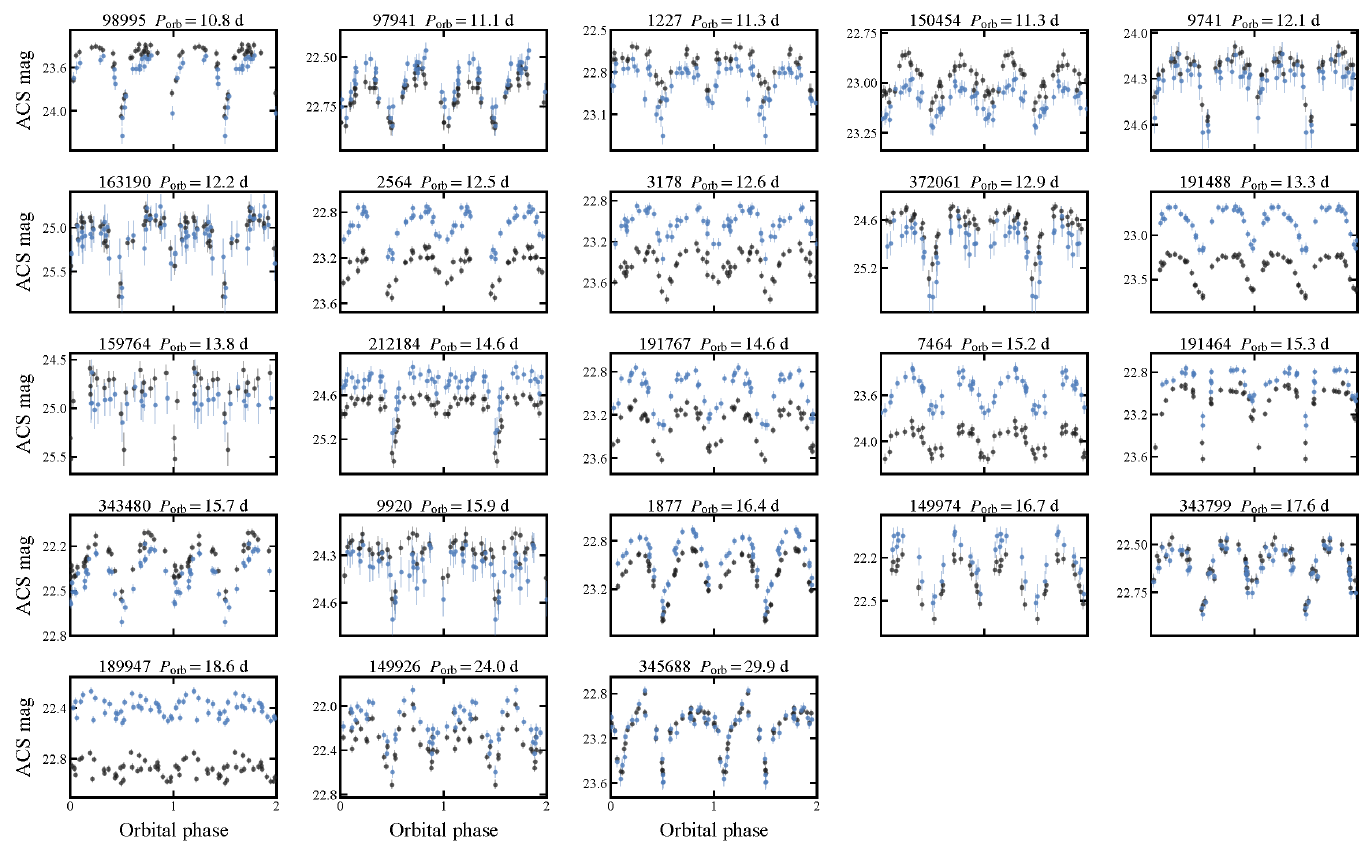}
    \caption{continued}
    \label{fig:appendix_lcs_4}
\end{figure*}

\clearpage

\bibliography{references}

\end{document}